\pdfoutput=1
\documentclass[12pt]{revtex4}

\usepackage{graphicx}
\usepackage{bm}
\usepackage{amsmath}
\usepackage{flushend}
\usepackage{color}
\usepackage{hyperref}

\DeclareMathOperator{\sinc}{sinc}

\begin{document}

\title{Superradiant and stimulated-superradiant emission of bunched electron beams
}

\author{A. Gover, University of Tel-Aviv, Tel-Aviv, Israel\\
        R. Ianconescu, University of Tel-Aviv, Tel-Aviv and Shenkar College, Ramat Gan, Israel\\
        A. Friedman, Ariel University, Ariel\\
        C. Emma, N. Sudar, P. Musumeci, C. Pellegrini, UCLA, Los Angeles}

\begin{abstract}
We outline the fundamental coherent radiation emission processes from a bunched charged particles beam. In contrast to spontaneous emission of radiation from a random electron beam that is proportional to the number of particles, a pre-bunched electron beam can emit spontaneously coherent radiation proportional to the number of particles - squared, through the process of (spontaneous) superradiance (SP-SR) (in the sense of Dicke's). The coherent SP-SR emission of a bunched electron beam can be even further enhanced by a process of stimulated-superradiance (ST-SR) in the presence of a seed injected radiation field. In this review, these coherent radiation emission processes for both single bunch and periodically bunched beams are considered in a model of radiation mode expansion.

We also analyze here the SP-SR and ST-SR processes in the nonlinear regime, in the context of enhanced undulator radiation from a uniform undulator (wiggler) and in the case of wiggler Tapering-Enhanced Stimulated Superradiant Amplification (TESSA).

The processes of SP-SR and TESSA take place also in tapered wiggler seed-injected FELs. In such FELs, operating in the X-Ray regime, these processes are convoluted with other effects. However these fundamental emission concepts are useful guidelines in efficiency enhancement strategy of wiggler tapering.

Based on this model we review previous works on coherent radiation sources based on SP-SR (coherent undulator radiation, synchrotron radiation, Smith-Purcell radiation etc.), primarily in the THz regime and on-going works on tapered wiggler efficiency-enhancement concepts in various frequency regimes.
\end{abstract}
\maketitle

\newpage
{\large List of abbreviations}

\bigskip
\bigskip

\begin{tabular}{| l | l |}
\hline
CSR    & Coherent Synchrotron Radiation \\
CTR    & Coherent Transition Radiation \\
EEHG   & Echo-Enabled Harmonic Generation \\
ERL    & Energy Retrieval LINAC \\
FEL    & Free electron laser \\
HGHG   & High Gain Harmonic Generation \\
IR     & Infrared \\
KMR    & Kroll Morton Rosenbluth \\
LINAC  & Linear accelerator \\
PEHG   & Phase-merging Enhanced Harmonic Generation \\
SASE   & Self-Amplified Spontaneous Emission \\
SP-SR  & Spontaneous Superradiance \\
ST-SR  & Stimulated Superradiance  \\
TES    & Tapering-Enhanced Superradiance \\
TESSA  & Tapering-Enhanced Stimulated Superradiant Amplification \\
TESSO  & Tapering Enhanced Stimulated Superradiant Oscillator \\
THz    & Terahertz \\
UR     & Undulator radiation \\
\hline
\end{tabular}

\bigskip
\bigskip

{\large Note:} The terms ``wiggler'' and ``undulator'' are interchangeably used along this manuscript, as are the terms
``superradiance'' and ``coherent spontaneous radiation''.

\section{Introduction}



Accelerated free electrons emit electromagnetic radiation when subjected to an external force (e.g. synchrotron radiation \cite{Carr_2002,Abo-Bakr_2002,Sannibale_2004,Byrd_2004,Adams_2004,Hirschmugl_1991,Wang_1998,Tamada_1993,Andersson_2000,Arp_2001,Carr_2001,Giovenale_1999,Nodvick_1954,Berryman_1996,Byrd_2002,Michel_1981,Krinsky,Green,Geloni}, Undulator radiation \cite{Doria_1993,Jaroszynsky_1993,Kuroda_2011,McNeil_1999,Ciocci_1993,Gover_1994,Faatz_2001,Jeong_1992,Asakawa_1994,Bonifacio_1984,Gover_1987,Pinhasi_2002,Neumann_2003,Arbel_2001,Arbel_2000,Neuman_2000,Huang_2014,Huang_2015,Huang_2007,Seo_2013,Lurie_2007,Watanabe_2007,Hama_2011,Musumeci_2013,Cohen,Mayhew}, Compton scattering \cite{gover_sprangle}). Radiation can also be emitted by currents that are induced by free electrons in polarizable structures and materials, such as in Cerenkov radiation \cite{Neighbors,Wiggins}, transition radiation \cite{Happek_1991,Lihn_1996,Piestrup,Shibata_1994,Leemans,Orlandi,Geloni_2009} and Smith-Purcell radiation \cite{Shin_2007,Korbly_2005,Brownell_1998,Ginzburg_2013}. Of most current interest nowadays are Free Electron lasers (FEL), a most potent intense coherent radiation source that can operate in a wide range of radiation wavelengths from microwaves to X-Rays (see recent review in this journal \cite{Pellegrini_2016, Bostedt_2016,Feng_2018}).

Here we use the laser physics terminology of stimulated interaction and spontaneous emission by atomic radiators - namely: stimulated emission/absorption is the radiation field amplification/attenuation of an incident radiation field, and spontaneous emission is the radiation emission of the particulate radiators in the absence of incident radiation field. The laser physics quantum description of free electron radiation sources reduces to the classical point-particle description of radiation emission by electrons in acceleration/deceleration structures, including analogous fundamental (Einstein) relations between spontaneous and stimulated emission \cite{Madey_1979,Friedman_1988,Pan_2018}. In the present context both spontaneous and stimulated interaction of electrons with radiation are treated in the classical point-particle limit of force equations and Maxwell equations.

Contrary to FEL, that by its definition as a laser is a stimulated radiation emission device, and is based on a continuous stream of accelerated electron, the focus of the present review is free electron radiation devices that emit intense coherent spontaneous (superradiant) radiation without the fundamental process of stimulated emission. This is possible in all the above mentioned radiation schemes, if the beam is pre-bunched before entering the radiative interaction region (in the case of a pre-bunched beam FEL - a magnetic undulator). Namely, such radiation sources emit coherent radiation without a coherent input radiation field (as required in a laser). However, as discussed later on, the coherent spontaneous radiation field can still be further amplified by stimulated emission if an external coherent input radiation field is inserted.

\begin{figure*}[!tbh]
\includegraphics[width=15cm]{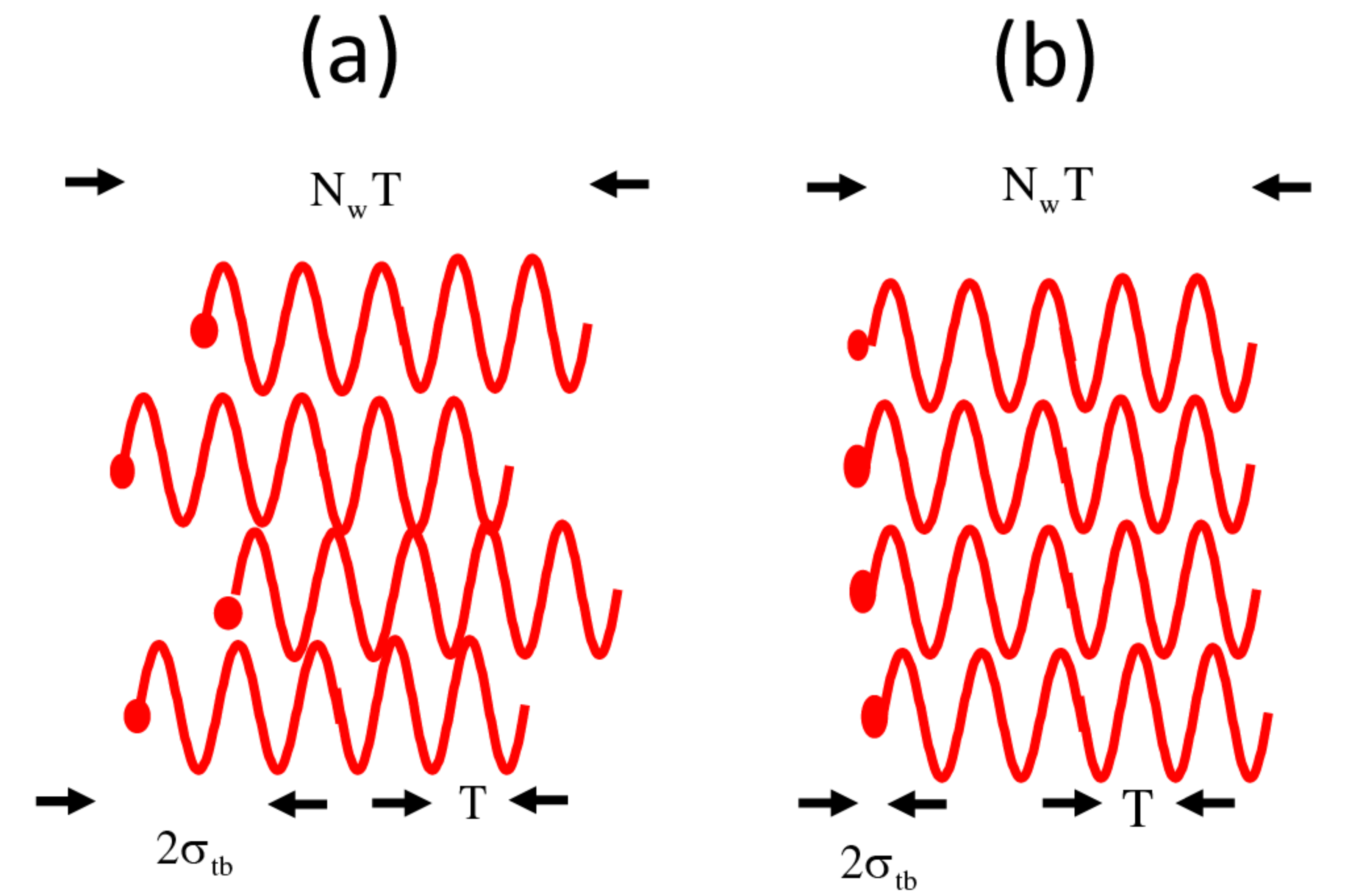}
\caption{(a) Spontaneous radiation emission (b) Superradiant emission (coherent spontaneous emission)}
\label{spontaneous_radiation_emission_and_superradiant_emission}
\end{figure*}

The condition for the generic coherent spontaneous superradiance process is:
\begin{equation}
2\sigma_{tb}<T=2\pi/\omega
\label{condition_generic_coherent}
\end{equation}
where $\omega$ is the radiation emission frequency, and $2\sigma_{tb}$ is the duration of the electron beam bunch. The process is visualized in Figure~\ref{spontaneous_radiation_emission_and_superradiant_emission}
as a time-interference of a train of radiation waves emitted by the electrons in a bunch, and observed, with some retardation and Doppler shift, at a long distance away from the emission point. Each electron emits in any specific direction radiation wavepackets of frequency $\omega=2\pi/T$ and duration $N_wT$, where $N_w$ is the number of wiggling oscillations in the interaction length. The spontaneously emitted radiation fields of the different electrons add coherently in phase if the electron beam bunch is shorter than the emitted radiation period (Eq.~\ref{condition_generic_coherent}, Figure~\ref{spontaneous_radiation_emission_and_superradiant_emission}b), and the resultant field is proportional to the number of electrons in the bunch - $N$. Consequently, the intensity of the radiation of a pre-bunched beam is proportional to $N^2$. This is in contrast to spontaneous emission from a randomly distributed beam in a long pulse (opposite of Eq.~\ref{condition_generic_coherent}), where the radiation intensity is proportional to the number of electrons in the beam ($N$). This coherent radiation process is analogous to Dicke's superradiance from an ensemble of stationary atoms located within a volume smaller than their spontaneous emission radiation wavelength, and excited so that their dipole moments emit in phase with each other \cite{dicke, Gross_1982}. While Dicke's analysis starts from a fundamental Quantum-Electrodynamics (QED) formulation, he showed that this process is valid also in the classical limit. The difference between a bunched electron beam and Dicke's ensemble of oscillating dipoles is only the movement of the electron bunch in the axial dimension. This provides in the relativistic beam velocity limit, large Doppler frequency shift of the radiation emitted in the forward direction.

As mentioned, superradiant emission from a single electron bunch beam takes place when the beam enters the interaction region of the radiative emission scheme with duration  shorter than the period of the radiation wave (Eq.~\ref{condition_generic_coherent}). Superradiance of a periodically bunched beam takes place when a train of tightly bunched electron bunches enters the interaction region at a rate  equal to the radiation wave frequency. This generic coherent spontaneous radiation process can take place in any kind of free electron radiation emission scheme \cite{superradiant, Gover_2005, schnitzer,Dattoli_1997,Penn_2006}, including synchrotron radiation (where it is also termed Coherent Synchrotron Radiation - CSR or Edge Radiation) \cite{Carr_2002,Abo-Bakr_2002,Sannibale_2004,Byrd_2004,Adams_2004,Hirschmugl_1991,Wang_1998,Tamada_1993,Andersson_2000,Arp_2001,Carr_2001,Giovenale_1999,Nodvick_1954,Berryman_1996,Byrd_2002,Michel_1981,Krinsky,Green,Geloni}, Coherent Transition Radiation CTR \cite{Happek_1991,Lihn_1996,Piestrup,Shibata_1994,Leemans,Orlandi,Geloni_2009}, Undulator Radiation \cite{Doria_1993,Jaroszynsky_1993,Kuroda_2011,McNeil_1999,Ciocci_1993,Gover_1994,Faatz_2001,Jeong_1992,Asakawa_1994,Bonifacio_1984,Gover_1987,Pinhasi_2002,Neumann_2003,Arbel_2001,Arbel_2000,Neuman_2000,Huang_2014,Huang_2015,Huang_2007,Seo_2013,Lurie_2007,Watanabe_2007,Hama_2011,Musumeci_2013,Cohen,Mayhew}, Smith-Purcell Radiation \cite{Shin_2007,Korbly_2005,Brownell_1998,Ginzburg_2013}, Cerenkov Radiation \cite{Neighbors,Wiggins}, dielectric waveguide radiation and more.

Another interesting related coherent emission effect is exhibited by the same kind of single or periodically bunched electrons when they are subjected to a coherent radiation field of a co-propagating wave in any kind of radiation emission scheme. If such a beam is tightly bunched relative to the wave period, or periodically bunched at the wave frequency, and if properly phased, then all electrons would experience the same deceleration force, and emit in phase Stimulated-Superradiance radiation. This process is analogous to the same process of enhanced coherent radiation emission by an ensemble of two-quantum-level atoms that are subjected to a strong coherent radiation field. In the nonlinear regime all atoms undergo phase correlated Rabi oscillation between the two quantum levels, and simultaneously can emit coherent Stimulated-Superradiance radiation \cite{Ismailov_1999}. The analogue of the quantum Rabi oscillation, in the case of a bunched electron beam, is the Synchrotron oscillation of a trapped bunched electron beam under the time harmonic force of a synchronous coherent radiation wave (ponderomotive wave in the case of undulator radiation).

Superradiant emission from a bunched beam may have important application in development of coherent radiation sources at wavelength regimes and operating conditions where a stimulated emission radiation source is not practical, because the accelerated beam current is too low to provide sufficient gain within a practicable interaction length.  We identify the THz frequency regime as a range where compact superradiant radiation sources are being developed \cite{Gover_2005,Huang_2015,Hama_2011,Shin_2007,Korbly_2005,Ginzburg_2013,Ciocci_1993,Gover_1994,Faatz_2001,Lurie_2015,Shibata_2002,Yasuda_2015,Huang_2010,Su} based on moderately accelerated sub-picoSecond bunched beams, generated in photo-cathode injector electron guns \cite{Akre_2008}. We also assert that future compact coherent EUV radiation sources based on Dielectric Laser Accelerator (DLA) schemes \cite{Peralta_2013} are likely to be developed as superradiant sources because of the low current and short interaction length expected to be attainable with such schemes. Beam bunching at the femtosecond and sub-femtosecond duration range has been demonstrated \cite{Marceau_2015,Hilbert_2009,Wong_2015,Hommelhoff_2006,Hoffrogge_2014,Zholents_2008}, and may be useful for Superradiant radiation emission in the optical to EUV range. Various schemes for micro-prebunching the electron beam, including HGHG, EEHG, PEHG have been developed for superradiant generation of coherent UV and X-Ray radiation \cite{Freund_2012,Graves_2013,YU_1991,YU_2000,Stupakov_2009,Feng_2014,Qika,Qika_2008,Reiche_2008}. Most interestingly, recently tens of attoseconds duration e-beam bunching was demonstrated at the electron quantum wavefunction level \cite{Feist_2015}, and it may exhibit superradiant emission in the modulated quantum electron wavepacket level.

In the first part of this article, Sections II-III, we present an analysis of superradiant emission in a general radiation emission scheme, but subsequently specify particularly to the case of undulator radiation. In general, the analysis of a radiative emission process requires simultaneous solution of Maxwell equations for a particulate charge current source together with the force equations that govern the particles trajectories. However, in the case of spontaneous emission (contrary to stimulated emission), the effect of the emitted radiation on the electron that had generated it, is usually neglected (namely, self-radiative interaction effect is not considered). 

In this case, after evaluating the trajectories of the bunched beam in a force field in the absence of radiation, superradiant emission can be calculated based on a solution of Maxwell equations alone. This is presented in Sections II-III, as follows: in Sect. II we derive the general expressions for random spontaneous emission, superradiant emission and stimulated-superradiant emission from either single bunch or finite duration pulse of periodic bunches (``bunches train''). This analysis is carried out in a general spectral (Fourier transform)
presentation of Maxwell equations. In both cases the current source is finite in time, the emitted radiation has finite energy, and therefore the continuous multi-frequency spectral formulation is proper. In Section~III we reiterate the analysis of spontaneous superradiance (SP-SR) and ``zero-order'' (in terms of the radiation field) stimulated superradiance (ST-SR) (namely the effect of the radiation on the electron trajectories is negligible) for the case of an infinite (long) periodically bunched beam. The analysis in this chapter is carried out in a single frequency (phasor) formulation for the steady
state case of undulator radiation (UR) by a periodically bunched electron beam (namely, an infinite train of bunches). In this case the radiation is composed only of the fundamental bunching frequency and its harmonics, and a single frequency model is proper. In this chapter we still use the approximation of negligible energy loss of the interacting e-beam, namely the radiation field is not intense enough to modify the electron trajectories, and explicit zero-order expressions for SP-SR and ST-ST emission are derived from Maxwell equations only. Using this zero-order approximation, we evaluate analytically the contribution of each term, in the case of undulator radiation, and weigh the ratio between them and its scaling.

Confining the analysis to Undulator radiation schemes, we extend in Section IV the zero-order analysis of superradiance and stimulated-superradiance to nonlinear regime interaction (namely, the effect of the radiation on the electron trajectories is non-negligible) in a uniform and tapered wiggler. This is the case where an intense radiation wave is injected externally into the interaction region together with a bunched e-beam, and the interaction between the radiation and the beam is strong enough to produce non-negligible e-beam energy loss and a consequent slowing down of the beam. Interestingly enough, this includes also a special case of ``self interaction'' (discussed in detail in Section~VI.G), where a periodically bunched beam interacts nonlinearly with the spontaneous superradiant radiation it had generated in the first place. We review there the bunched beam dynamics qualitatively in terms of the mathematical pendulum equation and tilted pendulum equation models for the uniform and tapered wiggler respectively. The characteristics of the pendulum equations are outlined in Appendix~\ref{pend_append}.

A nonlinear analysis is required for studying the dynamics of the bunched beam with the radiation field and understanding the role of the fundamental processes of SP-SR, ST-SR and TESSA (Tapering Enhanced Stimulated Superradiant Amplification). For this purpose we present in Chapter~V a simple model for the beam-radiation interaction. This model is a self-consistent, energy conserving formulation for the simultaneous solution of Maxwell equations and the force equations. The conservation of energy relation is proved for general free electron radiation schemes in Appendix~\ref{cons_energy_append}. The formulation is employed for the idealized case of an infinite, periodically tightly-bunched cold beam, interacting with a single transverse radiation mode in a uniform or tapered wiggler. Expectedly, this model is consistent with the tilted pendulum equation model of KMR \cite{kroll} for FEL saturation, but rather than starting from a random beam, we assume a beam with initial conditions of \underline{tight electron bunches}, and study their full nonlinear dynamics in the ponderomotive wave traps of the radiation mode as they evolve along the wiggler. 

In Chapter~VI we present the solution of the coupled bunched-beam-radiation interaction equations based on numerical solution of a normalized master equations of the model for a uniform and tapered wiggler. The nonlinear dynamics of the fundamental SP-SR, ST-SR and TESSA processes are presented by numerical examples and video simulations, and are checked for consistence with the zero-order limits of the earlier chapters.

In Chapter~VII, the rigorous but ideal model of a perfect tightly bunched e-beam is replaced by an approximate but more practical multi-particle model of arbitrary beam bunching and energy spread. This model is used for estimating limits of efficiency enhancement in a tapered wiggler in realizable configurations.

In Chapter~VIII we review the applications of superradiant radiation sources in different realizations. These include review of development of various superradiant sources in the THz regime, new concepts of energy efficient schemes of TESSA and TESSO (Tapering Enhanced Stimulated Superradiant Oscillator), and relation to simulation and design work for optimization of energy extraction in a tapered wiggler FEL.

\subsection{Superradiance in the wide sense}

In the simplified model of superradiance processes presented in this review, we refer to processes in which the bunching amplitude of the electron beam is fixed. Whether we refer to a single short bunch beam or to an infinitely long periodically bunched beam, the assumption is that the bunch shape and bunching amplitude is constant throughout the interaction. The radiation emission is then characterized in the zero-order regime by the scaling $\propto N^2$ as in Dicke's superradiance. Also in the nonlinear regime, discussed from chapter IV on, the model assumption is of tight full bunching: The bunches have dynamic processes of energy exchange with an intense radiation wave, but they do not spread and remain tightly bunched.

Of course, this model is a simplified idealization of more elaborate processes in real free electron radiation sources. 
There are two main reasons that elaborate our clear-cut distinction between the processes of seeded FEL (FEL amplifier), SASE-FEL and superradiant FEL, and lead to alternative wider sense definitions of superradiance (beyond superradiance in Dicke's sense). We will explain and review here briefly the alternative definitions but will keep the terminology of superradiance in the rest of the article to be in the narrow sense of Dicke's superradiance.

The first reason that breaks the distinction between a superradiant undulator radiation source and a FEL amplifier is the fact that while a superradiant source is based on a pre-bunched e-beam, the conventional FEL radiation process also involves bunching. The stimulated emission process that is the fundamental radiation process in any laser, is carried out in the FEL through a bunching process of a random electron beam by an externally injected coherent radiation field that bunches the beam at its frequency. Thus, in the case of a FEL amplifier there is no coherent radiation emission in the first sections of the interaction region (wiggler), but as the random electron beam gets bunched by the external radiation field, it starts radiating ``superradiantly'' in phase with the ``Seed-injected'' radiation field. As the bunching and radiation emission processes continue along the interaction length, the radiation field starts growing exponentially by stimulated interaction, until the beam bunching saturates. The bunching stage in the FEL amplifier is the linear (low or high) gain regime of FEL theory. This stage is skipped in a pre-bunched superradiant FEL.

The situation is somewhat similar in SASE-FEL. In this case, there is no external radiation field that establishes coherent bunching in the beam, but the partially-coherent spontaneous synchrotron undulator radiation emitted in the first section of the undulator can produce bunching of short coherence length in the beam that can still lead to a linear (field) exponential stimulated emission gain. In single path interaction, this process is enabled owing to the establishment of partial coherence in the beam through the ``optical slippage effect'': the light wavepackets, emitted by the electrons, are faster than the electrons that generate them (propagating one wavelength $\lambda$ relative to the electron during any wiggling period $\lambda_w$ path of the electron in the wiggler). Consequently, partial coherence range is established between electrons within the so-called ``cooperation length'' $l_c$ which is the accumulated slippage $\lambda (l_g/\lambda_w)$ of the electrons, where $l_g$ is exponential gain length of the SASE-FEL.

Even though some bunching and ``superradiance'' processes are involved in the exponential light generation and amplification process of FEL amplifier and SASE-FEL, they would not be usually considered superradiant radiation sources. There are however some mixed cases of superradiance and stimulated emission gain. Such is the case of the microwave klystron, where the bunching of a continuous beam and the radiation process take place in separate cavities \cite{Collin}. The emission of the pre-bunched beam in the second cavity is superradiance in the narrow sense. A similar example is the ``optical klystron oscillator'' \cite{Girard}. Here the energy of the electron beam is bunched in an undulator by an input laser radiation field, after a process of density bunching in a dispersive magnetic section (Chicane). Because the gain is small the emission is enhanced by letting the bunched beam interact again in phase with the same laser beam in a second wiggler. The interaction in this second step is certainly stimulated-superradiance in the narrow sense.

Another case of mixed superradiance and stimulated emission is when the electron beam is partially bunched and then, at short interaction lengths, it emits superradiantly in the narrow sense ($\propto b_n^2N^2$) where $b_n<1$ is the bunching factor of harmonic $n$. However, before saturation, if the beam is not fully bunched, it can continue to increase exponentially its bunching and radiation emission by stimulated emission in the linear gain regime as in a regular FEL amplifier \cite{schnitzer,Qika}. This principle is used in ``High Gain Harmonic Generation'' (HGHG) process \cite{Yu_2000_science} in which a beam is energy bunched by an intense laser at optical (IR) frequency, and after passage through a dispersive magnet (chicane) it gets tightly bunched spatially, and its density contains high harmonics at small amplitude. The beam is then injected into a second undulator, synchronous with this small amplitude high harmonic current, where it radiates and gets amplified in an exponential ``stimulated superradiant'' process, producing coherent radiation at extreme UV frequencies \cite{YU_1991,Allaria_2012}.

Another case where the superradiance and stimulated emission processes are mixed, and lead to alternative wider-sense definitions of radiation is the case of finite pulse beam. In this case the SASE exponential growth process gets mixed with the short pulse superradiance process when the random beam pulse length $l_b$ is shorter or near equal to the cooperation length: $l_b \lesssim l_c$. In this case, the partially coherent SASE process may eventually yield a ``single spike'' coherent radiation pulse, that may be termed ``superradiant'' in a wider sense, but the scaling of the radiation with the beam density is not always ($\propto N^2$) as in Dicke's superradiance because of the involvement of the exponential SASE processes. These kind of wider sense superradiance processes were thoroughly studied mostly by Bonifacio et al, and others \cite{Bonifacio_1984_proceedings,Bonifacio_1991,Bonifacio_1994,Watanabe_2007,Krinsky_1999} who also identified similar ``superradiance'' processes in the leading and trailing regions of a long pulse ($l_b>l_c$) \cite{Bonifacio_1990_358-367,Bonifacio_1991}. Also numerous publications of Ginzburg and co-workers operating at the long wavelength (THz) regime \cite{Ginzburg_2015} may be considered in this same category of superradiance in the wider sense.

As indicated, in the rest of this review we will use the term of superradiance in the narrow (Dicke's) sense.
	 
\section{Superradiance and Stimulated Superradiance of Bunched Electron Beam}

As a starting point we present the theory of superradiant (SP-SR) and stimulated superradiant (ST-SR) emission
from free electrons in a general radiative emission process \cite{superradiant}. In this section we use a spectral formulation,
namely, all fields are given in the frequency domain as Fourier transforms of the real time-dependent fields:
\begin{equation}
\breve{A}(r,\omega)=\int_{-\infty}^{\infty}A(r,t)e^{i\omega t} dt
\label{A_r_omega}
\end{equation}
We use the radiation modes expansion formulation of \cite{superradiant}, where the radiation field is expanded in terms of an orthogonal set of eigenmodes in a waveguide structure or in free space (eg. Hermite-Gaussian modes):
\begin{equation}
\{\tilde{\boldsymbol{E}}_q(\mathbf{r}), \tilde{\boldsymbol{H}}_q(\mathbf{r})\} = \{\tilde{\boldsymbol{\mathcal{E}}}_q(\mathbf{r}_{\perp}), \tilde{\boldsymbol{\mathcal{H}}}_q(\mathbf{r}_{\perp})\} e^{ik_{qz}z}
\label{EqHq}
\end{equation}
\begin{equation}
\breve{\mathbf{E}}(\mathbf{r},\omega)=\sum_{\pm q} \breve{C}_q(z,\omega)\tilde{\boldsymbol{\mathcal{E}}}_q(\mathbf{r}_{\perp})e^{ik_{qz}z}
\label{E_r_omega}
\end{equation}
\begin{equation}
\breve{\mathbf{H}}(\mathbf{r},\omega)=\sum_{\pm q} \breve{C}_q(z,\omega)\tilde{\boldsymbol{\mathcal{H}}}_q(\mathbf{r}_{\perp})e^{ik_{qz}z}
\label{H_r_omega}
\end{equation}
The electric/magnetic fields representing the structure of the mode are named $\tilde{\boldsymbol{\mathcal{E}}}_q$ and $\tilde{\boldsymbol{\mathcal{H}}}_q$ and
are usually nearly frequency independent. Their units are [V/m] and [A/m] respectively. The actual fields
$\breve{\mathbf{E}}$ and $\breve{\mathbf{H}}$ are Fourier transforms and hence are in units of [sec~V/m] and [sec~A/m] respectively.
Therefore, the amplitude coefficients $\breve{C}_q$ have dimensions of time, hence units of [sec].

The excitation equations of the mode amplitudes is:
\begin{equation}
\frac{d\breve{C}_q(z,\omega)}{dz}=\frac{-1}{4\mathcal{P}_q} \int\mathbf{\breve{J}}(\mathbf{r},\omega)\cdot \tilde{\boldsymbol{\mathcal{E}}}^*_q(\mathbf{r}_{\perp}) e^{-ik_{qz}z} d^2\mathbf{r}_{\perp}.
\label{dCdz}
\end{equation}
where the current density $\mathbf{\breve{J}}(\mathbf{r},\omega)$ is the Fourier transform
of $\mathbf{J}(\mathbf{r},t)$.

The above is formally integrated and given in terms of the initial mode excitation amplitude and the currents
\begin{equation}
\breve{C}_q(z,\omega)-\breve{C}_q(0,\omega) = -\frac{1}{4\mathcal{P}_q} \int \mathbf{\breve{J}}(\mathbf{r},\omega)\cdot \tilde{\boldsymbol{\mathcal{E}}}^*_q(\mathbf{r}_{\perp}) e^{-ik_{qz}z} dV,
\label{mode_exp}
\end{equation}
where $\mathcal{P}_q$ is the power normalization parameter:
\begin{equation}
\mathcal{P}_q=\frac{1}{2}Re\iint(\tilde{\boldsymbol{\mathcal{E}}}_q \times \tilde{\boldsymbol{\mathcal{H}}}_q^*)\cdot\hat{e}_z d^2 \mathbf{r}_{\perp}=\frac{|\tilde{\boldsymbol{\mathcal{E}}}_q(\mathbf{r}_{\perp}=0)|^2}{2Z_q}A_{em\, q},
\label{P_q}
\end{equation}
where $Z_q$ is the mode impedance (in free space $Z_q=\sqrt{\mu_0/\epsilon_0}$), and for a narrow beam, passing on axis
near $\mathbf{r}_{\perp}=0$, Eq.~(\ref{P_q}) defines the mode effective area $A_{em\, q}$ in terms of the field of the
mode on axis $\tilde{\boldsymbol{\mathcal{E}}}_q(\mathbf{r}_{\perp}=0)$.

For the Fourier transformed fields we define the total spectral energy (per unit of angular frequency) based on
Parseval theorem as
\begin{equation}
\frac{dW}{d\omega}=\frac{2}{\pi}\sum_q\mathcal{P}_q |\breve{C}_q(\omega)|^2,
\label{dW_domega_all_modes}
\end{equation}
This definition corresponds to positive frequencies only: $0<\omega<\infty$. Considering now one single mode $q$,
\begin{equation}
\frac{dW_q}{d\omega}=\frac{2}{\pi}\mathcal{P}_q |\breve{C}_q(\omega)|^2,
\label{dW_d_omega_P_q}
\end{equation}

For a particulate current (an electron beam):
\begin{equation}
J(\mathbf{r},t)=\sum_{j=1}^N -e \mathbf{v}_j(t) \delta(\mathbf{r}-\mathbf{r}_j(t))     
\label{J_r}
\end{equation}
The field amplitude increment appears as a coherent sum of contributions (energy wavepackets) from all the electrons in the beam:
\begin{equation}
\breve{C}_q^{out}(\omega)-\breve{C}_q^{in}(\omega)\equiv\sum_{j=1}^N \Delta\breve{C}_{qj}(\omega)= -\frac{1}{4\mathcal{P}_q} \sum_{j=1}^N \Delta \breve{\mathcal{W}}_{qj}
\label{Cout_Cin}
\end{equation}
\begin{equation}
\Delta \breve{\mathcal{W}}_{qj}=-4\mathcal{P}_q\Delta\breve{C}_{qj} =-e \int_{-\infty}^{\infty} \mathbf{v}_j(t)\cdot \tilde{\boldsymbol{\mathcal{E}}}_q^* (\mathbf{r}_j(t)) e^{i\omega t}dt
\label{DeltaWqj}
\end{equation}
The contributions can be split into a spontaneous part (independent of the presence of radiation field) and
stimulated (field dependent) part:
\begin{equation}
\Delta\breve{\mathcal{W}}_{qj}= \Delta\breve{\mathcal{W}}_{qj}^0 + \Delta\breve{\mathcal{W}}_{qj}^{st}.
\label{DeltaWqj_1}
\end{equation}
We do not deal in this section with stimulated emission, but indicate that in general the second term $\Delta\breve{\mathcal{W}}_{qj}^{st}$
is a function of $\breve{C}_q(z)$ through $\mathbf{r}_j(t)$ and $\mathbf{v}_j(t)$ therefore the integral in Eq.~(\ref{DeltaWqj}) cannot be calculated explicitly. Its calculation requires solving the electron force equations and the differential
equation (\ref{dCdz}). In the context of the linear gain regime of conventional FEL, $\Delta\breve{C}_{qj}^{st}$ is proportional
to the input field, i.e. proportional to $\breve{C}_q^{in}$, and in this case the solution of (\ref{dCdz}) results in the exponential
gain expression of conventional FEL \cite{gover_sprangle}.

Assuming a narrow cold beam where all particles follow the same trajectories, we may write
$\mathbf{r}_j(t)=\mathbf{r}_j^0(t-t_{0j})$ and $\mathbf{v}_j(t)=\mathbf{v}_j^0(t-t_{0j})$,
change variable $t'=t-t_{0j}$ in Eq.~(\ref{DeltaWqj}) \cite{MOP078}, so that the spontaneous emission
wavepacket contributions are identical, except for a phase factor corresponding to their injection
time $t_{0j}$:
\begin{equation}
\Delta\breve{\mathcal{W}}_{qj}^0 = \Delta\breve{\mathcal{W}}_{qe}^0 e^{i\omega t_{0j}}
\label{DeltaWqj0}
\end{equation}
where
\begin{equation}
\Delta\breve{\mathcal{W}}_{qe}^0 = -e \int_{-\infty}^{\infty} v_e^0(t)\cdot \tilde{\boldsymbol{\mathcal{E}}}_q^* (r_e^0(t)) e^{i\omega t}dt .
\label{DeltaWqe0}
\end{equation}
The radiation mode amplitude at the output is composed of a sum of wavepacket contributions including
the input field contribution (if any):
\begin{align}
&\breve{C}_q^{out}(\omega)= \breve{C}_q^{in}(\omega) + \Delta\breve{C}_{qe}^0(\omega)\sum_{j=1}^N e^{i\omega t_{0j}} + \sum_{j=1}^N \Delta\breve{C}_{qj}^{st} = \notag \\
&\breve{C}_q^{in}(\omega)-\frac{1}{4\mathcal{P}_q}\Delta\breve{\mathcal{W}}_{qe}^0 \sum_{j=1}^N e^{i\omega t_{0j}} -\frac{1}{4\mathcal{P}_q}\sum_{j=1}^N \Delta\breve{\mathcal{W}}_{qj}^{st}
\label{C_out}
\end{align}
so that the total spectral radiative energy from the electron pulse is
\begin{align}
&\frac{dW_q}{d\omega}=\frac{2}{\pi}\mathcal{P}_q\left|\breve{C}_q^{out}(\omega)\right|^2=\frac{2}{\pi}\mathcal{P}_q\left\{\left|\breve{C}_q^{in}(\omega)\right|^2+  \right. \notag \\
&\left|\Delta C_{qe}^{(0)}(\omega)\right|^2\left|\sum_{j=1}^N e^{i\omega t_{oj}}\right|^2 + \left[\breve{C}_q^{in\,*}(\omega)\Delta C_{qe}^{(0)}(\omega) \sum_{j=1}^N e^{i\omega t_{oj}}+c.c.\right] + \notag \\
& \left. \left[\breve{C}_q^{in\,*}(\omega) \sum_{j=1}^N \Delta C_{qj}^{st}(\omega) +c.c.\right] +
 \left| \sum_{j=1}^N \Delta C_{qj}^{st}(\omega) \right|^2  \right\} \equiv  \notag \\
& \left(\frac{dW_q}{d\omega}\right)_{in}+\left(\frac{dW_q}{d\omega}\right)_{SP-SR}+\left(\frac{dW_q}{d\omega}\right)_{ST-SR}+\left(\frac{dW_q}{d\omega}\right)_{st}.
\label{dW_domega}
\end{align}
The first term in the $\{\}$ parentheses represents the input wave spectral energy, given the subscript ``in''.
The second term is the spontaneous emission, which may also be superradiant in case that all contributions
add in phase, hence given the subscript ``SP-SR''. The third term has a very small value (averages to 0)
if the contributions add randomly. Thus it is relevant only if the electrons of the beam enter in phase
with the radiated mode. It is therefore dependent on the coherent mode complex amplitude $\breve{C}_q^{in}$, and 
hence it is marked by the subscript ``ST-SR'', i.e. ``zero-order''stimulated superradiance. The last 2 terms in the $\{\}$ parentheses represent stimulated emission.
\begin{figure*}[!tbh]
\includegraphics[width=15cm]{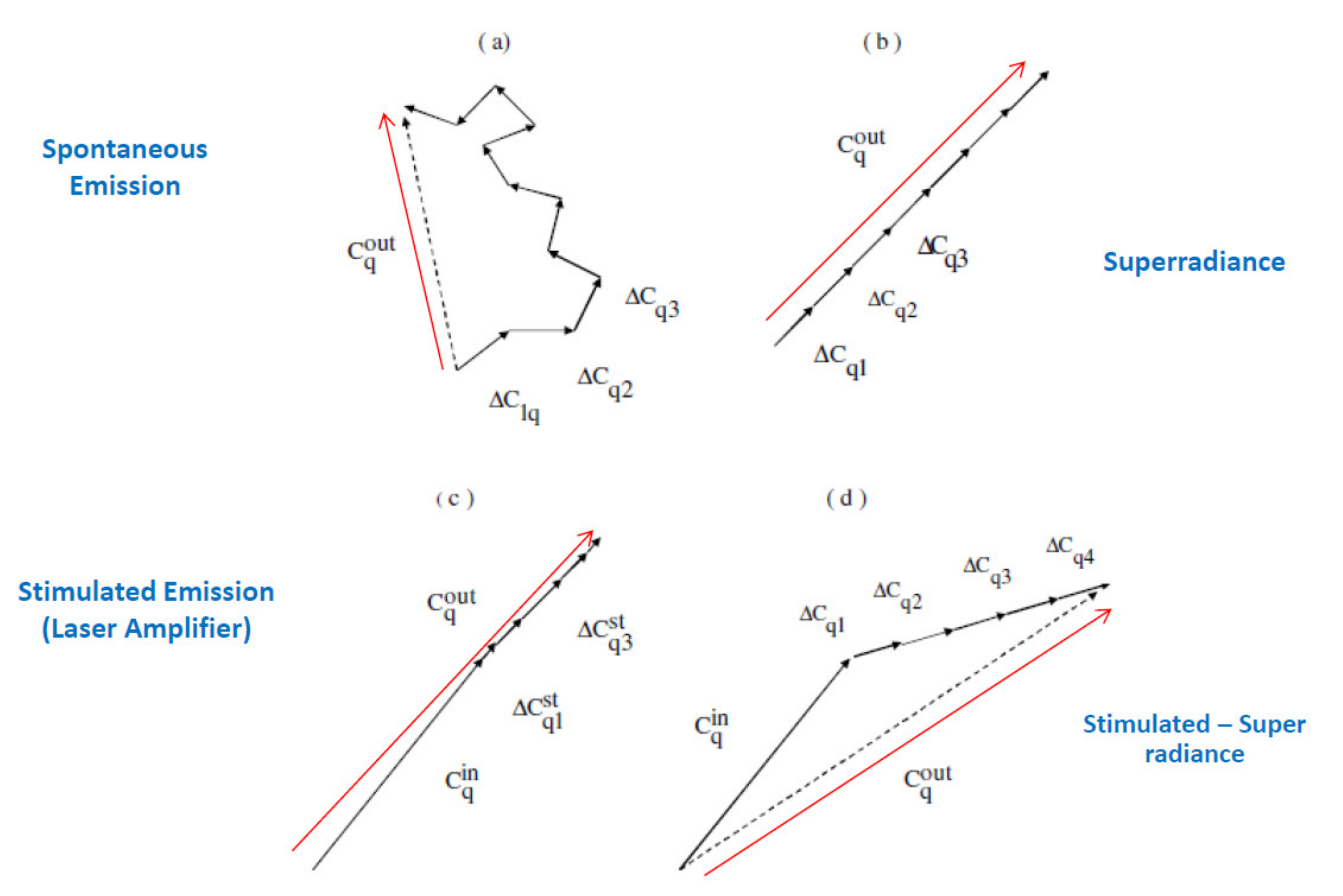}
\caption{Different cases of radiation: (a) spontaneous emission, (b) superradiance, (c) stimulated emission and (d) stimulated superradiance.}
\label{radiations}
\end{figure*}

Figure~\ref{radiations}(a) and (b) represent in the $\breve{C}_q$ complex plane the conventional spontaneous
emission and superradiant emission
that correspond to the second term in Eq.~(\ref{dW_domega}) where in \ref{radiations}(a) the wavepackets
interfere randomly and in \ref{radiations}(b), constructively in phase. Figure~\ref{radiations}(d) represents the third
term in Eq.~(\ref{dW_domega}) where the coherent constructive interference of a prebunched beam
interferes with the input field with some phase offset. When the electrons in the
beam are injected at random in a long pulse, then in averaging the second term in Eq.~(\ref{dW_domega}), the uncorrelated
mixed terms cancel out, and one obtains the conventional shot-noise driven spontaneous emission \cite{superradiant,MOP078}.
\begin{equation}
\left(\frac{dW_q}{d\omega}\right)_{sp}=\frac{1}{8\pi \mathcal{P}_q} \left|\Delta\breve{\mathcal{W}}_{qe}
^{(0)}\right|^2 N
\label{DWdq_st}
\end{equation}
Only when the electrons are bunched into a pulse shorter than an optical period $\omega|t_{0i}-t_{0j}|\ll\pi$
one gets enhanced superradiant spontaneous emission, in which case all the terms in the bracket of the third term
of Eq.~(\ref{dW_domega}) add up constructively in phase $\sum_{j=1}^N e^{i\omega t_{oj}}=Ne^{i\omega t_{o}}$ resulting
\begin{equation}
\left(\frac{dW_q}{d\omega}\right)_{SP-SR}=\frac{1}{8\pi \mathcal{P}_q} \left|\Delta\breve{\mathcal{W}}_{qe}
^{(0)}\right|^2 N^2=\left\langle\left(\frac{dW_q}{d\omega}\right)\right\rangle_{sp}N
\label{DWdq_SR}
\end{equation}
Figure~\ref{radiations}(d) displays a process of of stimulated superradiance: all electrons oscillate in
phase, but because a radiation mode of distinct phase is injected in, the third term in Eq.~(\ref{dW_domega})
will contribute positive or negative radiative energy, depending whether the electron bunch oscillates in phase or
out of phase with the input radiation field. If the phase of the electron bunch relative to the wave is
$\varphi$, then the third term in Eq.~(\ref{dW_domega}) represents stimulated superradiance spectral energy:
(consistent with \cite{superradiant} except for a missing factor of 2):
\begin{equation}
\left(\frac{dW_q}{d\omega}\right)_{ST-SR}=-\frac{1}{\pi} \left|\breve{C}_q^{in}\right| \left|\Delta\breve{\mathcal{W}}_{qe}
^{(0)}\right| N\cos\varphi.
\label{DWdq_ST_SR}
\end{equation}
For purpose of comparison, we also display in Figure~\ref{radiations}(c) the process of conventional stimulated
emission (fourth term in Eq.~(\ref{dW_domega})), e.g. for the case of FEL amplifier that we do not further consider here.

At this point we extend the analysis to include partial bunching, namely electron beam bunches of finite duration
and arbitrary bunch-shape function. One can characterize the distribution of electron entrance times $t_{0j}$ of the
electron bunch by means of a normalized bunch-shape function $f(t^{\prime}_0-t_0)=i(t^{\prime}_0-t_0)/(eN)$, where $i(t)$
is the e-beam bunch current, and $t_0$ is bunch center entrance time:
\begin{equation}
\int_{-\infty}^{\infty}f(t^{\prime}_0-t_0)dt^{\prime}_0 = 1.
\label{normalization}
\end{equation}
Then the summation over $t_{0j}$ may be substituted by integration over entrance times $t^{\prime}_0$:
\begin{equation}
\sum_{j=1}^N e^{i\omega t_{oj}}=N \int f(t^{\prime}_0-t_0)e^{i\omega t^{\prime}_0}dt^{\prime}_0 = Ne^{i\omega t_0}M_b(\omega),
\label{sum_to_int}
\end{equation}
where
\begin{equation}
M_b(\omega)=\frac{1}{N}\left\langle\sum_{j=1}^{N} e^{i\omega t_{0j}}\right\rangle=\int f(t)e^{i\omega t}dt,
\label{M_b1}
\end{equation}
is the Fourier transform of the bunch-shape function, i.e. the bunching amplitude at frequency $\omega$.
It modifies Eqs.~(\ref{DWdq_SR}) and (\ref{DWdq_ST_SR}) to 
\begin{equation}
\left(\frac{dW_q}{d\omega}\right)_{SP-SR}=\frac{1}{8\pi \mathcal{P}_q} \left|\Delta\breve{\mathcal{W}}_{qe}
^{(0)}\right|^2 |M_b|^2 N^2
\label{DWdq_SR_M_b}
\end{equation}
and
\begin{equation}
\left(\frac{dW_q}{d\omega}\right)_{ST-SR}=-\frac{1}{\pi} \left|\breve{C}_q^{in}\right| \left|\Delta\breve{\mathcal{W}}_{qe}
^{(0)}\right| |M_b| N\cos\varphi.
\label{DWdq_ST_SR_M_b}
\end{equation}
In conditions of perfect bunching $f(t)=\delta(t)$ (and consequently $M_b=1$), Eqs.~(\ref{DWdq_SR}) and (\ref{DWdq_ST_SR})
are restored. For a finite size bunch, modeled by a Gaussian electron beam bunch distribution 
\begin{equation}
f(t)=\frac{1}{\sqrt{2\pi}\sigma_{tb}}e^{-t^2/(2\sigma_{tb}^2)},
\label{gauss_dist}
\end{equation}
the bunching factor is
\begin{equation}
M_b(\omega)=\int_{-\infty}^{\infty} e^{i\omega t} f(t) dt=e^{-\omega^2 \sigma_{tb}^2/2}.
\label{M_b_1}
\end{equation}

The ``zero-order'' analysis, so far is valid for any interaction scheme for which the electron trajectories in the radiating
structure are known explicitly to zero order, namely in the absence of radiation field, or where the change in the particles
velocity and energy due to interaction with an external radiation field or their self-generated
radiation field is negligible.

\subsection{Superradiant undulator radiation}

For the case of interest of undulator radiation we specify for each electron:
\begin{equation}
\mathbf{v}_i(t)=Re\left[\mathbf{\tilde{v}}_{\perp\, i} e^{-ik_w z_i(t)}\right]
\label{und_perp_vel1}
\end{equation}
where
\begin{equation}
\mathbf{\tilde{v}}_{\perp\, i}=\frac{c\mathbf{\tilde{a}}_w}{\gamma_i}=\frac{e\hat{z}\times\mathbf{\tilde{B}}_w}{\gamma_i m k_w}
\label{v_perp1}
\end{equation}
where $\mathbf{\tilde{B}}_w$ is the complex amplitude of the undulator periodic magnetic field $\mathbf{B}(z)=Re[\mathbf{\tilde{B}}_we^{-ik_wz}]$.
Assume that the electron beam is narrow enough so that all electrons experience the same field when interacting
with the mode
\begin{equation}
\tilde{\boldsymbol{E}}_q(\mathbf{r}_j^0(t))=\tilde{\boldsymbol{\mathcal{E}}}_q(\mathbf{r}_{\perp}=0)e^{ik_{qz}z_j^0(t)}
\label{narrow_beam}
\end{equation}
where $z_j^0(t)=v_z(t-t_{0j})$, and $\mathbf{r}_{\perp}$ is the transverse coordinates vector of the electron beam, then
substituting this and Eq~(\ref{und_perp_vel1}) in (\ref{DeltaWqe0}) one obtains
\begin{equation}
\Delta\breve{\mathcal{W}}_{qj}^0 = -e\frac{\mathbf{\tilde{v}}_{\perp 0}\cdot\mathbf{\tilde{\boldsymbol{\mathcal{E}}}}_q^*}{2v_z}L_w\sinc(\theta L_w/2)e^{i\theta L_w/2}e^{i\omega t_{0j}},
\label{DeltaWqj_01}
\end{equation}
where $L_w=N_w\lambda_w$ is the interaction length ($\lambda_w=2\pi/k_w$), $\sinc(x)=\sin x/x$, and
$\theta(\omega)$, the detuning parameter, is defined by
\begin{equation}
\theta(\omega)=\frac{\omega}{v_z}-k_{zq}(\omega)-k_w .
\label{detuning_par1}
\end{equation}
The detuning function $\sinc(\theta L/2)$ attains its maximum value at the synchronism frequency $\omega_r$
defined by
\begin{equation}
\theta(\omega_r)=\frac{\omega_r}{v_z}-k_{zq}(\omega_r)-k_w=0 .
\label{omega_0}
\end{equation}
Near synchronism
\begin{equation}
\theta(\omega)L_w\simeq (\omega-\omega_r)t_s=2\pi\frac{\omega-\omega_r}{\Delta\omega}.
\label{near_sync}
\end{equation}
where
\begin{equation}
t_s=\frac{2\pi}{\Delta\omega}=\frac{L_w}{v_z}-\frac{L_w}{v_{gq}}
\label{t_s}
\end{equation}
is the wave packet slippage time and $v_{gq}=d\omega/dk_{zq}$ at $\omega_r$ is the group velocity
of the mode. In free space $k_{zq}=\omega/c$, $v_{gq}=c$, and the solution of (\ref{omega_0}) results in
\begin{equation}
\omega_r=\frac{ck_w}{1/\beta_z-1}\simeq 2\gamma^2_z ck_w
\label{omega_0_expr}
\end{equation}
\begin{equation}
\Delta\omega=\frac{\omega_r}{N_w}
\label{delta_omega}
\end{equation}
where the second part of Eq.~(\ref{omega_0_expr}) applies for an ultra-relativistic beam ($\beta\simeq 1$), and
\begin{equation}
\gamma_z^2=\frac{\gamma^2}{1+\overline{a}_w^2}
\label{gamma_z2}
\end{equation}
where $\overline{a}_w\equiv e\overline{B}_w/(mck_w)$ is the one period r.m.s. average of $a_w(z)$. It is equal to the amplitude
$a_w$ in the case of a helical (circularly polarized) wiggler and to $a_w/\sqrt{2}$ in a linear (linearly polarized) magnetic wiggler.

When substituting (\ref{DeltaWqj_01}) and (\ref{v_perp1}) into (\ref{DWdq_SR}) and (\ref{DWdq_ST_SR}) one obtains the expressions of UR
superradiance and stimulated-superradiance from a tight single bunch into a single mode $q$ \cite{superradiant}
\begin{equation}
\left(\frac{dW_q}{d\omega}\right)_{SP-SR}=\frac{N^2e^2Z_q}{16\pi}\left(\frac{\overline{a}_w}{\beta_z\gamma}\right)^2\frac{L_w^2}{A_{em\, q}}\sinc^2(\theta L_w/2)
\label{dWdomega_SR_2}
\end{equation}
\begin{equation}
\left(\frac{dW_q}{d\omega}\right)_{ST-SR}=\frac{N}{\pi}\left(\frac{\overline{a}_w}{\beta_z\gamma}\right)e|\breve{E}(0,\omega)|  L_w  \sinc(\theta L_w/2)\cos(\varphi_{qb0}-\theta L_w/2)
\label{dWdomega_ST_SR_2}
\end{equation}
where $\breve{E}(0,\omega)=\breve{C}_q(0,\omega)\tilde{\boldsymbol{\mathcal{E}}}_q(\mathbf{r}_{\perp}=0)$ is the Fourier transform of the input injected radiation mode (Eqs.~\ref{A_r_omega},\ref{E_r_omega}) and $\varphi_{qb0}(\omega)$ is the phase between the radiation field and the bunch at the entrance to the wiggler.

While in the present paper we stay, for the sake of transparency, with a single radiation mode analysis,
we point out that the general expression for
radiation into all modes is found from summation over the contributions of all modes
(Eq.~\ref{dW_domega_all_modes}). This
expression can be extended also to the case of free space radiation \cite{MOP078} where the
far field spectral energy intensity is found to have a similar frequency functional dependence as
Eq.~\ref{dWdomega_SR_2} with the substitution $k_z=(\omega/c)\cos\Theta$ in the detuning parameter
$\theta(\omega)$ (Eq.~\ref{detuning_par1}), where $\Theta$ is the far field observation angle off
the wiggler axis.

\begin{figure*}[!tbh]
\includegraphics[width=18cm]{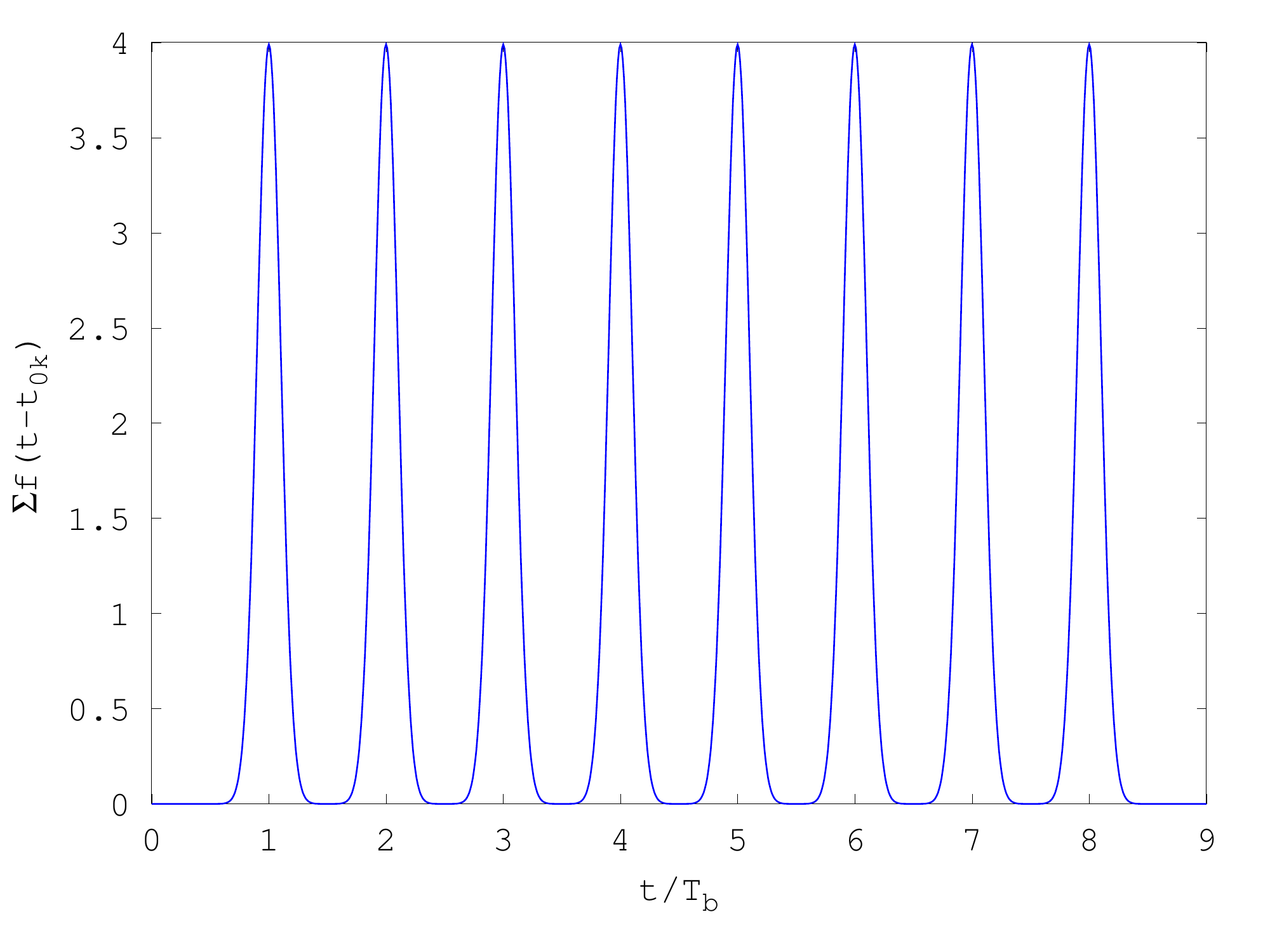}
\caption{The charge distribution function of a ``macropulse'' finite train of bunches arriving at times $t_{0k}$, with period $T_b=2\pi/\omega_b$. Each bunch is a normalized Gaussian (Eq.~\ref{gauss_dist}) with $\sigma_{tb}\ll T_b$.}
\label{train_fig}
\end{figure*}

\begin{figure*}[!tbh]
\includegraphics[width=18cm]{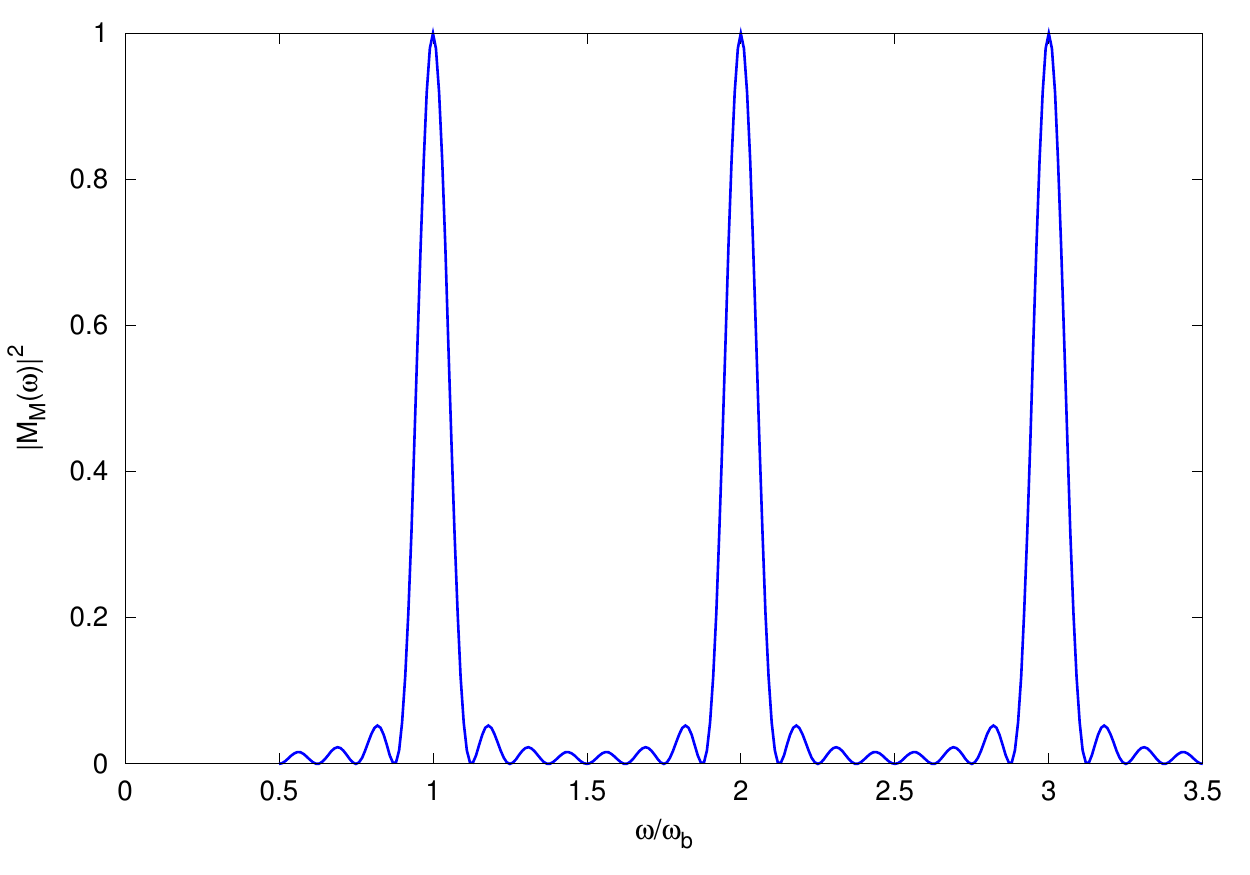}
\caption{The absolute square of the macropulse form-factor function in Eq.~(\ref{periodic_bunch6}), for $N_M=8$}
\label{M_M_fig}
\end{figure*}

We now extend the analysis to the case of spontaneous emission from a finite train of bunches.
Following the formulation of \cite{superradiant}, we consider a train of $N_M$ identical bunches
(neglecting shot noise) separated in time $T_b\equiv 2\pi/\omega_b$ apart. The arrival times of bunch $k$ is
\begin{equation}
t_{0k}=[k-(N_M/2)]2\pi/\omega_b
\label{t_0k}
\end{equation}
The summation of the phasors $e^{i\omega t_{0j}}$ in the second term of (\ref{C_out}) is now reorganized into
summation over the $N_M$ bunches and summation over the $N_b$ particles in each bunch as shown in Figure~\ref{train_fig}.
Given that electron $n$ ($n$ is between 1 to $N=N_M N_b$) is the electron $j$ of bunch $k$, and $|\Delta t_{0i}|<T_b$, we have
$t_{0n}=t_{0k}+\Delta t_j+t_0$ ($t_0$ being a pulse origin reference, e.g. the arrival time of the center of the train
pulse), we may write:
\begin{equation}
\sum_{n=1}^N e^{i\omega t_{0n}}=\sum_{n=1}^N e^{i\omega t_{0k}} e^{i\omega \Delta t_j}e^{i\omega t_{0}}=e^{i\omega t_{0}}\sum_{k=1}^{N_M}\sum_{j=1}^{N_b}e^{i\omega t_{0k}} e^{i\omega \Delta t_j}=e^{i\omega t_{0}} \left(\sum_{k=1}^{N_M}e^{i\omega t_{0k}}\right) \left(\sum_{j=1}^{N_b} e^{i\omega \Delta t_j}\right).
\label{periodic_bunch}
\end{equation}
We define the microbunch bunching factor:
\begin{equation}
M_b(\omega)=\frac{1}{N_b}\left\langle\sum_{j=1}^{N_b} e^{i\omega \Delta t_j}\right\rangle,
\label{M_b}
\end{equation}
where $\langle\rangle$ mean averaging on the random $\Delta t_j$. We also define the macrobunch (pulse) form factor:
\begin{equation}
M_M(\omega)=\frac{1}{N_M}\sum_{k=1}^{N_M} e^{i\omega t_{0k}}.
\label{M_M}
\end{equation}
Setting Eq.~(\ref{M_b}) and (\ref{M_M}) into (\ref{periodic_bunch})
we obtain
\begin{equation}
\left\langle\sum_{n=1}^N e^{i\omega t_{0n}}\right\rangle =N M_b(\omega)M_M(\omega)e^{i\omega t_0}
\label{periodic_bunch1}
\end{equation}
Note that the assumption that all microbunches in the macrobunch have equal number of particles $N_b$
amounts to neglecting shot-noise due to random variance of particles along the macrobunch.
If one assumes that the distribution of the electron particles within the bunches is tight enough
$|\Delta t_{0j}|\ll T_b$,  Eq.~\ref{M_b} can be written in terms of the particles distribution function within one period
$M_b(\omega)=\frac{\omega_b}{2\pi}\int_{-\pi/(2\omega_b)}^{\pi/(2\omega_b)}f_b(\Delta t_0)d(\Delta t_0)$ and
approximated by Eq.~\ref{M_b1}. For a Gaussian distribution (\ref{gauss_dist}),
$M_b$ is then given by Eq.~(\ref{M_b_1}). The macrobunching form factor (\ref{t_0k}) is calculated using
(\ref{M_M}), as a geometric series sum:
\begin{equation}
M_M(\omega)=\frac{\sin(N_M\pi\omega/\omega_b)}{N_M\sin(\pi\omega/\omega_b)}.
\label{periodic_bunch6}
\end{equation}
This form factor contains the basic bunching frequency $\omega_b$ peak and an infinite number of high harmonics,
as shown in Figure~\ref{M_M_fig}.
Consequently the superradiant spectral energy of the bunch train (the second term in (\ref{dW_domega}) is
\begin{equation}
\left(\frac{dW_q}{d\omega}\right)_{SP-SR}=\frac{N^2}{8\pi \mathcal{P}_q}\left|\Delta\breve{\mathcal{W}}_{qe}^{(0)}\right|^2 \left|M_b(\omega)\right|^2 \left|M_M(\omega)\right|^2,
\label{dWdomega_SR}
\end{equation}
and the stimulated superradiance at zero order approximation (the third term in Eq.(\ref{dW_domega}) is
\begin{equation}
\left(\frac{dW_q}{d\omega}\right)_{ST-SR}=\frac{N}{\pi}|\breve{C}_q^{in}(\omega)| \left|\Delta\breve{\mathcal{W}}_{qe}^{(0)}\right| \left|M_b(\omega)\right| \left|M_M(\omega)\right| \cos\varphi_{qb0}(\omega),
\label{dWdomega_ST_SR}
\end{equation}
where $\varphi_{qb0}(\omega)$ is the phase between the radiation field and the periodically bunched beam,
determined at the entrance to the wiggler.

For the case of interest of UR we substitute Eq.~(\ref{DeltaWqj_01}) into Eqs.~(\ref{dWdomega_SR}) and
(\ref{dWdomega_ST_SR}) and obtain the general expression for  spontaneous SP-SR and ST-SR spectral energy
of a finite train of periodic bunches:
\begin{equation}
\left(\frac{dW_q}{d\omega}\right)_{SP-SR}=\frac{N^2e^2Z_q}{16\pi}\left(\frac{\overline{a}_w}{\beta_z\gamma}\right)^2\frac{L_w^2}{A_{em\, q}}\left|M_b(\omega)\right|^2\left|M_M(\omega)\right|^2 \sinc^2(\theta L_w/2)
\label{dWdomega_SR_1}
\end{equation}
and the stimulated superradiant term is
\begin{align}
\left(\frac{dW_q}{d\omega}\right)_{ST-SR}&=\frac{N}{\pi}\left(\frac{\overline{a}_w}{\beta_z\gamma}\right)e|\breve{E}(0,\omega)| L_w\notag \\
&\left|M_b(\omega)\right|\left|M_M(\omega)\right|\sinc(\theta(\omega) L_w/2)\cos(\varphi_{qb0}(\omega)-\theta L_w/2)
\label{dWdomega_ST-SR_1}
\end{align}
The spectrum of superradiant and stimulated-superradiant UR is composed of harmonics of narrow linewidth
$\Delta\omega\simeq \omega_b/N_M$ (Figure~\ref{M_M_fig}) within the low frequency filtering band of the
bunching factor $\omega<1/\sigma_{tb}$ (Eq.~\ref{M_b_1}) and the finite interaction length
bandwidth (\ref{delta_omega}).

\section{Single Frequency Formulation}

In the limit of a continuous train of microbunches or a long macropulse $N_M\gg 1$, the grid
function $M_M(\omega)$ behaves like a comb of delta functions and narrows the spectrum of the
prebunched beam SP-SR and ST-SR Undulator Radiation to harmonics of the bunching frequencies
$\omega=n\omega_b$. Instead of spectral energy, one can then evaluate the average radiation power
output by integrating the spectral energy expressions (\ref{dWdomega_SR}) and (\ref{dWdomega_ST_SR})
over frequency and dividing the integrated spectral energy by the pulse duration:
$T_M=N_M2\pi/\omega_b$. Alternatively, one may have analyzed the continuous bunched beam problem
from the start in a single frequency model using ``phasor'' formulation, concentrating for now on
a single frequency $\omega_0$:
\begin{equation}
A(\mathbf{r},t)=Re[\tilde{A}(\mathbf{r},\omega_0)e^{-i\omega_0 t}]
\label{phasor}
\end{equation}
It is to be mentioned that in this case the radiation frequency $\omega$ must be
equal to the bunching frequency or one of its harmonics $\omega=\omega_0=n\omega_b$, otherwise there will not be any steady-state interaction
between them.
The radiation mode excitation equations in the phasor formulation of the radiation fields
$\{\mathbf{\tilde{E}}(\mathbf{r}), \mathbf{\tilde{H}}(\mathbf{r})\}$
is the same as Eqs.~(\ref{EqHq})-(\ref{dCdz}) with $\tilde{C}_q(z,\omega_0)\equiv\tilde{C}_q(z)$ replacing
$\breve{C}_q(z,\omega)$, and the spectral energy radiance expression (\ref{dW_domega_all_modes}) replaced
by the total steady state radiation power
\begin{equation}
P=\sum_q\mathcal{P}_q |\tilde{C}_q|^2,
\label{power_all_modes}
\end{equation}

As in \cite{schneidmiller,arbel,schnitzer}, we take a model of a periodically modulated e-beam current of a single frequency $\omega_0$:
\begin{equation}
I(z,t)=I_0\{1+Re[\tilde{M} e^{-i\omega_0(t-z/v_z)}]\}
\label{periodic_model}
\end{equation}
This current represents one of the harmonics of a periodically bunched beam $\omega_0=n\omega_b$.

The parameter $M$ can be calculated for each of the harmonics $\omega_n=n\omega_b$ from the Fourier series expansion of an infinite train of identical microbunches (shot-noise is neglected):
\begin{equation}
I(z,t)=I_0T_b\sum_{n=-\infty}^{\infty}f(t-z/v_z-nT_b)
\label{periodic_model_1}
\end{equation}
where $T_b=2\pi/\omega_b$ and the bunch profile is normalized according to $\int_{-T_b/2}^{T_b/2}f(t)dt=1$. The Fourier expansion is
\begin{equation}
I(z,t)=I_0\{1+2 Re[b_n e^{in\omega_b(t-z/v_z)}]\}
\label{periodic_model_fourier}
\end{equation}
where
\begin{equation}
b_n=\int_{-T_b/2}^{T_b/2}f(t)e^{-in\omega_b(t-z/v_z)} dt
\label{b_n}
\end{equation}

Thus, Eq.~\ref{periodic_model} represents one of the harmonic components of frequency $\omega_0=\omega_{b\,n}=n\omega_b$ and phasor amplitude $M=2b_n$.

The bunching parameter $b_n$ depends on the profile function of the microbunch. If the microbunches can be represented by the Gaussian function (\ref{gauss_dist}), such that $\sigma_{tb}\ll T_b/n$, then the integration in (\ref{b_n}) can be carried to infinity, and then (see Eq.~\ref{M_b_1}):
\begin{equation}
b_n=M_b(\omega_n)=e^{-\omega_n^2 \sigma_{tb}^2/2}.
\label{b_n_narrow}
\end{equation}

The Gaussian approximation is not always most fitting to describe the bunch distribution function. A most useful scheme of bunching a continuous or long pulse beam is modulating its energy with a high intensity laser beam in a wiggler (or another interaction scheme), and then turning its energy modulation to density modulation by passing it through a dispersive section (DS), such as a ``chicane'' (see Appendix~\ref{bunching_append}). This scheme of bunching is useful for a variety of short wavelength radiation schemes, including HGHG \cite{YU_1991,YU_2000}, EEHG \cite{Stupakov_2009,Xiang_2009}, Phase-merging Enhanced Harmonic Generation \cite{Feng_2014,Qika_2008} and e-SASE \cite{Zholents_2005}. Following the notation of Stupakov \cite{Stupakov_2009}, the bunching parameter after the DS is determined in this case by the initial energy spread of the beam $\sigma_{\gamma 0}/\gamma_0$, the compression parameter $B=\omega_b\sigma_t=\omega_b (R_{56}/c)(\sigma_{\gamma 0}/\gamma_0)$ and the energy modulation parameter $A=\Delta\gamma_{mod}/\sigma_{\gamma 0}$, where $\sigma_{\gamma 0}$ is the intrinsic energy spread of the beam before modulation. For optimized bunching of harmonic $n$, (given $n>4$), the dispersion is adjusted so that $AB=1$. In this case a useful expression for the bunching coefficient is (see Appendix~\ref{bunching_append})
\begin{equation}
b_n=\frac{0.67}{n^{1/3}}e^{-n^2B^2/2}
\label{bunching_harmonic_n_no_gaussian_no_append}
\end{equation}

Assuming the beam has a normalized transverse profile distribution $f(\mathbf{r}_{\perp})$. The transverse current density in the wiggler is:
\begin{equation}
\mathbf{J}_{\perp}(\mathbf{r},\omega_0)=\frac{\tilde{I}_{m\, \perp}\hat{\mathbf{e}}_{\perp}}{2} f(\mathbf{r}_{\perp})e^{i(\omega_0/v_z-k_w)z}
\label{periodic_model_J}
\end{equation}
Where:
\begin{equation}
\tilde{I}_{m\, \perp}\hat{\mathbf{e}}_{\perp}=I_0 \tilde{M} \frac{\tilde{\boldsymbol{\beta}}_w}{\beta_z}
\label{periodic_model_Im}
\end{equation}
Writing now the excitation equation in phasor formulation:
\begin{equation}
\tilde{C}_q(z)=\tilde{C}_q(0) -\frac{1}{4\mathcal{P}_q} \int \mathbf{\tilde{J}}_{\perp}(\mathbf{r},\omega_0)\cdot \tilde{\boldsymbol{\mathcal{E}}}^*_q(\mathbf{r_{\perp}})e^{-ik_{zq}z} dV,
\label{mode_exp2}
\end{equation}
One obtains:
\begin{equation}
\tilde{C}_q(z)=\tilde{C}_q(0) -\frac{\tilde{I}_{m\, \perp}}{8\mathcal{P}_q}|\tilde{\boldsymbol{\mathcal{E}}}_q(0)|F_q z e^{i\theta z/2}\sinc(\theta z/2)
\label{mode_exp3}
\end{equation}
where
\begin{equation}
\tilde{I}_{m\, \perp}=|\tilde{I}_{m\, \perp}|e^{i\varphi_{b0}}
\label{angle_of_current}
\end{equation}
and $F_q$ is a field ``filling factor'':
\begin{equation}
F_q=\frac{1}{|\tilde{\boldsymbol{\mathcal{E}}}_q(0)|}\left|\int_{-\infty}^{\infty}\hat{\tilde{\mathbf{e}}}_{\perp}\cdot\tilde{\boldsymbol{\mathcal{E}}}^*_q(\mathbf{r}_{\perp})f(\mathbf{r}_{\perp})d^2 r_{\perp}\right|.
\label{F_func}
\end{equation}
This parameter is close to 1 when the beam is narrow relative to the transverse variation of the mode and diffraction effect is negligible, or in the case of a transversely uniform beam and radiation field (1D model).

The time averaged radiation power will then be given by:
\begin{equation}
P_q(z)=\mathcal{P}_q|\tilde{C}_q(z)|^2=P_q(0)+P_{SP-SR}(z)+P_{ST-SR}(z)
\label{P_q1}
\end{equation}
With the definition (\ref{P_q}) for the effective area of the radiation mode $A_{em\, q}$, the superradiant and stimulated superradiant powers are:
\begin{equation}
P_{SP-SR}(z)=\frac{1}{32}Z_q|\tilde{I}_{m\, \perp}|^2F_q^2\frac{z^2}{A_{em\, q}}\sinc^2(\theta z/2)
\label{P_SR}
\end{equation}
and
\begin{equation}
P_{ST-SR}(z)=\frac{1}{4}E(0)|\tilde{I}_{m\, \perp}|F_q z \cos(\varphi_{qb0}-\theta z/2)\sinc(\theta z/2)
\label{P_ST_SR}
\end{equation}
where
\begin{equation}
E(0)=|\tilde{C}_q(0)||\tilde{\boldsymbol{\mathcal{E}}}_q(0)|
\label{E0}
\end{equation}
and
\begin{equation}
\varphi_{qb0}=\varphi_{q0}-\varphi_{b0}
\label{phi_rb0}
\end{equation}
is the phase difference between the radiation field phase $\varphi_{q0}$ and the bunching current phase
$\varphi_{b0}$ at the entrance to the wiggler.

Maximal power generation is attained for $\theta=0$ and $\varphi_{qb0}=0$ (phase matching between the bunched current and radiation field):
\begin{equation}
P_{SP-SR}(z)=\frac{1}{32}Z_q|\tilde{I}_{m\, \perp}|^2F^2\frac{z^2}{A_{em\, q}}
\label{P_TES_match}
\end{equation}
and
\begin{equation}
P_{ST-SR}(z)=\frac{1}{4}|\tilde{I}_{m\, \perp}|\sqrt{\frac{2 Z_q}{A_{em\, q}}}\sqrt{P_{in}} F_q z 
\label{P_TESSA_match}
\end{equation}
The ratio between the two contributions to the radiation power is 
\begin{equation}
\frac{P_{ST-SR}}{P_{SP-SR}}=8\frac{A_{em\, q}}{Z_q \tilde{I}_{m\, \perp} F_q z}\sqrt{\frac{2 Z_q}{A_{em\, q}}}\sqrt{P_{in}}=8\frac{A_{em\, q}}{Z_q \tilde{I}_{m\, \perp} F_q z}E(0) 
\label{ratio}
\end{equation}

\begin{figure}[!tbh]
\includegraphics[width=8cm]{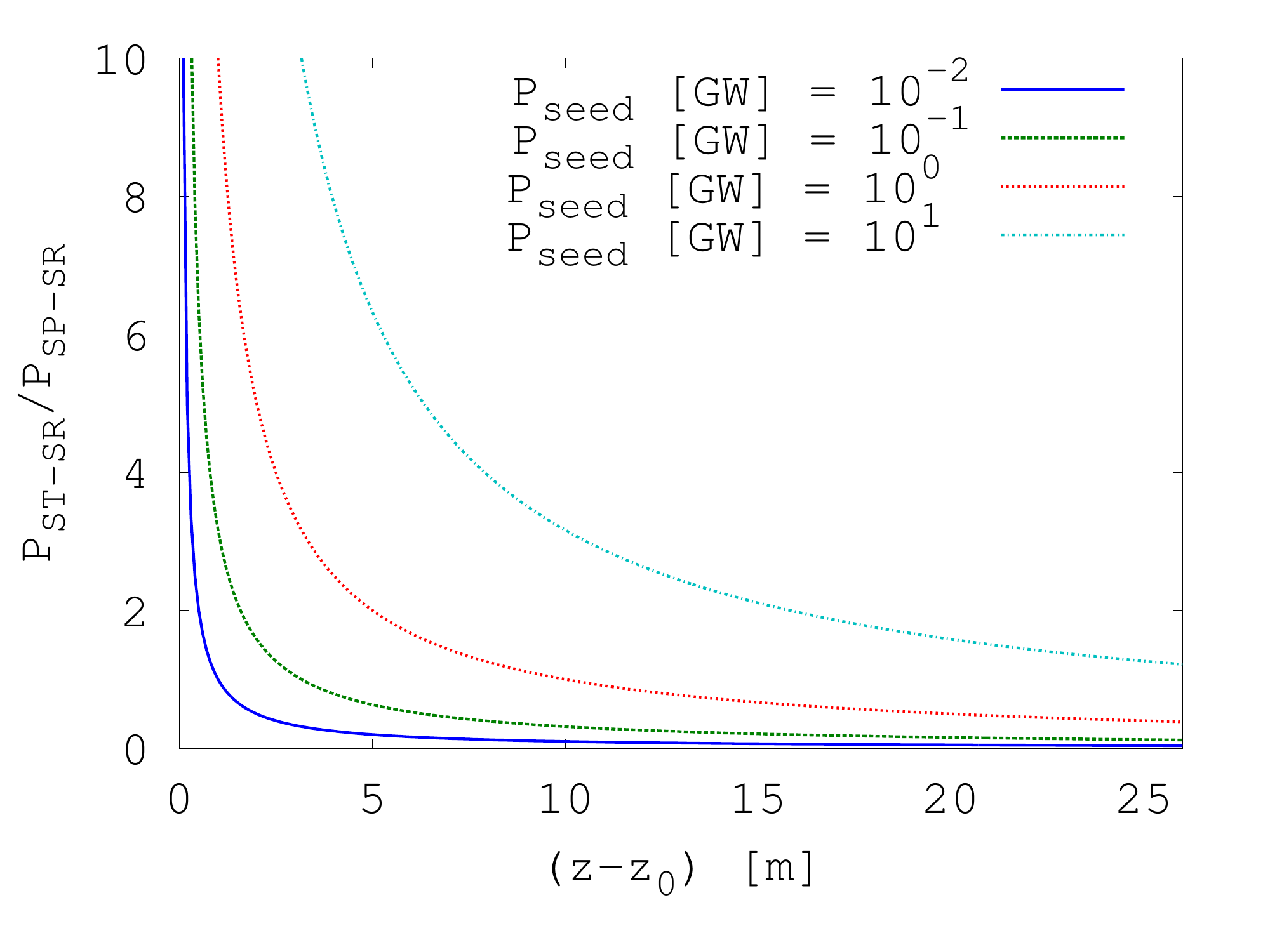}
\caption{Ratio of 0-order ST-SR to SP-SR generated power for different initial input power.}
\label{power_ratio}
\end{figure}

In Figure~\ref{power_ratio} we show the Ratio of 0-order ST-SR to SP-SR for different initial power levels at $z=z_0$.
Initially the ST-SR power dominates the SP-SR power, but evidently, for long interaction length the SP-SR power
that grows like $z^2$, exceeds the ST-SR power that grows like $z$.
At the beginning stages of interaction in the wiggler the ST-SR power may be significantly higher
than the SP-SR power if the initial radiation power $P_{in}$ injected is large enough.
This balance is demonstrated in Figure~\ref{power_ratio} for the parameters of LCLS \cite{emma} (without
tapering).

We point out that in the case of a long wiggler, diffraction effects of the radiation beam become
significant (see section~VIII.C), and a single mode analysis
would be relevant only in the initial part of the wiggler up to a distance of a Rayleigh length. The
superradiant part of the radiation emission was analyzed, including diffraction
effects based on a Gaussian model for the radiation beam in \cite{schneidmiller}. The contribution of the stimulated
superradiance to the radiation emission has been usually ignored in analytic modeling.
Our conclusion on the dominance of this contribution in the initial section of the
wiggler is valid at least up to the distance of a Rayleigh length, within which the single radiation
mode model is valid, and would be valid then only if tight bunching is realizable. More complete review of the
modeling of radiation in the tapered wiggler section of an FEL and the limitations of the 1-D modeling is postponed
to Section~C of Chapter~VIII.

\section{Superradiance and stimulated superradiance in the nonlinear regime}

The underlying assumption in the calculation of spontaneous emission, superradiant spontaneous emission and (zero order)
stimulated superradiant emission is that the beam energy loss as a result of radiation emission is negligible. When
this is not the case, the problem becomes a nonlinear evolution problem. We now extend our model to the case of
a continuously bunched electron beam interacting with a strong radiation field in an undulator, so that the electron
beam loses an appreciable portion of its energy in favor of the radiation field. In this case, the electrons experience
the dynamic force of the radiation wave and change their energy according to the force equation (\ref{force_balance1}) in
Appendix~\ref{cons_energy_append}.
As derived in the conventional theory of FEL \cite{kroll,Pellegrini_2016} the scalar product of the radiation field
$\mathbf{E}(r,t)$ and the wiggling velocity $\mathbf{v}_i(t)$ (Eq.~\ref{und_perp_vel1}) produce a periodic ``beat wave''
force (the ``ponderomotive force''), propagating with phase velocity
\begin{equation}
v_{ph}=\frac{\omega_0}{k_{zq}+k_w}
\label{phase_velocity}
\end{equation}
This force wave can be synchronous with the electron beam near the synchronism condition in Eq.~(\ref{omega_0}) or
(\ref{omega_0_expr}).

Near synchronism, electrons interact efficiently with the sinusoidal ponderomotive force. The dynamics of
this interaction is analyzed and presented in the next section~V. In the present chapter we discuss
qualitatively the conceptual transition from the zero-order regime (no e-beam dynamics) to the non linear
regime. For this we use the ``pendulum'' model that had been earlier developed for FEL theory
\cite{colson,Pellegrini_2016,kroll}. According to this model, the incremental energy of the electron
(off the synchronism energy) and its phase (relative to the sinusoidal ponderomotive wave) satisfy the
``pendulum equations''. The characteristics of the solution of this well-known mathematical equations
and of the ``tilted pendulum equations'' are presented briefly in Appendix~\ref{pend_append}. In analogy to the physical
pendulum, the dynamics of the electron in the ponderomotive wave potential is described in terms of its
trajectories in the phase-space of its detuning parameter $\theta$ (Eq.~\ref{detuning_par1}) and its phase $\psi$ relative to the
ponderomotive wave. The non linear regime saturation process of FEL is explained in Appendix~\ref{pend_append} in terms of the phase-space
trajectories of Figure~\ref{simple_pendulum} that consist of two kinds of trajectories: open and closed (trapped).
The maximal loss of energy of an electron within the trap (and respectively its maximal deviation off synchronism -
$\Delta\theta$ due to the interaction) depends on the height of the trap $\Delta\theta=2\theta_m$. A well bunched
electron beam will release maximum energy (transformed to radiation), if inserted into the trap near
synchronism at phase $\psi=0$ and detuning parameter near $-\theta=\theta_m$ corresponding to the top of the trap
(Figure~\ref{simple_pendulum}), and winds-up at the bottom of the trap at the end of the interaction length
(the wiggler).

\begin{figure*}[!tbh]
\includegraphics[width=18cm]{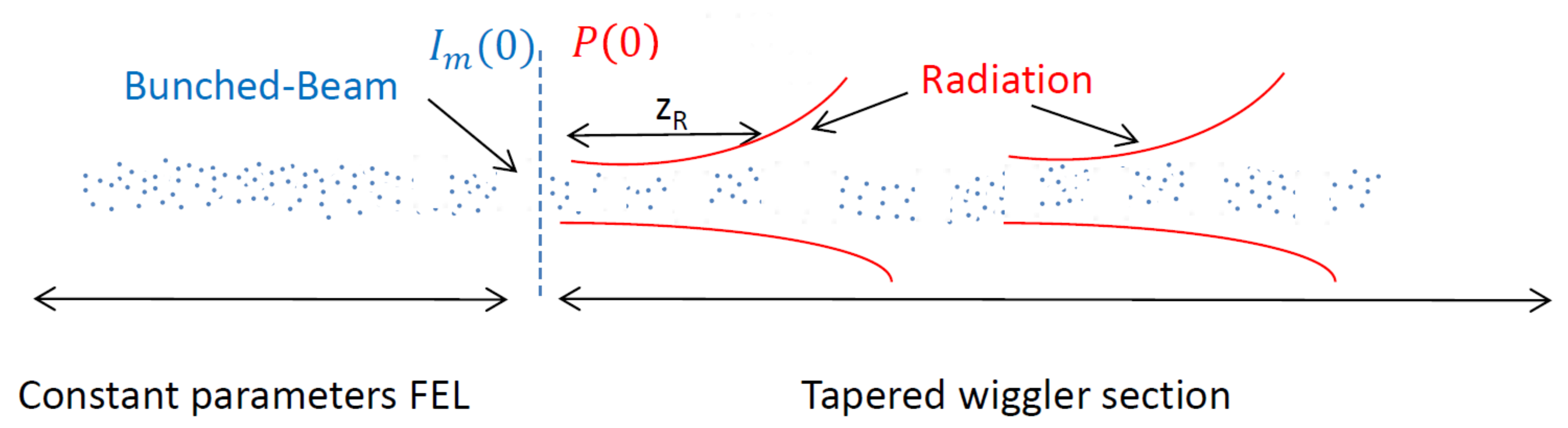}
\caption{Schematics of seed-injected FEL followed by a tapered wiggler: at the end
of the constant parameters wiggler, the partially bunched e-beam and the amplified radiation wave are
injected into a tapered wiggler section, where further radiation energy is extracted out of the bunched beam.}
\label{const_vs_tap}
\end{figure*}

Extending the technology of uniform wiggler FEL, a scheme of ``tapered wiggler'' has been developed in the
field of free electron lasers for extracting higher radiation energy
from the beam \cite{kroll} beyond the maximal energy extraction efficiency of a uniform FEL. The experimental realization
of this scheme is
described schematically in Figure~\ref{const_vs_tap} for the case of a tapered wiggler FEL. After a uniform wiggler,
the waisted beam that is partly bunched due to the interaction in the first section, continues to interact along a tapered wiggler with the coherent
radiation wave that was generated in the first section, and is further amplified in the second section. In such a scheme, the wiggler
wavenumber $k_w(z)$ is increased along the tapered wiggler section, so that the ponderomotive phase
velocity $v_{ph}$ (Eq.~\ref{phase_velocity}) goes down gradually, keeping synchronism with the correspondingly
slowing down electrons, trapped in the ponderomotive wave, so that the synchronism condition in Eq.~(\ref{omega_0})
can be kept all along:
\begin{equation}
\theta(z)=\int_0^z\left[\frac{\omega_0}{v_z(z')}-k_{zq}(\omega_0)-k_w(z')\right]dz' \simeq 0
\label{near_synchronism}
\end{equation}

The trapped electrons dynamics and energy extraction process in this scheme can be presented in terms of
the ``tilted pendulum'' model (Appendix~\ref{pend_append}). They are described quantitatively in terms of their trajectories
in phase-space in Figure~\ref{kmr}, that shows that the electrons can stay trapped (though the trap is
somewhat shrinked) and still keep decelerating along the tapered wiggler, keeping near synchronism with the slowing down
ponderomotive wave. Note that in (\ref{near_synchronism}) $v_z$ is the axial velocity of the beam averaged over
the wiggler period. In a linear wiggler with $\overline{a}_w \gg 1$ the linear transverse wiggling gives rise to
longitudinal periodic quiver of $v_z(z)$ and a consequent radiative interaction at odd harmonic frequencies
\cite{Pellegrini_2016,encyclopedia}. For simplicity we ignore here these harmonic interactions.
Also, in using the pendulum equation model to describe the dynamics of the electron inside the trap
(Synchrotron oscillation), it is implicitly assumed that the pendulum oscillation period
(Synchrotron period - $\lambda_S$) is much longer than the wiggler period: $\lambda_S\gg \lambda_w$.

For completion of this short review of tapered wiggler FEL, we point out that besides tapering the wiggler
wavenumber $k_w(z)$, enhanced energy extraction efficiency
of saturated FEL is possible also in an alternative scheme of magnetic field wiggler parameter tapering. In this scheme,
the period of the tapered wiggler stays constant, but the strength of
the magnetic field and correspondingly the wiggler parameter $a_w(z)=|\mathbf{\tilde{a}}_w(z)|$
(Eq.~\ref{v_perp1}) is reduced gradually along the wiggler so that
$\gamma_z^2(z)=\gamma^2(z)/[1+a_w^2(z)]$ remains constant in Eq.~\ref{gamma_z2}. Since
$\beta_z=[1-1/\gamma_z^2]^{1/2}$, the detuning synchronism condition (\ref{omega_0}) or (\ref{omega_0_expr})
can be maintained along the interaction length despite the decline of the beam energy.

For a free-space wave, propagating on-axis: $k_{zq}=k_0=\omega_0/c$, and then from Eq.~(\ref{detuning_par1}):
\begin{equation}
\theta(z)=k_0\left[\beta_z^{-1}(z)-1\right]-k_w(z).
\label{theta_gamma_exact}
\end{equation}
where $k_0=\omega_0/c$. The synchronism condition $\theta(z)=0$ defines an energy synchronism condition between
the electron beam and the wave for a general case of either period or field tapered wiggler:
\begin{equation}
\gamma_r(z)=\sqrt{\frac{1+\overline{a}_w^2(z)}{1-(1+k_w(z)/k_0)^{-2}}}\simeq\sqrt{\frac{1+\overline{a}_w^2(z)}{2} \frac{k_0}{k_w(z)}},
\label{gamma_r}
\end{equation}
where we used the identities $\gamma_z=(1-\beta_z^2)^{-1/2}$ and $\gamma=\gamma_z/(1+\overline{a}_w^2)^{1/2}$, and
the second part simplification of the equation corresponds to the ultra-relativistic beam limit.

Assuming that the electrons are trapped, so that in the presence of the radiation field, they stay with
energy close to the synchronism energy $\gamma_r$ we write
\begin{equation}
\gamma=\gamma_r+\delta\gamma
\label{delta_gamma}
\end{equation}
and therefore the connection between the dynamic energy exchange of the electron within the trap $\delta\gamma(z)$ and
the detuning parameter relative to the slowing down ponderomotive wave $\theta(z)$ is:
\begin{equation}
\theta=\left.\frac{d\theta}{d\gamma}\right|_{\gamma_r}\delta\gamma=-\frac{k_0}{\beta_{zr}^3\gamma_{zr}^2\gamma_r}\delta\gamma\simeq -2k_w(z)\frac{\delta\gamma}{\gamma_r(z)},
\label{theta_gamma}
\end{equation}
(the approximate expression is for the ultra-relativistic case where $\beta_{zr}\simeq 1$, $k_w=k_0/(2\gamma_{zr}^2)$).
The synchronism energy $\gamma_r(z)$ is the energy of an electron moving at exact synchronism with the ponderomotive
wave phase velocity (``fully trapped'').

In the following chapters we analyze the dynamic processes of a tightly bunched electron beam trapped in the
ponderomotive potential of a uniform or tapered wiggler. Tight bunching of the beam relative to the period of
the ponderomotive wave would allow determination of the bunching phase relative to the ponderomotive phase
and corresponding optimization of superradiant and stimulated superradiant processes. However, such tight
bunching is hard to come in the present technological state of the art.

The tight bunching model presented in the next chapter can describe quite well recent experiments of Tapering-Enhanced
Superradiant Amplification TESSA and inverse FEL bunched beam acceleration demonstrated on the
RUBICON and NOCIBUR set-ups in ATF/BNL \cite{nocibur,rubicon}, that are described in Section~B of Chapter~VIII. Here very tight pre-bunching was attained using a high
intensity 10.6$\mu$ CO$_2$ laser. In the case of seed injected tapered wiggler FEL the tight bunching model is presently only partly relevant to describe the dynamics in the
tapered wiggler section. 
In this case (see Figure~\ref{const_vs_tap}), both coherent radiation field and a bunched e-beam are inserted into
the tapered wiggler section from the uniform wiggler (linear gain) section. Both superradiant (SP-SR)
and stimulated superradiant (ST-SR) radiation would have been emitted from the tapered section, if the bunching
produced in the uniform wiggler section of the FEL is tight enough and if phase-shifter technology
 \cite{Ratner_2007,Curbis_2017} can be harnessed to adjust the phase of the bunching relative to the
radiation field, it could be used to enhance the ST-SR emission process. However, tight bunching is hard to
get, and the input field intensity and phase
of the bunching 
are not independently controlled in present day tapered wiggler X-Ray FEL facilities, as is necessary for optimizing ST-SR. These,
as well as the maintenance of small enough energy spread of the bunched beam when it enters
the tapered section, are hard to control in present X-Ray FEL facilities. Furthermore, transverse effects, primarily 
wave diffraction (see Figure~\ref{const_vs_tap}) are significant in the long section of the tapered undulator, despite
the short wavelength of the radiation: they require extension of the single mode analysis to a multimode
formulation or full 2D or 3D solution of Maxwell equations \cite{Chen_2014,Emma_2014,Tsai_2018}. 
These limitations are discussed in more
detail here in Section C of Chapter VIII. Thus the presented ideal model of the
distinct SP-SR and ST-SR radiation extraction schemes serve only for qualitative identification of tapering optimization strategies.

\section{Formulation of the Dynamics of a periodically bunched electron beam interacting with radiation field in a general wiggler}

In this section we extend the analysis of SP-SR and ST-SR in undulator radiation of a periodically bunched
beam, that was presented in section III based on radiation mode excitation and phasor formulations,
and we add the dynamics of the electrons under interaction with the radiation wave. Beyond the qualitative
introduction of the pendulum equation in Section~IV, we develop here master equations for the
coupled radiation field and periodically bunched beam.

Solving now for the axial ($z$ coordinate) evolution of the
bunched beam in steady-state, we assume that the infinite periodically bunched
beam is composed of all identical bunches (namely, shot-noise and finite pulse effects are
neglected). The bunches are tightly bunched, hence they can be modeled as Dirac delta functions
(see Appendix~\ref{cons_energy_append}, Eqs.~\ref{J_t_single}, \ref{J_t_single_z}). They
all experience the same force equation and have the same trajectories as macro-particles of charge
$Q_b=-eN_b$ and the time interval between two consecutive injected bunches is $T_b\equiv2\pi/\omega_0$,
therefore
\begin{equation}
\mathbf{J}(\mathbf{r},t)=Q_b\mathbf{v}_e(t)f(\mathbf{r}_{\perp})\sum_{n=-\infty}^{\infty}\delta[z-z_e(t-nT_b-t_0)]
\label{J_r_t}
\end{equation}
where we use $f(\mathbf{r}_{\perp})$ in order to represent a beam of finite transverse distribution, as in
Eq.~\ref{periodic_model_J}.

With these simplifying assumptions, the phasor mode excitation equation (\ref{dCdz}) can be employed to any harmonic
frequency of the radiation emitted by the current (\ref{J_r_t}) for calculating the radiation power
(\ref{power_all_modes}). As we show in Appendix~\ref{cons_energy_append}, this radiation power expression, combined with the beam energy exchange rate, derived from the force equation on the bunches:
\begin{equation}
N_bmc^2\frac{d\gamma}{dt}=Q_b\mathbf{v}\cdot \mathbf{E}(\mathbf{r},t),
\label{force_balance}
\end{equation}
result in exact conservation of power exchange between the radiation power $P(z)$ and the beam power
$P_e=N_bmc^2(\gamma-1)/T_b$, so that
\begin{equation}
\frac{dP}{dz}=-\frac{dP_e}{dz}
\label{power_cons}
\end{equation}

Quite remarkably, this result is shown in generality for bunched beam interaction with the
radiation field (either external or self generated by the beam) in any kind of radiation mechanism. It
demonstrates the rigurousity of the mode expansion formulation of Maxwell equation
(Eqs.~\ref{EqHq}-\ref{H_r_omega}) and its consistency
with the simplified bunched beam dynamics model.

The excitation equation for interaction of a tightly bunched periodic beam (Eq.~\ref{J_r_t}) for interaction
in a wiggler is derived in Appendix~\ref{cons_energy_append} (Eq.~\ref{dCqdz}) (for simplicity we assume from now on $F_q=1$):
\begin{equation}
\frac{d\tilde{C}_q(z)}{dz}=-\frac{Q_b\omega_0\tilde{\boldsymbol{\beta}}_w(z)\cdot\tilde{\boldsymbol{\mathcal{E}}}_q^*(0)}{8\pi \mathcal{P}_q \beta_{zr}}e^{i\varphi_b(z)},
\label{dCqdz_non_app}
\end{equation}
where the beam bunching phase relative to the ponderomotive wave dynamically changes as a function of $z$
because of the tapering and because of the energy change in the nonlinear regime:
\begin{equation}
\varphi_b(z)=\int_0^z \left(\frac{\omega_0}{v_z(z')}-k_w(z')-k_{zq}\right) dz' + \varphi_{b0},
\label{phi_non_app}
\end{equation}

We define the dynamic detuning parameter, consistent with Eq.~(\ref{detuning_par1})
\cite{encyclopedia}
\begin{equation}
\theta(z)\equiv\frac{d\varphi_b}{dz}=\frac{\omega_0}{v_z(z)}-k_w(z)-k_{zq}.
\label{dpsi_dz}
\end{equation}
The rate of change of the bunches energy is found in generality based on Appendix~\ref{cons_energy_append} by substituting
\begin{equation}
\mathbf{E}(\mathbf{r},t_e(z))=Re\left[\tilde{C}_q(z)\tilde{\boldsymbol{\mathcal{E}}}(\mathbf{r}_{\perp})e^{-i\int_0^z(\omega_0/v_z(z')-k_z)dz'-i\varphi_{b0}}\right],
\label{E_r_te}
\end{equation}
and (\ref{v_perp}) in Eq.~(\ref{force_balance2}), which for a thin beam ($\mathbf{r}_{\perp}=0$) results in
\begin{equation}
mc^2\frac{d\gamma}{dz}=\frac{1}{2\beta_{zr}}(-e)\eta_p|\tilde{\boldsymbol{\beta}}_w||\tilde{\boldsymbol{\mathcal{E}}}_q(0)||\tilde{C}_q(z)|\cos[\varphi_{qb}(z)],
\label{dP_b_dz1}
\end{equation}
where
\begin{equation}
\varphi_{qb}(z)=\varphi_q(z)-\varphi_b(z).
\label{varphi_qb_z}
\end{equation}
It is evident that the rate of beam energy change (\ref{dP_b_dz1}) depends both on the amplitude of the field
$|\tilde{C}_q(z)|$ and on the $z$-dependent relative phase $\varphi_{qb}(z)$, which is the dynamic phase of the
radiation field relative to the bunching current (consistent with (\ref{phi_rb0})). The phase $\varphi_b(z)$ is defined
in (\ref{phi}) and $\varphi_q(z)$ is the z-dependent phase of the radiation complex amplitude, so that:
\begin{equation}
\tilde{C}_q(z)=|\tilde{C}_q(z)|e^{i\varphi_q(z)}.
\label{C_q_amp_phase}
\end{equation}

The polarization match factor $\eta_p$ is defined by
\begin{equation}
\eta_p=\frac{|\tilde{\boldsymbol{\beta}}_w^*\cdot\tilde{\boldsymbol{\mathcal{E}}}_q(0)|}{|\tilde{\boldsymbol{\beta}}_w||\tilde{\boldsymbol{\mathcal{E}}}_q(0)|}
\label{wigg_mode_match}
\end{equation}

It is useful at this point to redefine the interaction coordinate z-dependent varying phase of the ponderomotive wave relative to the varying phase of the bunches as:
\begin{equation}
\psi\equiv \varphi_{qb}(z)+\pi/2=\psi(0)+[\varphi_q(z)-\varphi_q(0)]-\int_0^z\theta(z')dz',
\label{psi}
\end{equation}
so that $\psi(0)=\varphi_{qb0}+\pi/2$, where $\varphi_{qb0}=\varphi_{q0}-\varphi_{b0}$ is the phase of the radiation mode relative to the
bunching at $z=0$ (see Eq.~\ref{phi_rb0}). The $\pi/2$ phase shift corresponds to relating the bunches to the radiation vector potential or the ponderomotive wave potential, rather
than to the electric field phase $\varphi_q(z)$ ($\mathbf{\tilde{E}}=-i\omega \mathbf{\tilde{A}}$).

Since at present we confine the analysis to interaction with a single mode, we simplify the notation for the field amplitude:
\begin{equation}
\tilde{E}(z)=\tilde{C}_q(z)|\tilde{\boldsymbol{\mathcal{E}}}_q(0)|
\label{tilde_E_z}
\end{equation}
where $\tilde{E}(z)$ is the complex radiation field amplitude on and along the beam axis.
This results in (similarly to KMR \cite{kroll}) the coupled beam and wave equations:
\begin{equation}
\frac{d\gamma}{dz}=-a|\tilde{E}(z)|\sin\psi
\label{dgamma_dz_psi}
\end{equation}
\begin{equation}
\frac{d\tilde{E}(z)}{dz}=b e^{i\varphi_b(z)}=ibe^{i[\varphi_q(z)-\psi]},
\label{dCqdz_psi}
\end{equation}
where
\begin{equation}
a=\frac{e\eta_p}{2\beta_{zr}\gamma_r mc^2}\overline{a}_w(z)
\label{AAA1}
\end{equation}
\begin{equation}
 b=\frac{|Q_b|\omega_0\eta_p\overline{a}_w(z)|\tilde{\boldsymbol{\mathcal{E}}}_q(0)|^2}{8\pi\mathcal{P}_q\beta_{zr}\gamma_r}=
\frac{I\eta_p\overline{a}_w(z)Z_q}{2A_{em\,q}\beta_{zr}\gamma_r} ,
\label{BBB}
\end{equation}
where we used $|\tilde{\boldsymbol{\beta}}_w|=\overline{a}_w(z)/\gamma$, $|Q_b|=eN_b$ and $I$ is the current.
Here $\psi(z)$ (Eq.~\ref{psi})
is the phase of the bunch relative to the vector potential of the wave $\mathbf{A}(t)$ (or relative to the electric field of
the wave with $\pi/2$ phase shift), and $\varphi_q(z)$ is the phase of the radiation wave (\ref{C_q_amp_phase}) that may also change dynamically due to the interaction.

The complex equation (\ref{dCqdz_psi}) can be broken into two equations for the modulus and the phase of the radiation mode:
\begin{equation}
\frac{d|\tilde{E}|}{dz}=b\sin\psi,
\label{dabsCqdz2_dev}
\end{equation}
\begin{equation}
\frac{d\varphi_q}{dz}=\frac{b}{|\tilde{E}|}\cos\psi,
\label{dphi_E_dz2}
\end{equation}

The detailed solution of the problem includes an iterative calculation of the beam energy ($\gamma$) (Eq.~(\ref{dgamma_dz_psi}))
and the radiation mode amplitude $|\tilde{E}|$ (Eq.~\ref{dabsCqdz2_dev}) that are coupled to each other through the phase
$\psi(z)$ (Eq.~\ref{psi}) and the definition of the detuning parameter $\theta(z)$ (Eqs.~\ref{delta_gamma},\ref{theta_gamma}).

Note that only
the initial phase of the mode relative to the bunching $\varphi_{qb0}=\varphi_q(0)-\varphi_b(0)$ is required for the determination
of the initial condition $\psi(0)=\varphi_{qb0}+\pi/2$. After the solution of the coupled equations the phase variation
of the mode can always be calculated by integration
\begin{equation}
\varphi_q(z)=\phi_q(0)+\int_0^z\frac{b}{|\tilde{E}(z')|}\cos\psi(z') dz'.
\label{dphi_E_dz2_int}
\end{equation}

\subsection{Uniform wiggler}

In this subsection we specify to a uniform wiggler, and therefore $\gamma_r$ (see Eq.~\ref{gamma_r}) is independent of $z$,
so that
\begin{equation}
\frac{d\theta}{dz}=-\frac{k_0}{\beta_{zr}^3\gamma_{zr}^2\gamma_r}\frac{d\delta\gamma}{dz}=-\frac{k_0}{\beta_{zr}^3\gamma_{zr}^2\gamma_r}\frac{d\gamma}{dz}.
\label{theta_gamma1}
\end{equation}

The total power of the electron beam can be expressed as
\begin{equation}
P_{el}=\frac{1}{T_b}N_bmc^2(\gamma-1)\simeq \frac{1}{T_b}N_bmc^2\gamma,
\label{P_el}
\end{equation}
and using $\gamma=\gamma_r+\delta\gamma$, it is written as
\begin{equation}
P_{el}=\frac{1}{T_b}N_bmc^2\gamma=\frac{1}{T_b}N_bmc^2(\gamma_r+\delta\gamma)=
\frac{1}{T_b}N_bmc^2\left(\gamma_r-\frac{\beta_{zr}^3\gamma_{zr}^2\gamma_r}{k_0}\theta\right),
\label{P_el1}
\end{equation}

Using Eq.~\ref{theta_gamma1}, the energy equation (\ref{dgamma_dz_psi}) can be written in terms of the detuning parameter:
\begin{equation}
\frac{d\theta}{dz}=K_s^2(z)\sin\psi,
\label{dtheta_dz}
\end{equation}
where
\begin{equation}
K_s^2(z)=\frac{k_0e\eta_p}{2\beta_{zr}^4\gamma_{zr}^2\gamma_r^2 mc^2}\overline{a}_w|\tilde{E}(z)|
\label{K_s_2}
\end{equation}
is the synchrotron oscillation wavenumber.

We summarize here the equations to be solved in terms of $\theta(z)$ or $\delta\gamma(z)$:
\begin{equation}
\frac{d|\tilde{E}|}{dz}=b\sin\psi,
\label{dabsCqdz2}
\end{equation}
\begin{equation}
\frac{d\theta}{dz}=K_s^2(z)\sin\psi \,\,\,\,\,\,\,\,\,\,\text{or}\,\,\,\,\,\,\,\,\,\, \frac{d\delta\gamma}{dz}=-\frac{\beta_{zr}^3\gamma_{zr}^2\gamma_r}{k_0}K_s^2(z)\sin\psi,
\label{dtheta_dz2}
\end{equation}
\begin{equation}
\frac{d\psi}{dz}=-\theta+\frac{b}{|\tilde{E}|}\cos\psi \,\,\,\,\,\,\,\,\,\,\text{or}\,\,\,\,\,\,\,\,\,\, \frac{d\psi}{dz}=\frac{k_0}{\beta_{zr}^3\gamma_{zr}^2\gamma_r}\delta\gamma+\frac{b}{|\tilde{E}|}\cos\psi
\label{dpsi_dz2}
\end{equation}
Equations~(\ref{dphi_E_dz2}) and (\ref{dpsi_dz2}) seem to be singular for the special case of
$\tilde{E}(0)=0$, corresponding to a case of spontaneous emission and self interaction of the e-beam
bunch train with its own generated radiation. As we explain in Section VI.G, this
singularity is removable, and the formulation is valid also for the case of spontaneous emission and
self interaction.

Note that in the case that $\tilde{E}(z)\simeq$const these equations reduce to regular pendulum equations
for the bunches (Appendix~\ref{pend_append}).

\subsection{Tapered wiggler}

In the case of a tapered wiggler , the synchronism energy $\gamma_r$ is a function of $z$ (Eq.~\ref{gamma_r}).
Using Eqs.~(\ref{delta_gamma})
and (\ref{theta_gamma}), one obtains for the dynamics of the detuning parameter:
\begin{equation}
\frac{d\theta}{dz}=-\frac{k_0}{\beta_{zr}^3\gamma_{zr}^2\gamma_r}\left[\frac{d\gamma}{dz}-\frac{d\gamma_r}{dz}\right]-\delta\gamma \frac{d}{dz}\left[\frac{k_0}{\beta_{zr}^3\gamma_{zr}^2\gamma_r}\right]\simeq -\frac{k_0}{\beta_{zr}^3\gamma_{zr}^2\gamma_r}\left[\frac{d\gamma}{dz}-\frac{d\gamma_r}{dz}\right],
\label{theta_gamma1_tap}
\end{equation}
where the last equality is obtained by assuming that the energy tapering rate is slow relative to the synchrotron
oscillation dynamics near synchronism, hence neglecting the second term in (\ref{theta_gamma1_tap}).
Using Eq.~(\ref{dP_b_dz1}) we obtain
\begin{equation}
\frac{d\theta}{dz}=K_s^2(z)\sin\psi+\frac{k_0}{\beta_{zr}^3\gamma_{zr}^2\gamma_r}\frac{d\gamma_r}{dz}.
\label{dtheta_dz_tap}
\end{equation}
Hence Eqs.~(\ref{dabsCqdz2})-(\ref{dpsi_dz2}) remain unchanged, except for Eq.~(\ref{dtheta_dz2})
which becomes
\begin{equation}
\frac{d\theta}{dz}=K_s^2(z)\left[\sin\psi + \frac{k_0}{\beta_{zr}^3\gamma_{zr}^2\gamma_r K_s^2}\frac{d\gamma_r}{dz}\right],
\label{dtheta_dz2_tap1}
\end{equation}
The second term in (\ref{dtheta_dz2_tap1}) adds a slope to the pendulum equation potential, and if this slope is too big there cannot
be trapped trajectories. This puts a limit on the tapering strength, so that the absolute value of the
term which adds to $\sin\psi$ in Eq.~(\ref{dtheta_dz2_tap1}) must be smaller than 1, and therefore it is
useful to define it as:
\begin{equation}
\sin\psi_r \equiv -\frac{k_0}{\beta_{zr}^3\gamma_{zr}^2\gamma_r K_s^2}\frac{d\gamma_r}{dz}.
\label{sin_psi_r}
\end{equation}
With the simplifying assumptions $\beta_{zr}$, $\eta_p=1$ and using (\ref{K_s_2}) with $|\tilde{E}(z)|=E$=const,
the tapering resonant phase can be expressed as:
\begin{equation}
\sin\psi_r \equiv -2\frac{mc^2}{eE}\frac{\gamma_r}{\overline{a}_w}\frac{d\gamma_r}{dz}.
\label{sin_psi_r1}
\end{equation}

Hence we only need to add a term to Eq.~\ref{dtheta_dz2} and we rewrite here the master equations for the tapering
case
\begin{equation}
\frac{d|\tilde{E}|}{dz}=b\sin\psi,
\label{dabsCqdz2_tap}
\end{equation}
\begin{equation}
\frac{d\theta}{dz}=K_s^2(z)\left[\sin\psi - \sin\psi_r\right] \,\,\,\,\,\,\,\,\,\,\text{or}\,\,\,\,\,\,\,\,\,\, \frac{d\delta\gamma}{dz}=-\frac{\beta_{zr}^3\gamma_{zr}^2\gamma_r}{k_0}K_s^2(z)\left[\sin\psi - \sin\psi_r\right],
\label{dtheta_dz2_tap}
\end{equation}
\begin{equation}
\frac{d\psi}{dz}=-\theta+\frac{b}{|\tilde{E}|}\cos\psi \,\,\,\,\,\,\,\,\,\,\text{or}\,\,\,\,\,\,\,\,\,\, \frac{d\psi}{dz}=\frac{k_0}{\beta_{zr}^3\gamma_{zr}^2\gamma_r}\delta\gamma+\frac{b}{|\tilde{E}|}\cos\psi
\label{dpsi_dz2_tap}
\end{equation}
In a radiation emitting wiggler (as opposed to an accelerator scheme \cite{inv_FEL}), the electrons
lose energy, hence one will usually design $d\gamma_r/dz<0$, so that $0<\psi_r<\pi/2$.

Except for Eq.~(\ref{dtheta_dz2}) that is replaced by Eq.~(\ref{dtheta_dz2_tap}),
the other master
equations to be solved are unchanged, but note that
the coefficients of $|\tilde{E}|$ in Eq.~(\ref{K_s_2}) ($\gamma_r(z)$, $\gamma_{zr}(z)$, $\beta_{zr}(z)$,
$\overline{a}_w(z)$) are $z$-dependent in the case of tapering, and so is the parameter $b$ (Eq.~(\ref{BBB}))
if $\overline{a}_w=\overline{a}_w(z)$.

The power of the electron bunches is still according to Eq.~(\ref{P_el1}), but here $\gamma_r$ is a function
of $z$, therefore the kinetic power exchanged is composed of the contribution of the tapered deceleration of the
trap (first term) plus the contribution of the dynamics of the bunch within the trap (second term):
\begin{align}
\frac{dP_{el}}{dz}=&\frac{N_b}{T_b}mc^2\frac{d\gamma}{dt}=\frac{1}{T_b}N_bmc^2 \frac{\beta_{zr}^3\gamma_{zr}^2\gamma_r}{k_0}\left[\frac{k_0}{\beta_{zr}^3\gamma_{zr}^2\gamma_r} \frac{d\gamma_r}{dz}  - \frac{d\theta}{dz}\right]= \notag \\
& -\frac{1}{T_b}N_bmc^2 \frac{\beta_{zr}^3\gamma_{zr}^2\gamma_r}{k_0}\left[K_s^2\sin\psi_r+\frac{d\theta}{dz}\right] ,
\label{dP_el_dz_tap}
\end{align}

\section{Analysis of the interaction dynamics of a bunched electron beam with radiation in the trapping regime}

\subsection{The fundamental radiation processes in phase-space}

\begin{figure}[!tbh]
\includegraphics[width=18cm]{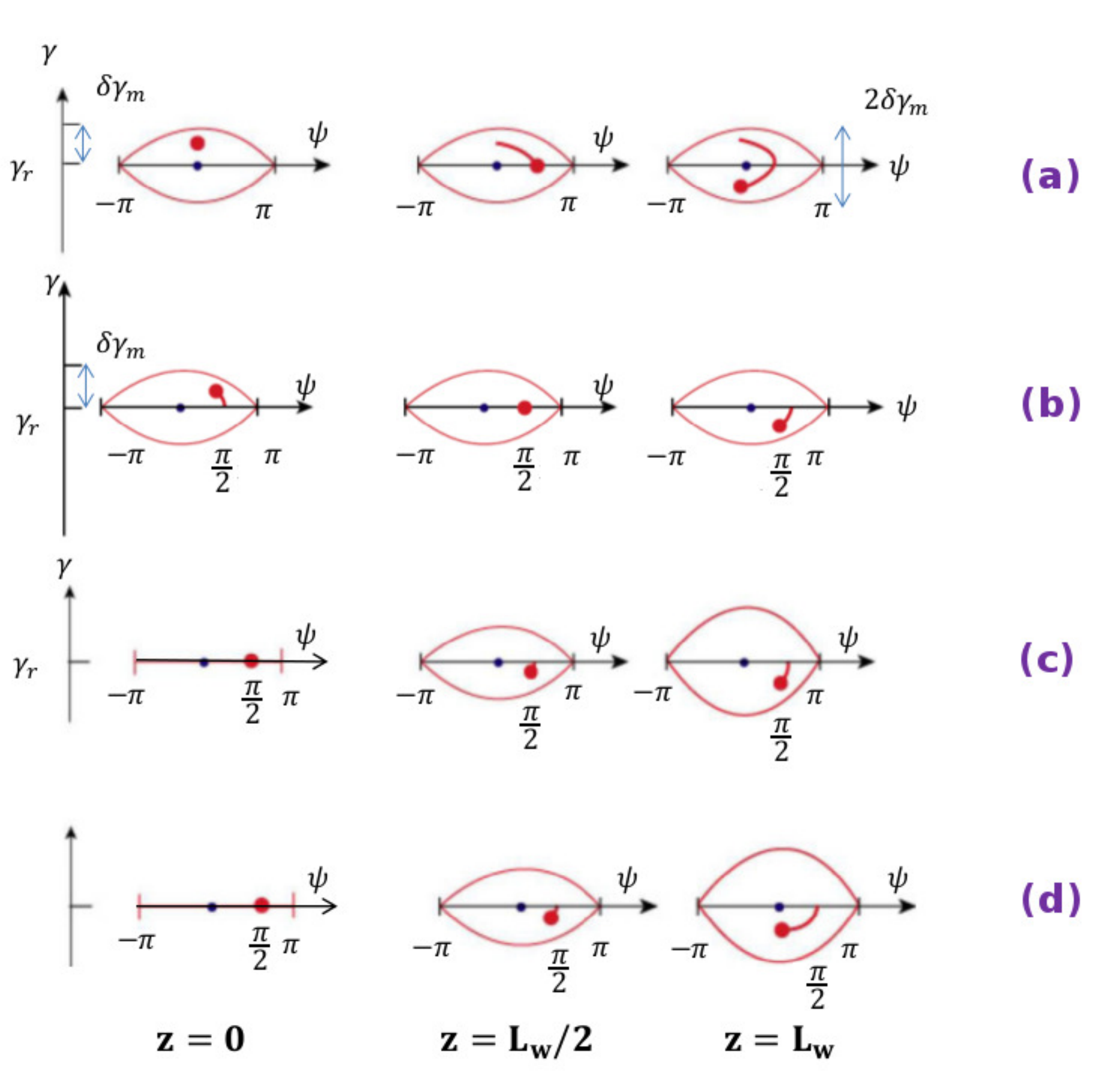}
\caption{Phase-space diagrams for uniform wiggler.}
\label{phase_space_general_non_tap}
\end{figure}

\begin{figure}[!tbh]
\includegraphics[width=18cm]{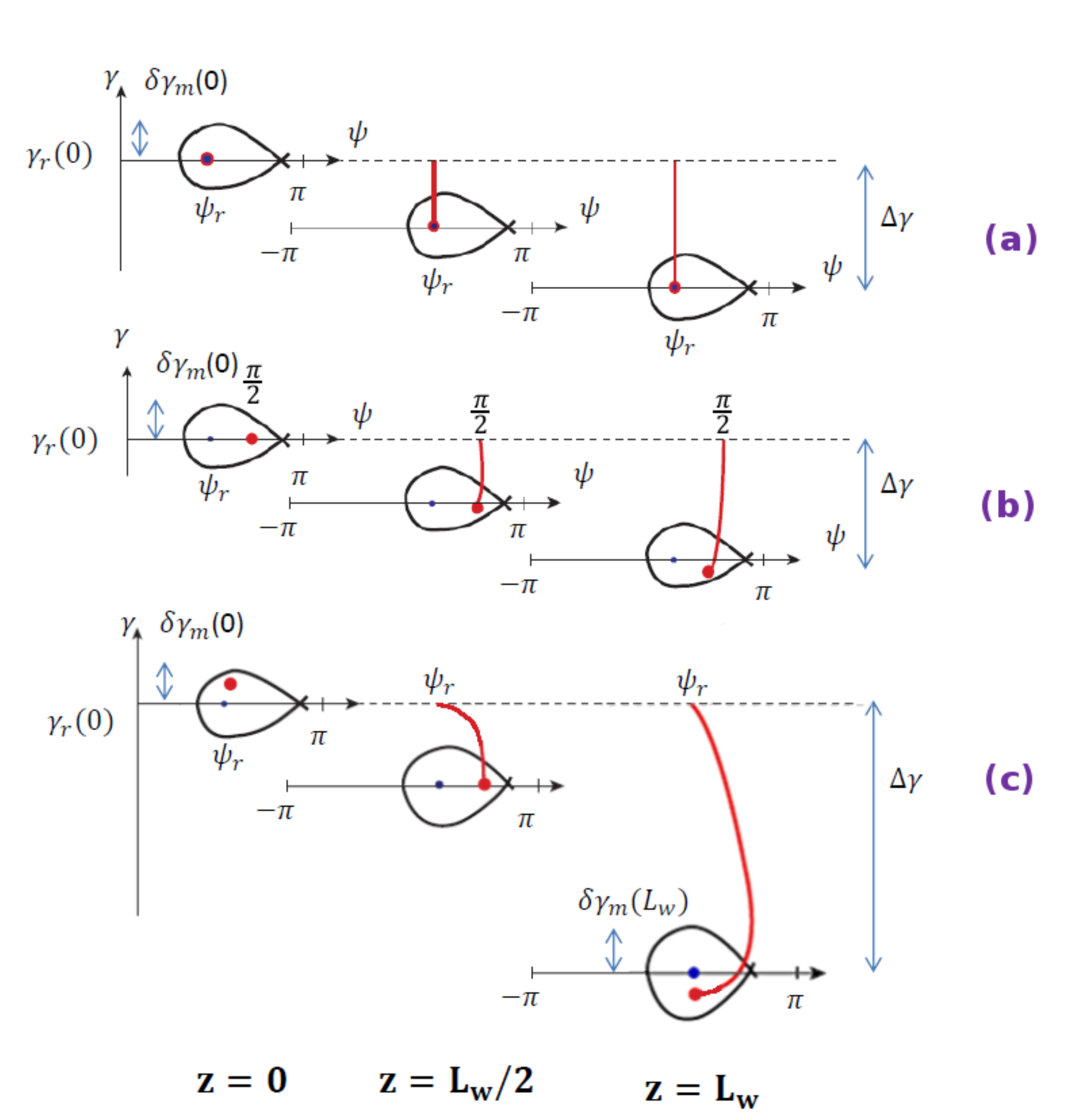}
\caption{Phase-space diagrams for a tapered wiggler.}
\label{phase_space_general_tap}
\end{figure}

We analyze in this section the phase-space dynamics of the bunched electron beam
that come out of the solution of the coupled equations of section~V, and relate them to the
fundamental coherent spontaneous radiation emission processes presented in the first sections (Chapters~II, III). Qualitatively, we expect
specific phase-space dynamic processes as depicted in Figures~\ref{phase_space_general_non_tap}
and \ref{phase_space_general_tap}.
For a uniform wiggler the trap height in the $\theta-\psi$ plane is $2\theta_m=4K_s$ (see Appendix~\ref{pend_append},
Figure~\ref{simple_pendulum} and Eq.(\ref{trap_height})). The generalization for a tapered wiggler is
$2\theta_m = 4 K_s\sqrt{\cos\psi_r+(\psi_r-\pi/2)\sin\psi_r}$ (see Appendix~\ref{pend_append},
Figure~\ref{tap_separatrix} and Eq.(\ref{theta_m})).

Using Eq.~(\ref{K_s_2}) for $K_s^2$ and the connection between $\theta$ and $\delta\gamma$ in Eq.~(\ref{theta_gamma})
one finds $\delta\gamma_m$ at the wiggler's entrance:
\begin{equation}
\delta\gamma_m=\sqrt{2\frac{e\eta_p \beta_{zr}^2\gamma_{zr}^2\overline{a}_w E(0)}{mc^2 k_0}}\sqrt{\cos\psi_r+(\psi_r-\pi/2)\sin\psi_r}\simeq \sqrt{\frac{e\eta_p\overline{a}_w E(0)}{mc^2 k_w}}\sqrt{\cos\psi_r+(\psi_r-\pi/2)\sin\psi_r}
\label{delta_gamma_m}
\end{equation}
and the second part simplification of the equation corresponds to the ultra-relativistic beam limit.

In section VI-B we formulate a normalized version of the bunched-beam - radiation coupled equations, and in the subsequent sections we demonstrate the phase-space
evolution dynamics of these processes in a uniform and tapered wiggler, by presenting the
$z$ dependent numerical computation solutions of the normalized coupled mode equations and via the linked video displays.

\subsection{Simulation of the dynamics and radiation of a perfectly periodically bunched beam in the saturation regime}

For the purpose of demonstrating the fundamental dynamic processes of SP-SR and ST-SR in uniform and tapered wiggler, described in the previous
section, we present simulation results and video displays based on numerical solution of the master
equations (\ref{dabsCqdz2})-(\ref{dpsi_dz2}) and for the case of tapered wiggler
(\ref{dabsCqdz2_tap})-(\ref{dpsi_dz2_tap}), that we normalized in Appendix~\ref{parameters_append}.
The normalized equations (\ref{theta_bar_def})-(\ref{f_K})
can be solved for a general case of wiggler amplitude and period tapering with arbitrary varying beam parameters:
$\overline{a}_w(z)$, $\gamma_r(z)$, $\sin\psi_r(z)$, and for general initial conditions of the bunches and
the radiation $\psi(0)$, $\theta(0)$, $|\tilde{E}(0)|$. The development of the radiation wave power and the
beam power along the interaction length are then calculated explicitly from Eqs.~(\ref{P_em})-(\ref{P_elect}).

In the absence of tapering and in the case of linear wiggler phase tapering ($\psi_r=$const), and also assuming moderate
variation of the wiggler parameters along the interaction length, we can set the coefficients
$f_B(u)$, $f_K(u)\simeq$const (Eqs.~\ref{f_B},\ref{f_K}). We use this model in the following computations for the purpose of
illustrating the interaction processes discussed in the previous subsections.
In this case, the general equations (\ref{dabsCqdz2_tap}-\ref{dpsi_dz2_tap}) are cast into a simple compact form for
the normalized field $\bar{E}\equiv |\tilde{E}|/[b(0)L_w]$, the phase $\psi$ and the normalized detuning
parameter $\bar{\theta}\equiv\theta L_w$, in terms of a normalized interaction length $u=z/L_w$:
\begin{equation}
\frac{d\bar{E}}{du}=\sin\psi,
\label{dabsCqdz2_ul1_b}
\end{equation}
\begin{equation}
\frac{d\bar{\theta}}{du}=K^2_{s0}\bar{E}\left[\sin\psi - \sin\psi_r\right],
\label{dtheta_dz2_tap1_ul1_b}
\end{equation}
\begin{equation}
\frac{d\psi}{du}=-\bar{\theta}+\frac{1}{\bar{E}}\cos\psi,
\label{dpsi_dz2_ul1_b}
\end{equation}

Remarkably, only one parameter $K^2_{s0}$ (Eq.~\ref{K_s0}) is required in addition to the initial conditions
in order to solve the closed equations (\ref{dabsCqdz2_ul1_b}-\ref{dpsi_dz2_ul1_b}) as a function of the normalized axial coordinate $u$, and display its trajectories in the normalized phase-space
$(\bar{\theta},\psi)$. Other laboratory parameters are needed only for the calculation of the power exchange
(see Appendix~\ref{parameters_append}, (\ref{P_em}-\ref{P_elect})):
\begin{equation}
\bar{P}_{em}=\bar{E}^2(u)
\label{pem_pref}
\end{equation}
\begin{equation}
\Delta \bar{P}_{el}=\Delta \bar{P}_{tap}+\Delta \bar{P}_{dyn}
\label{pel_pref}
\end{equation}
where
\begin{equation}
\Delta \bar{P}_{tap}=-2 \int_0^u\bar{E}(u')\sin\psi_r(u')du'
\label{deltaPtap}
\end{equation}
\begin{equation}
\Delta \bar{P}_{dyn}=-2 [\bar{\theta}(u)-\bar{\theta}(0)]/K_{s0}^2
\label{deltaPdyn}
\end{equation}
where all overbar power parameters are normalized $\bar{P}=P/P_{REF}$, $P_{REF}$ being defined in (\ref{P_REF}) is:
\begin{equation}
P_{REF}=\frac{1}{16\pi^2}\frac{\eta_p^2\overline{a}^2_w(0)}{\beta^2_{zr}(0)\gamma^2_r(0)}\frac{Q_b^2\omega_0^2L_w^2Z_q}{A_{em\,q}}
\label{P_REF_from_appendix}
\end{equation}
Under the simplifying assumptions leading to Eqs.~\ref{dabsCqdz2_ul1_b}-\ref{dpsi_dz2_ul1_b} we arrive to the
interesting comprehension (Eq.~\ref{pel_pref}) that the electron energy loss is a sum of contributions due to the tapering ($\Delta P_{tap}$) and the
inner trap dynamics ($\Delta P_{dyn}$).

In the following simulations (Figures~\ref{max_max_exraction_no_tap}-\ref{psi_0_equal_pi_over_2} and five videos) we use a
numerical value $K^2_{s0}=1.59$ (see Appendix~\ref{parameters_append}). This parameter (corresponding
to the Nocibur experiment \cite{nocibur}) is sufficient for the $\bar{\theta}-\psi$ trajectories
display. In order to display the $\gamma-\psi$ phase-space trajectories, we use in the following examples
in Eq.~(\ref{theta_gamma}) the laboratory parameters $\gamma_r=127.2$, $N_w=L_w/\lambda_w=11$ \cite{nocibur}
assuming idealized tight bunching and moderate tapering.

\subsection{Untrapped trajectories in a uniform wiggler}

In order to show the consistency of the normalized nonlinear equations with the earlier results of SP-SR and ST-SR in the
zero-order approximation of Chapters~II, III (Eqs.~(\ref{P_SR}), (\ref{P_ST_SR})), we set 
$\psi_r=0$ in Eq.~\ref{dtheta_dz2_tap1_ul1_b}, and for an untrapped electron
we consider $\bar{\theta}$ to be almost constant, i.e. $\bar{\theta}(u)\simeq\bar{\theta}(0)$.

Expressing Eq.~(\ref{dCqdz_psi}) in terms of the normalized parameters $u\equiv z/L_w$, $\bar{\theta}\equiv\theta L_w$,
(using (\ref{psi})), and defining $\tilde{E}\equiv\bar{E}e^{i\varphi_q}=(\tilde{E}/(b L_w))e^{i\varphi_q}$, one obtains:
\begin{equation}
\frac{d\tilde{E}}{du}=e^{i\varphi_b},
\label{dtildeC}
\end{equation}
Using the definition of $\varphi_b$ from Eq.~\ref{phi} or
\ref{dpsi_dz}, with $\bar{\theta}(u)\simeq\bar{\theta}(0)$, we obtain:
\begin{equation}
\varphi_b \simeq \bar{\theta}(0) u + \varphi_{b0}.
\label{varphib}
\end{equation}
Integrating Eq.~\ref{dtildeC} we obtain
\begin{equation}
\tilde{E}(u)=\tilde{E}(0)+u e^{i[\varphi_{b0}+\bar{\theta}(0)u/2]} \sinc[\bar{\theta}(0)u/2].
\label{tildeC_u}
\end{equation}
With $\tilde{E}(0)=|\tilde{E}(0)|e^{i\varphi_q(0)}$, one gets
\begin{equation}
|\tilde{E}(u)|^2=|\tilde{E}(0)|^2+u^2 \sinc^2[\bar{\theta}(0)u/2]+2u|\tilde{E}(0)|\sinc(\bar{\theta}(0)u/2)\cos[\varphi_{qb0}-\bar{\theta}(0)u/2],
\label{abs_tildeC_u_sq}
\end{equation}
and using the definition of $\psi$ in Eq.~\ref{psi}, it can also be written as
\begin{equation}
P_{em}/P_{REF}=|\tilde{E}(u)|^2=|\tilde{E}(0)|^2+u^2 \sinc^2[\bar{\theta}(0)u/2]+2u|\tilde{E}(0)|\sinc(\bar{\theta}(0)u/2)\sin[\psi(0)-\bar{\theta}(0)u/2],
\label{abs_tildeC_u_sq1}
\end{equation}
This represents the normalized output power with full correspondence to the zero-order approximate expressions (\ref{P_q1}-\ref{P_ST_SR}) derived in Chapter~III for
superradiance (SP-SR) and stimulated-superradiance (ST-SR) respectively.

\subsection{Maximal energy extraction from a bunched beam in a uniform wiggler}

\begin{figure}[!tbh]
\includegraphics[width=8cm]{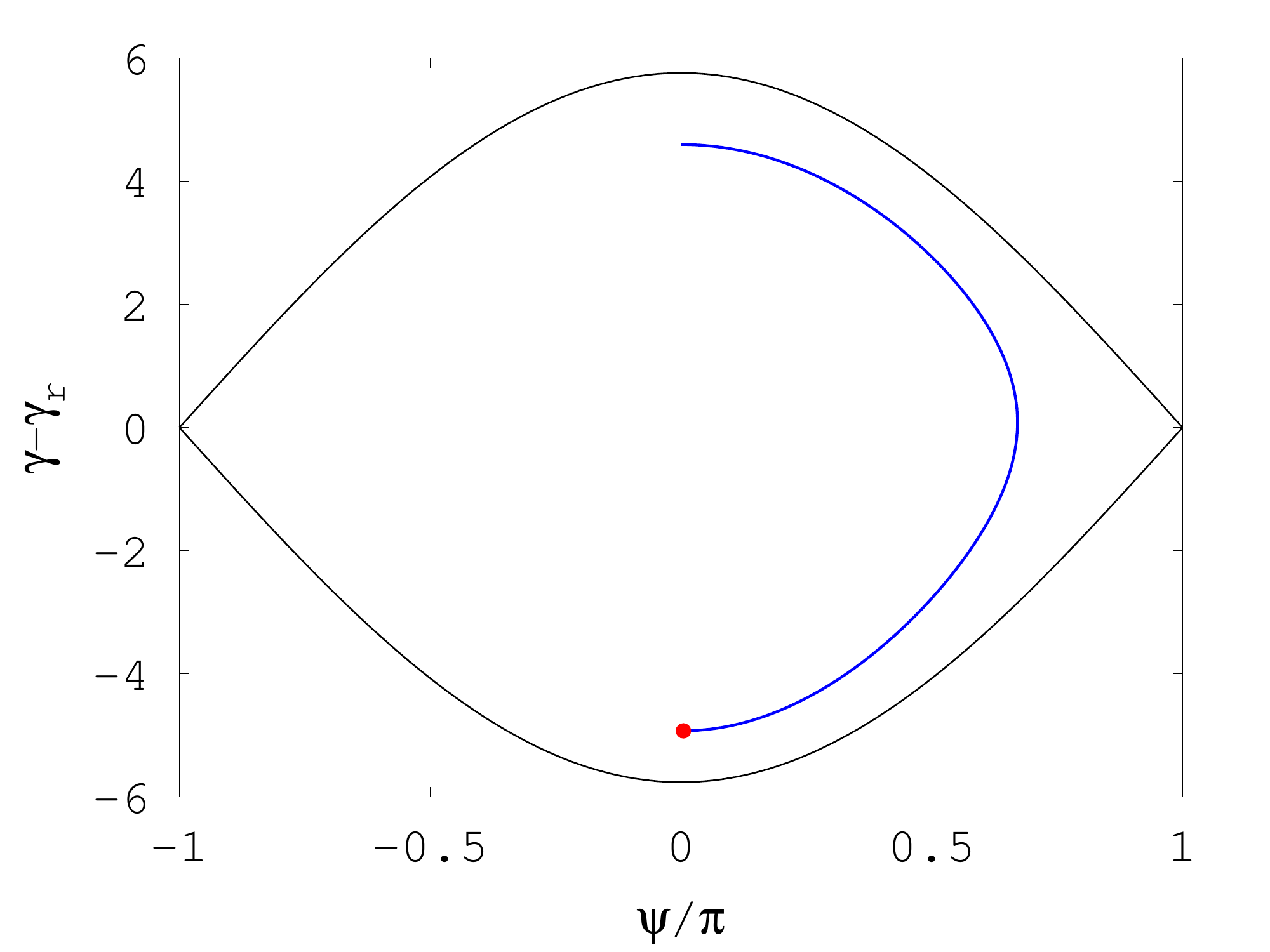}
\includegraphics[width=8cm]{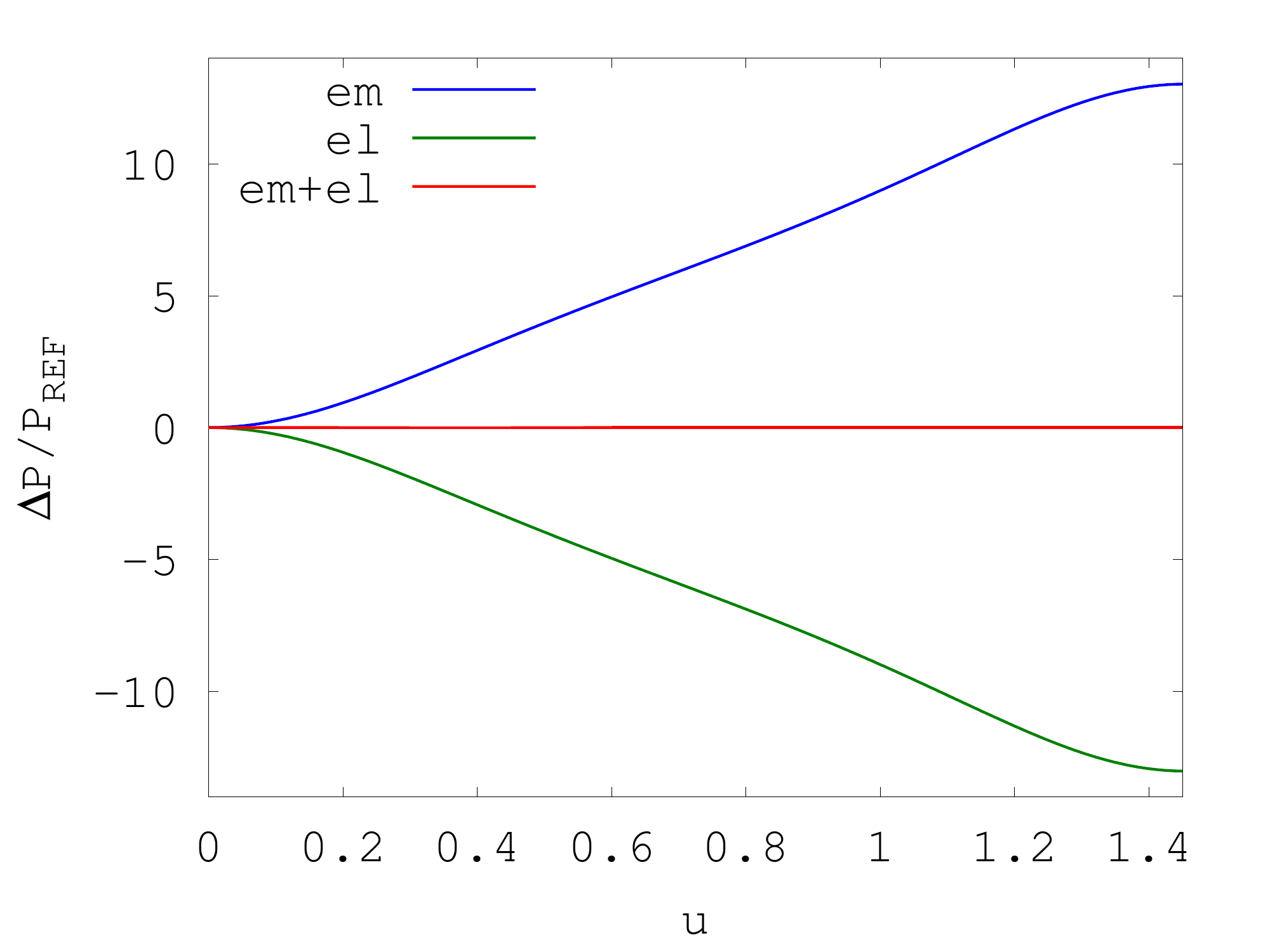}
\caption{Maximal energy extraction of a perfectly bunched beam in a uniform wiggler trap.
Panel (a) shows the phase-space diagram $\psi-\theta$, where the black line shows the separatrix at the
end of the trajectory. Panel (b) shows the radiation power change, the electron beam power change, and their sum, which
keeps 0. This figure corresponds to \href{https://youtu.be/Oz6rWS5xkEc}{[Uniform wiggler - maximum extraction video]}.}
\label{max_max_exraction_no_tap}
\end{figure}

In Figure~\ref{max_max_exraction_no_tap} and \href{https://youtu.be/Oz6rWS5xkEc}{[Uniform wiggler - maximum extraction video]} we display the dynamics of the electron beam in the
case of maximal energy extraction from a bunched beam (Figure~\ref{phase_space_general_non_tap}a), which
corresponds to the case of maximum power extraction from a perfectly bunched
beam in a saturated FEL. Maximal energy extraction from the e-beam - $2\delta\gamma_m$ - is attained when the bunch enters
the trap at phase $\psi(0)=0$ with energy detuning $\delta\gamma(0)\simeq\delta\gamma_m$, and winds up at the end of the
interaction length at the bottom of the trap $\delta\gamma(L_w)\simeq -\delta\gamma_m$, after performing half a period of synchrotron
oscillation. Note that in this case of maximal extraction, the initial gain is null: $d\delta\gamma/dz|_{z=0}=0$ (\ref{dabsCqdz2})
and the radiation build-up starts slow (quadratically as in Eq.~(\ref{DWdq_SR_M_b}) (see Figure~\ref{max_max_exraction_no_tap}b).

\subsection{Stimulated superradiance in a uniform wiggler}

\begin{figure}[!tbh]
\includegraphics[width=8cm]{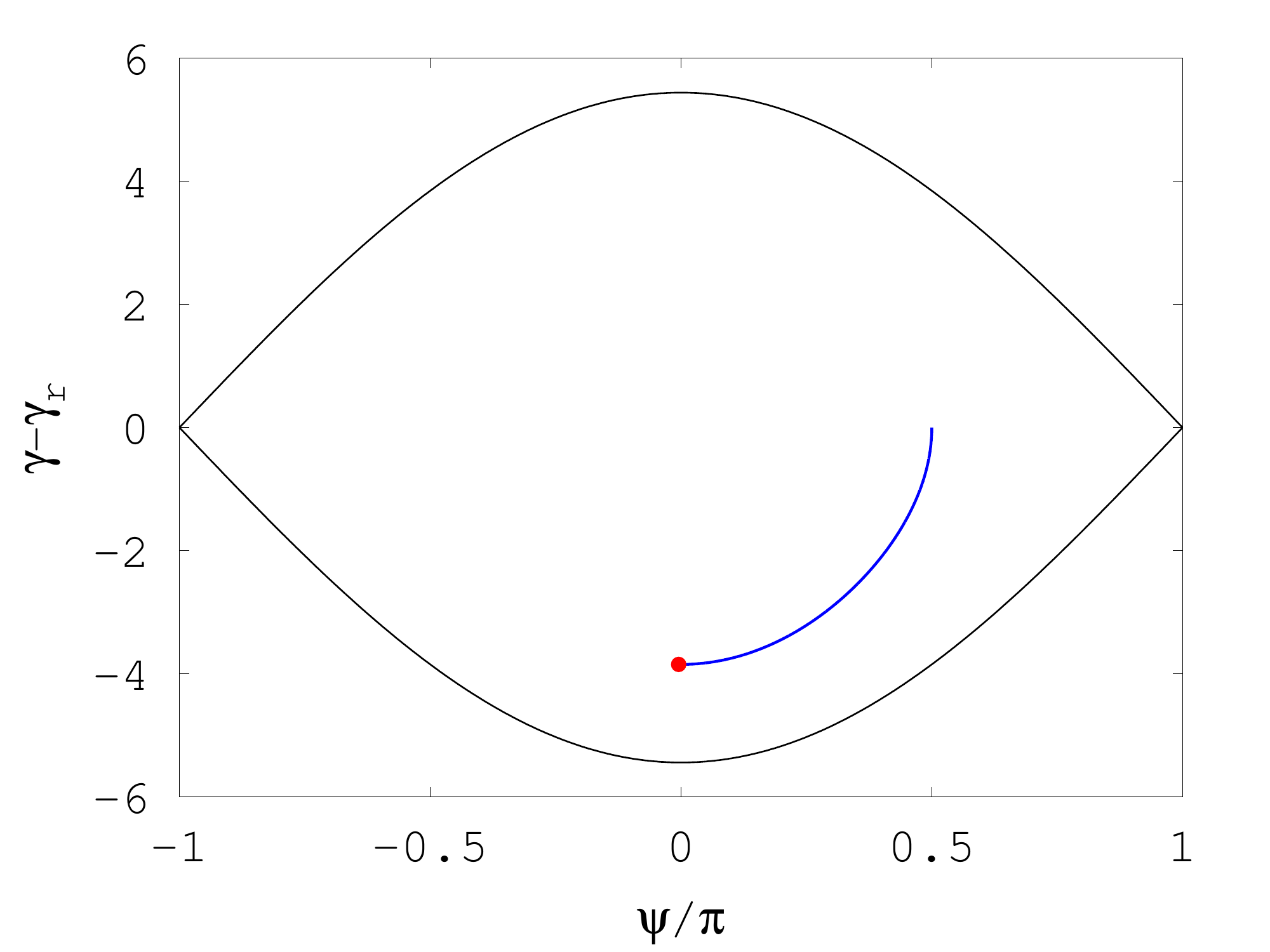}
\includegraphics[width=8cm]{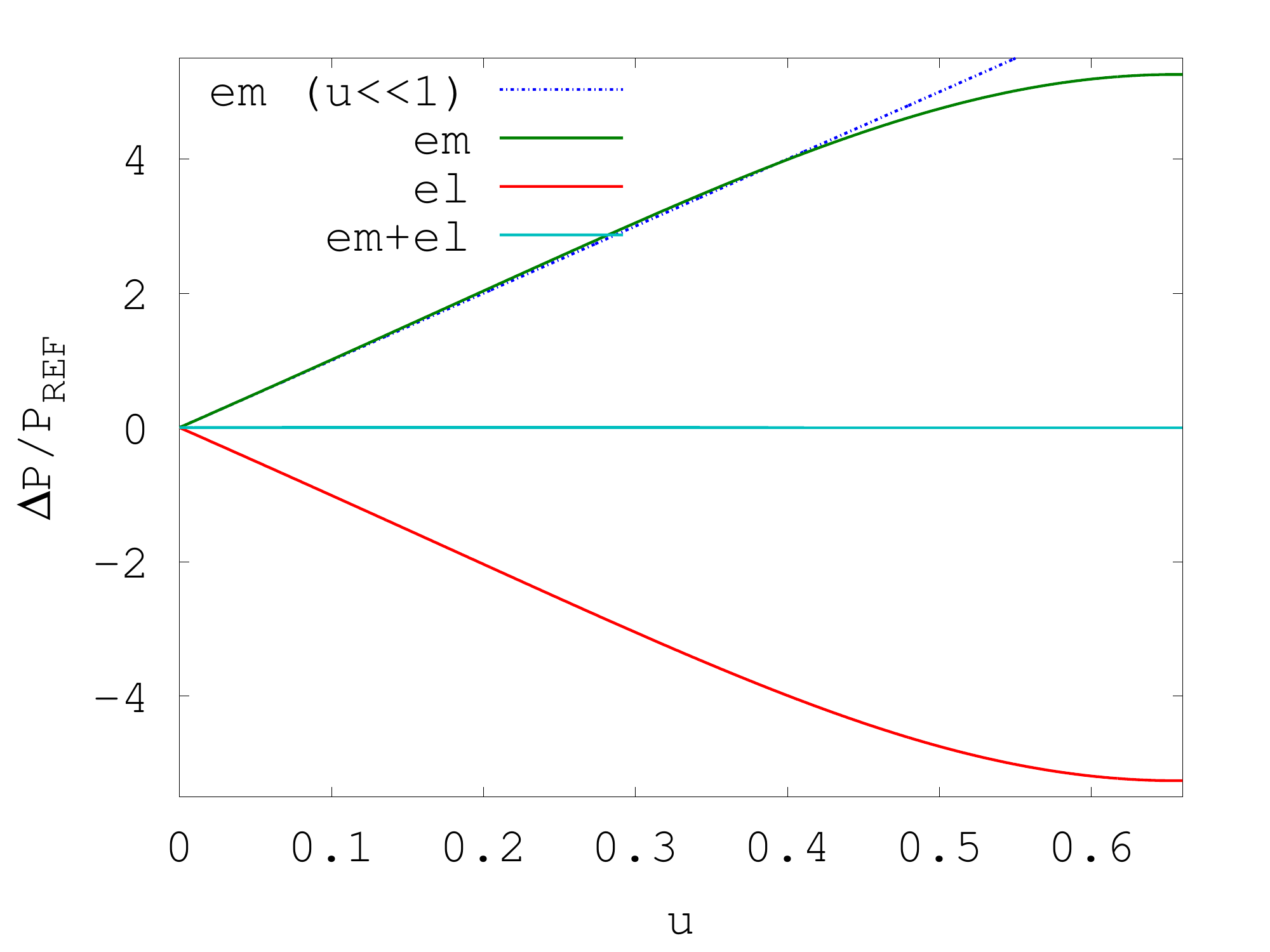}
\caption{Stimulated superradiance in a uniform wiggler with maximum gain bunching phase $\psi(0)=\pi/2$.
Panel (a) shows the phase-space diagram $\psi-\theta$, where the black line shows the separatrix at the
end of the trajectory. Panel (b) shows the radiation power change, the electron beam power change, and their sum, which
keeps 0. The broken line shows the initially linear radiation power growth as in Eq.~(\ref{delta_Pem_du}). This figure corresponds to
\href{https://youtu.be/yzJMVEzSDzU}{[Uniform wiggler - stimulated superradiance video]}.}
\label{stimulated_superradiance}
\end{figure}

Of special interest is the stimulated superradiance (ST-SR) (Figure~\ref{phase_space_general_non_tap}b) where maximum initial gain is expected
when starting from $\delta\gamma(0)\simeq 0$ and $\psi(0)=\pi/2$. The simulation result of this case is shown in
Figure~\ref{stimulated_superradiance} and \href{https://youtu.be/yzJMVEzSDzU}{[Uniform wiggler - stimulated superradiance video]}.

Note that direct differentiation of (\ref{pem_pref}), (\ref{pel_pref}) and (\ref{deltaPdyn}) results in:
\begin{equation}
\frac{1}{P_{REF}}\frac{dP_{em}}{du}=-\frac{1}{P_{REF}}\frac{dP_{el}}{du}=2\bar{E}(u)\frac{d\bar{E}(u)}{du}=2\bar{E}(u)\sin\psi(u)
\label{dPem_du}
\end{equation}
and therefore the initial power growth (in a range $0<u\ll 1$) is proportional to $u$ (see tangent dash-dotted line in
Figure~\ref{stimulated_superradiance})
\begin{equation}
\Delta P_{em}=-\Delta P_{el}=2P_{REF}\bar{E}(0)u\sin\psi(0)
\label{delta_Pem_du}
\end{equation}
This is exactly consistent with the zero-order approximation of ST-SR power growth - Eq.~(\ref{P_ST_SR}). Also it is
evident that maximum initial gain is attained with $\psi(0)=\pi/2$.

In second order in $u$, the integration of (\ref{dabsCqdz2_ul1_b}) results in
$\bar{E}(u)=\bar{E}(0)+u\sin\psi(0)$ and consequently from the integration of (\ref{dPem_du}) or directly from the definition
(\ref{pem_pref}):
\begin{equation}
\Delta P_{em}/P_{REF}=\bar{E}^2(u)-\bar{E}^2(0)=2u\bar{E}(0)\sin\psi(0)+u^2\sin^2\psi(0)
\label{pem_over_pref}
\end{equation}

We conclude that the initial emission process is always composed of both contributions of ST-SR (first term) and SP-SR (second term).
When the field is strong enough $\bar{E}(0)\gg [u\sin\psi(0)]/2$ then the ST-SR term is dominant and the power starts growing linearly
as in Figure~\ref{stimulated_superradiance}b.

\subsection{Tapered wiggler}

As shown in Figure~\ref{phase_space_general_tap}, in case of a tapered wiggler, the main contribution to the radiated power extraction from
the e-beam comes usually from the tapering process, but there is contribution also from the phase-space evolution of synchrotron oscillation
dynamics inside the decelerating trap. If the bunch is deeply trapped ($\psi(0)=\psi_r$, $\delta\gamma(0)=0$) (Figure~\ref{phase_space_general_tap}a)
the beam energy drops only with the trap deceleration. When the trapped bunch is not at the bottom of the trap potential as is the case in Figure~\ref{phase_space_general_tap}b ($\psi(0)=\pi/2$, $\delta\gamma(0)=0$)
 and Figure~\ref{phase_space_general_tap}c ($\psi(0)=0$, $\delta\gamma(0)=\delta\gamma_m(0)$), there is
also contribution of the inner trap synchrotron oscillation dynamics to the beam energy total drop. Maximal energy extraction is attained in the
case of Figure~\ref{phase_space_general_tap}c, however the initial energy drop rate is zero in this case ($\psi(0)=\psi_r$) and is maximal in the
case of Figure~\ref{phase_space_general_tap}b ($\psi(0)=\pi/2$). This may play a role in optimal tapering and bunch phasing
strategy.

\begin{figure}[!tbh]
\includegraphics[width=8cm]{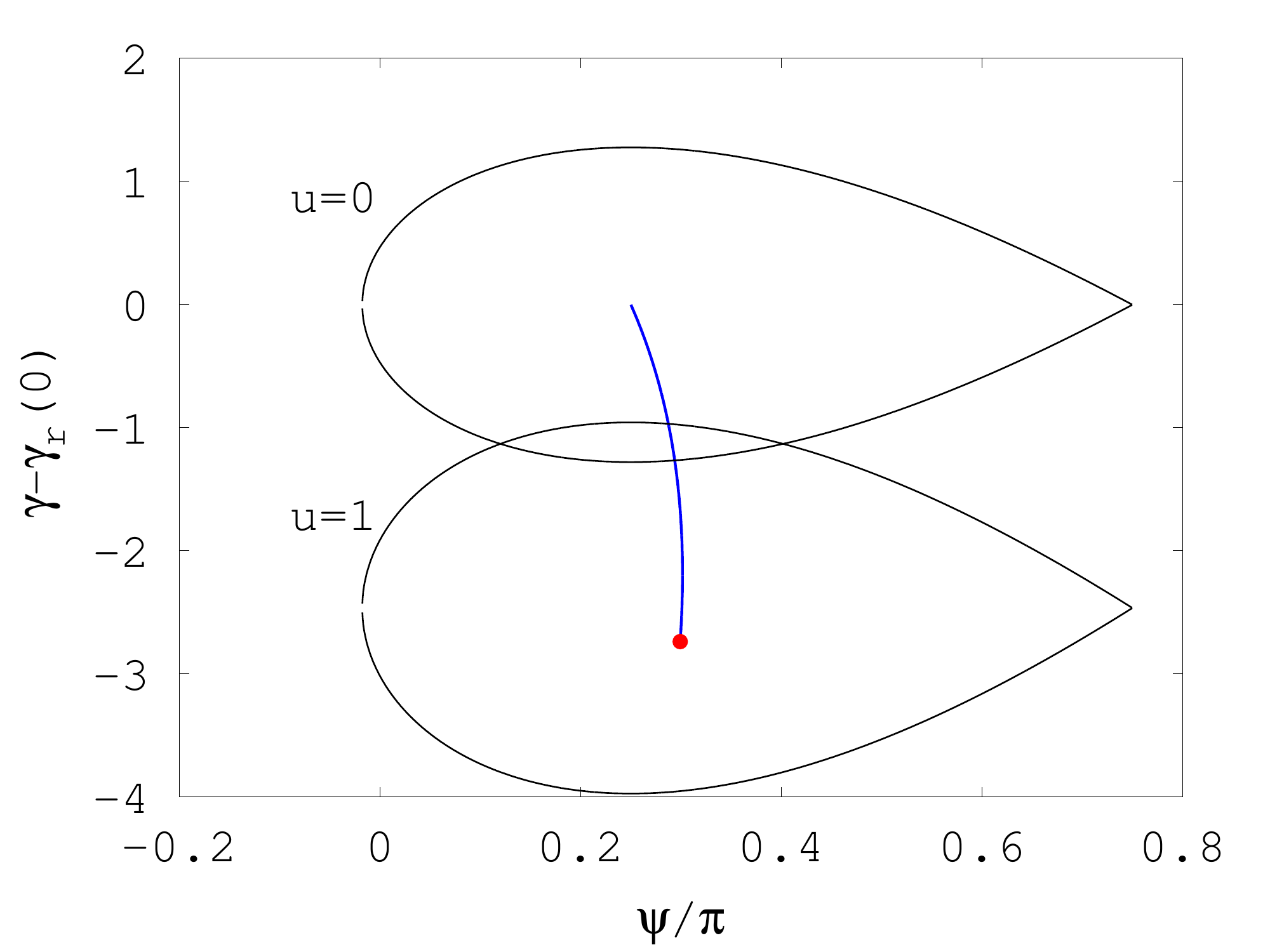}
\includegraphics[width=8cm]{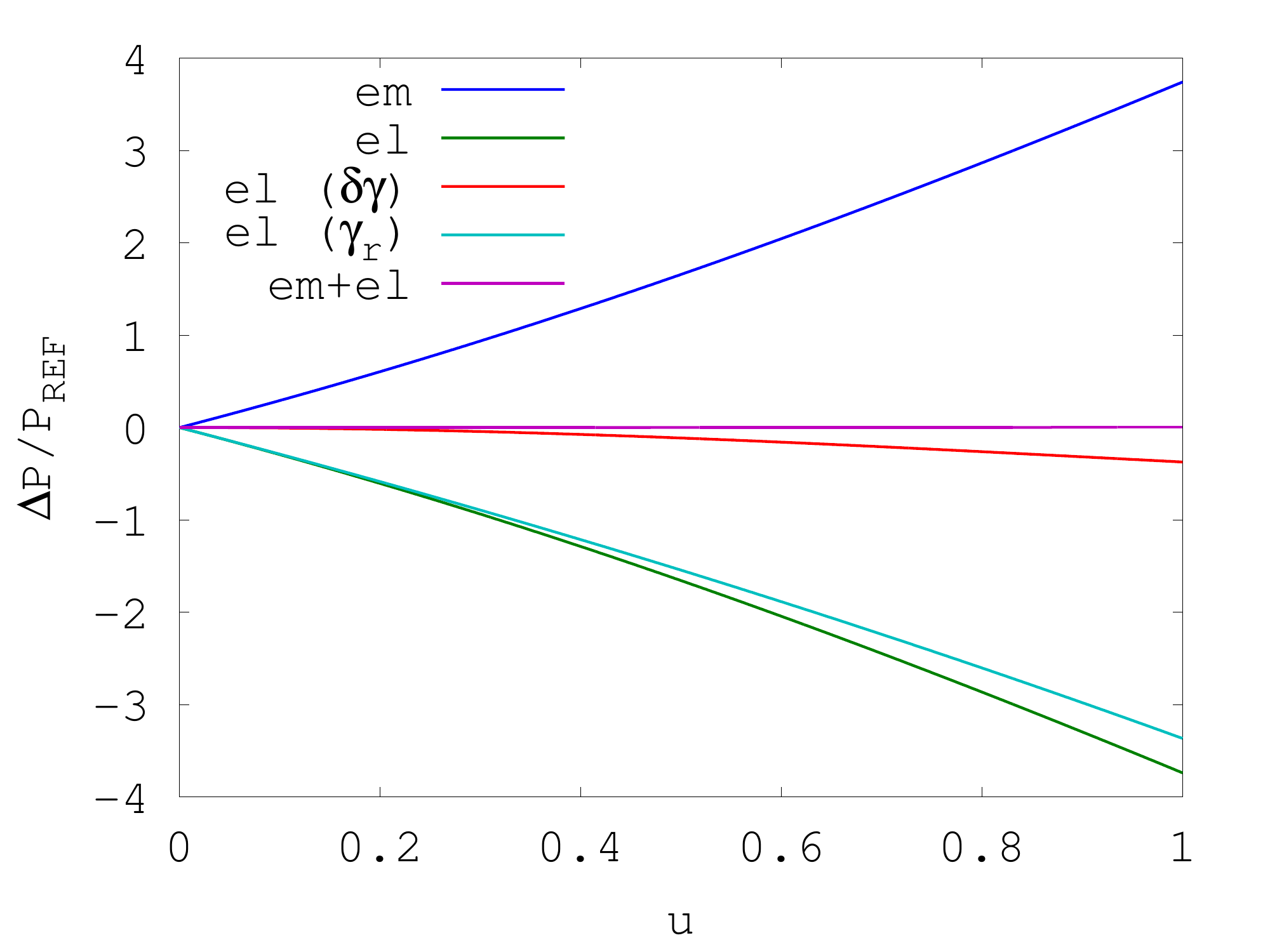}
\caption{Tapering enhanced superradiance (TESSA) of a perfectly bunched beam starting at the bottom of the trap $\delta\gamma(0)=0$,
$\psi(0)=\psi_r$.  Panel (a) shows the phase-space diagram $\psi$, $\gamma-\gamma_r(0)$, where the upper black line shows the separatrix
at the beginning of the trajectory ($u=0$) and the lower black line shows the separatrix at the end of the trajectory ($u=1$).
Panel (b) shows the radiation power change, the electron beam power change, and their sum which keeps 0. We show separately the
contributions of the tapering ($\Delta P_{el\,(\gamma_r)}$) and the synchrotron oscillation dynamics
($\Delta P_{el\,(\delta\gamma)}$) to the total beam power drop ($\Delta P_{el}$). This figure corresponds to \href{https://youtu.be/6u0HW_yeGsw}{[Tapered wiggler - $\psi(0)=\psi_r$ video]}.}
\label{psi_0_equal_psi_r}
\end{figure}

In Figure~\ref{psi_0_equal_psi_r} and \href{https://youtu.be/6u0HW_yeGsw}{[Tapered wiggler - $\psi(0)=\psi_r$ video]}, we show a bunch initially trapped in the middle of the trap (at $\psi(0)=\psi_r$, and $\theta(0)=0$ (i.e.
$\gamma(0)-\gamma_r(0)=0$) for normalized parameters example of initial input field $\bar{E}(0)=2$, using $\psi_r=\pi/4$. Panel (a) shows the phase-space diagram $\psi$, $\gamma-\gamma_r(0)$
in which the upper black line shows the separatrix at the beginning of the trajectory and the lower black line shows the separatrix at the end of the trajectory.
This shift in the separatrix location is due to the tapering, and gives the major contribution to the e-beam power decrement (deceleration). Panel (b) shows
the radiation power incremental growth (blue), the electron beam power decrement (green), and the sum of radiation and e-beam power increments which keeps 0.
To get a better insight into the different phenomena, we show separately the contributions of the tapering ($\Delta P_{el\,(\gamma_r)}$) (light blue) and the synchrotron oscillation dynamics
($\Delta P_{el\,(\delta\gamma)}$) (red) to the total beam power drop ($\Delta P_{el}$) (green) - see Eq.~(\ref{pel_pref}). The tapering
contribution here is around 9 times bigger than the synchrotron oscillation dynamics contribution.

\begin{figure}[!tbh]
\includegraphics[width=8cm]{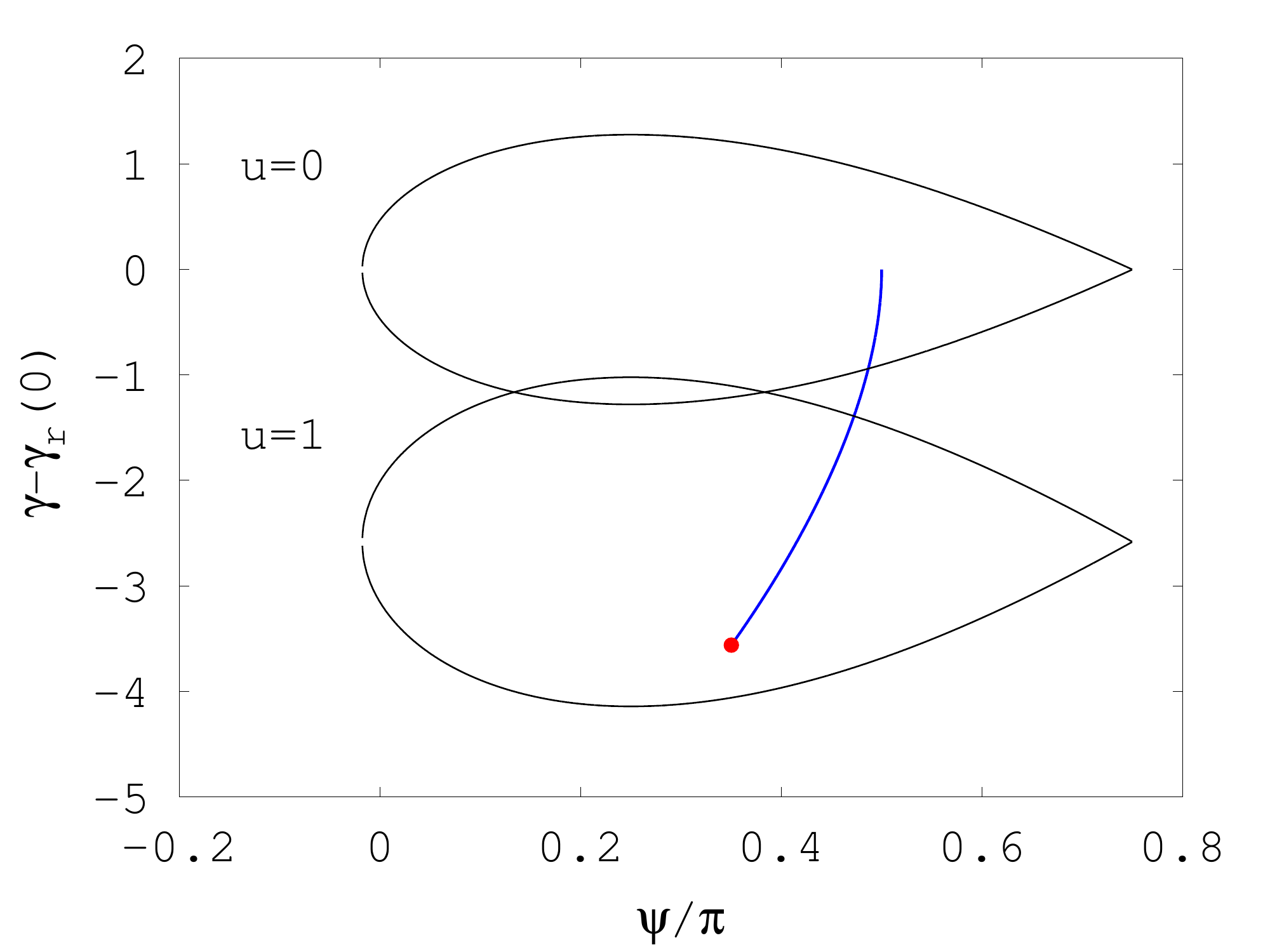}
\includegraphics[width=8cm]{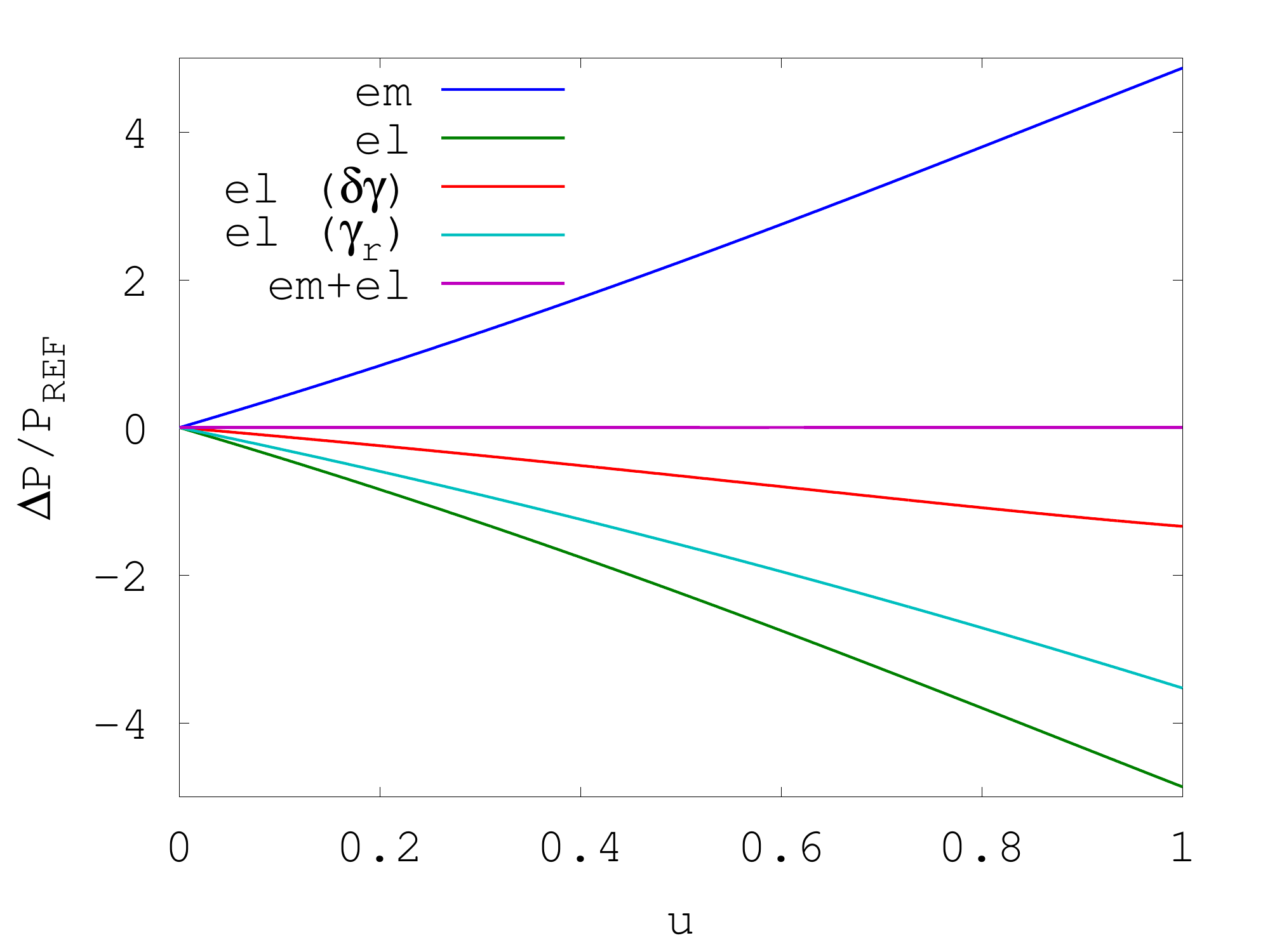}
\caption{TESSA radiation extraction by a perfectly bunched beam, phased initially at $\psi(0)=\pi/2$ for maximal
initial radiation growth rate (gain). Note that the total radiation
emission power in this case is 30\% higher than in the case of $\psi(0)=\psi_r$ (Figure~\ref{psi_0_equal_psi_r}). This figure corresponds to \href{https://youtu.be/JEEunwO6Qyw}{[Tapered wiggler - $\psi(0)=\pi/2$ video]}.}
\label{psi_0_equal_pi_over_2}
\end{figure}

Figure~\ref{psi_0_equal_pi_over_2} and \href{https://youtu.be/JEEunwO6Qyw}{[Tapered wiggler - $\psi(0)=\pi/2$ video]}, shows the same as Figure~\ref{psi_0_equal_psi_r}, only the initial bunch is at $\psi(0)=\pi/2$, instead of $\psi_r$.
 We find
that in this case the contribution of the tapering to the total e-beam power loss is still dominant, but less, being around 2.6 times bigger than the synchrotron oscillation
dynamics contribution. This is due to the fact that the synchrotron oscillation contribution increased significantly in the case of $\psi(0)=\pi/2$, relative to the case
$\psi(0)=\psi_r=\pi/4$. Therefore, the total radiation power enhancement is bigger in this case by 30\% than in the case of $\psi(0)=\psi_r=\pi/4$.

We draw attention to the initial power growth rate Eqs.~\ref{dPem_du}-\ref{pem_over_pref} that were derived
without use of the $\psi_r$ dependent Eq.~(\ref{dtheta_dz2_tap1_ul1_b}), and therefore are
valid also for the tapered wiggler case $\psi_r\neq 0$. To illustrate better the role of the tapering and
inner trap dynamics, as well as the spontaneous superradiance and stimulated superradiance process in the initial interaction stage $u\ll 1$, it helps to rewrite
Eq.~(\ref{pem_over_pref}) (that is valid also for a tapered wiggler $\sin\psi_r\neq 0$) in the following form:
\begin{equation}
\Delta P_{em}/P_{REF}=[2u\bar{E}(0)(\sin\psi(0)-\sin\psi_r)+u^2\sin^2\psi(0)]+2u\bar{E}(0)\sin\psi_r,
\label{pem_over_pref_tap}
\end{equation}
In this presentation the term in bracket (linear and square in $u$) represent the effect of the dynamics inside
the trap (synchrotron oscillations): ST-SR (linear) and SP-SR (quadratic) processes. The last term (linear) is the effect
of tapering (see Eq.~\ref{deltaPtap}).

We therefore conclude that also in the tapering case, highest gain is attained for initial phasing
$\psi(0)=\pi/2$, a factor of $\sqrt{2}$ relative to the case of deep trapping $\psi(0)=\psi_r=\pi/4$
shown in Figure~\ref{psi_0_equal_psi_r}.

However, the case of $\psi(0)=\psi_r$ is also important in practice, because the trapping is deeper, and
may be a preferred strategy for the case of imperfect bunching, where trapping efficiency is an issue (see Section~VII). For $\psi(0)=\psi_r$
Eq.~\ref{pem_over_pref_tap} reduces to:
\begin{equation}
\Delta P_{em}/P_{REF}=2u\bar{E}(0)\sin\psi_r+ u^2\sin^2\psi_r.
\label{pem_over_pref_tap_psi_0_is_psi_r}
\end{equation}

\subsection{Superradiance and self-interaction}
\label{self_int}

Of special interest is the case of pure superradiance where the radiation grows spontaneously in a uniform wiggler without
any input field ($\bar{E}(0)=0$) (Figure~\ref{phase_space_general_non_tap}c). If the built-up radiation grows up enough,
the beam may saturate by its own radiation (Figure~\ref{phase_space_general_non_tap}d).

The second term in Eq.~(\ref{pem_over_pref}) is independent of $\bar{E}(0)$ and for $\bar{E}(0)=0$ it results in
$\Delta P_{em}/P_{REF}=u^2\sin^2\psi(0)$. The phase $\psi(0)$ is ill defined because the null radiation field has arbitrary phase.
This is the reason for the seeming singularities in Eqs.~(\ref{dphi_E_dz2}) and (\ref{dpsi_dz2})
that can be removed only when $\psi(0)=\pi/2$. The physical explanation for this particular
determination of the radiation phase is that in the absence of initial radiation phase, the
phase of the excited radiation mode is determined by the phase of the bunched beam $\varphi_{b0}$,
as can be seen iteratively from Eq.~(\ref{mode_exp3}) by setting $z=\delta z\ll 2\pi/\theta$:
\begin{equation}
\tilde{E}(\delta z)=0^+ + |\tilde{E}|e^{i\varphi_{b0}}
\label{C_q_small_z}
\end{equation}
Setting then $\varphi_q(0)=\varphi_{b0}$, and using (\ref{phi}) and (\ref{phi_rb0}), i.e
$\varphi(0)=\varphi_{b0}$, we get from the definition (\ref{psi}) $\psi(0)=\pi/2$ and therefore:
\begin{equation}
P_{em}/P_{REF}=u^2
\label{pem_pref_SR}
\end{equation}
This is evidently a normalized parameters representation of the case of superradiance, where the power
grows quadratically from 0 with the interaction length $z$ - see Eq.~\ref{P_SR}.

\begin{figure}[!tbh]
\includegraphics[width=8cm]{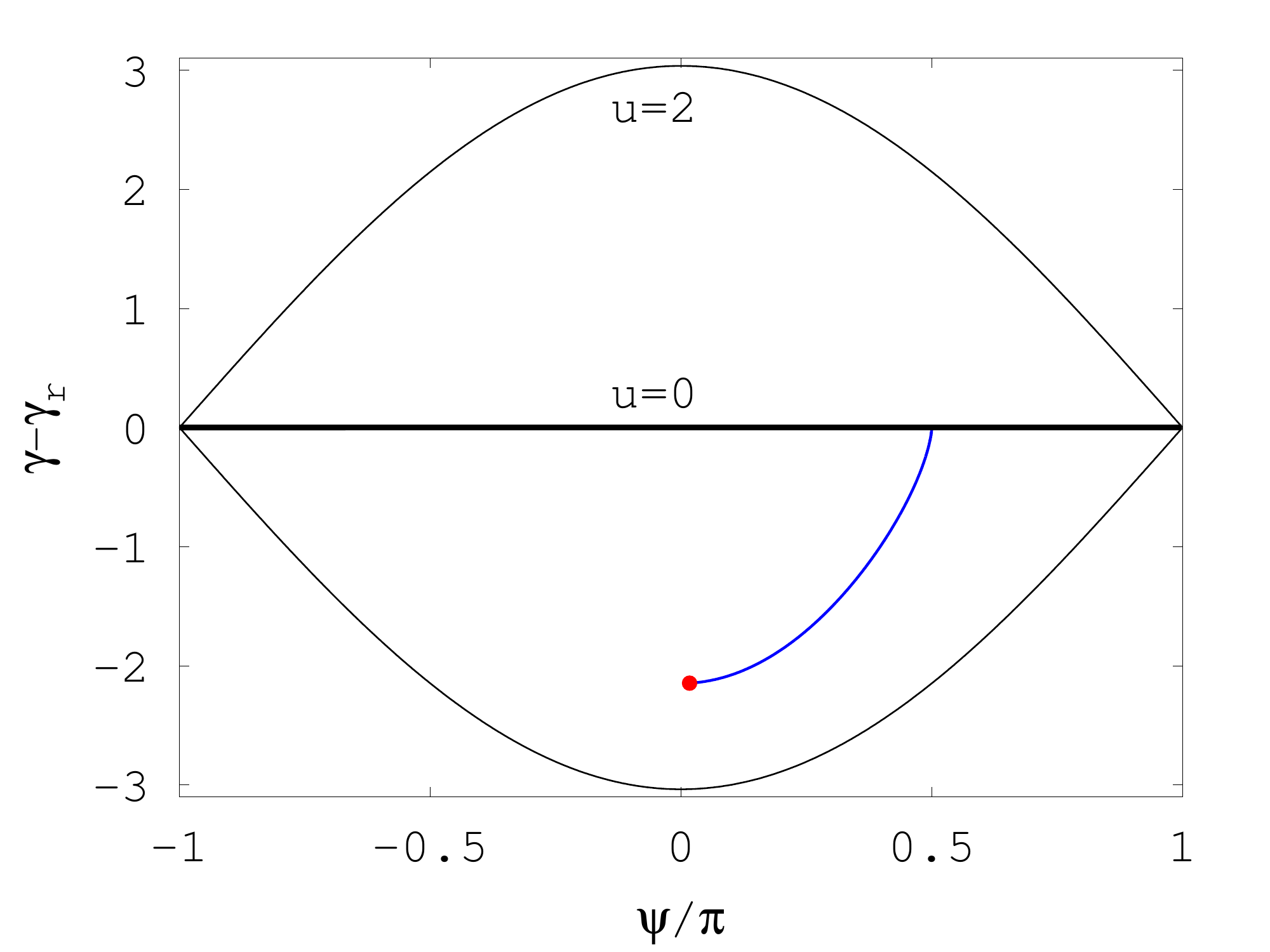}
\includegraphics[width=8cm]{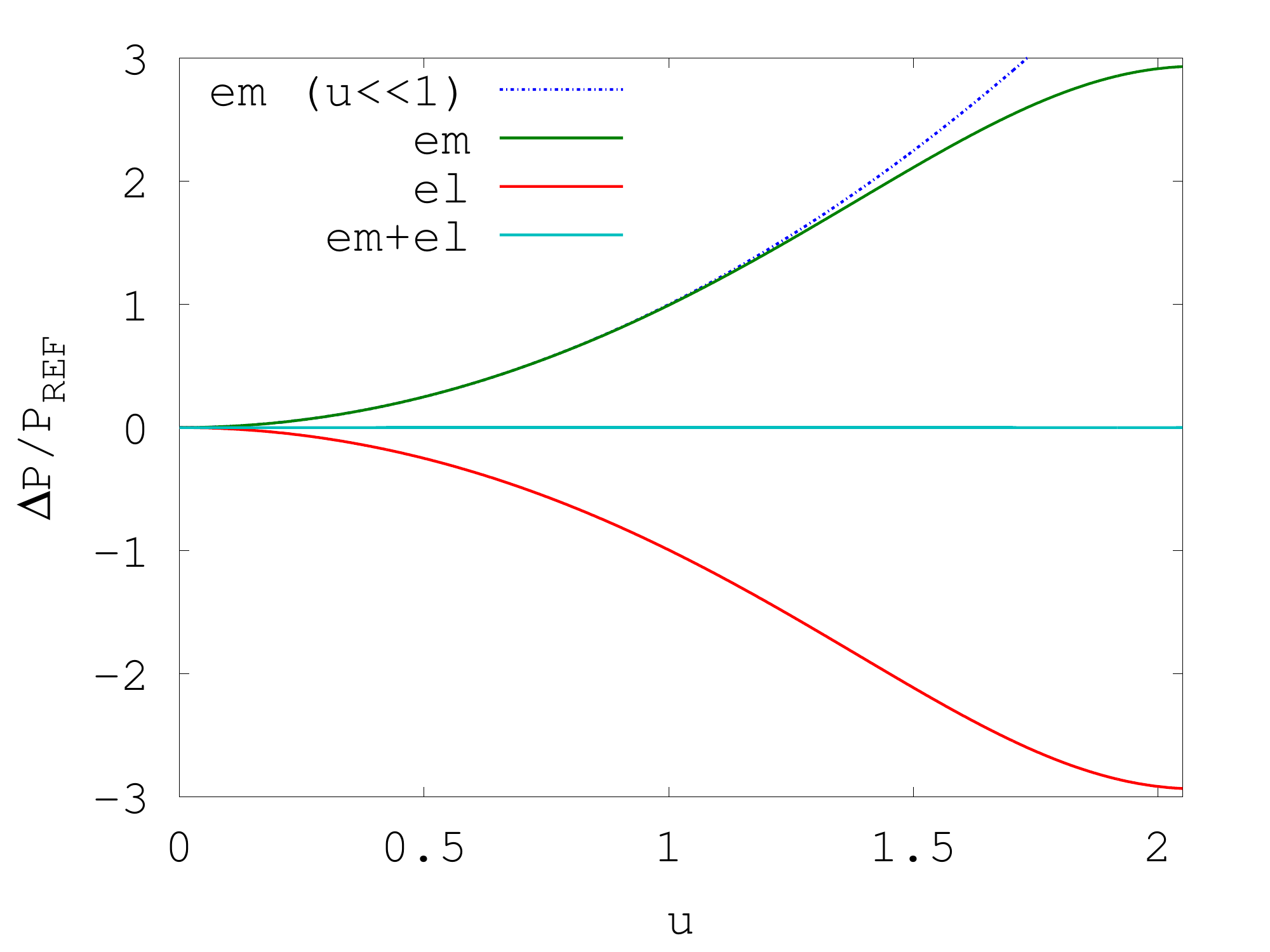}
\caption{Saturated self-interaction superradiance in a uniform wiggler. Panel (a) shows the phase-space
diagram $\psi-\theta$, where the middle black line shows the nonexistent
separatrix at the beginning of the trajectory ($u=0$) and the upper and lower black lines show the separatrix at the end of
the trajectory ($u=2$). Panel (b) shows the radiation power change, the electron beam power change, and their sum which keeps 0. This figure corresponds to \href{https://youtu.be/4hlc8QGTa1I}{[Uniform wiggler - self interaction video]}.}
\label{self_interaction}
\end{figure}

In Figure~\ref{self_interaction} and \href{https://youtu.be/4hlc8QGTa1I}{[Uniform wiggler - self interaction video]} we show the trajectories and power growth of the prebunched beam
radiation in a uniform wiggler, starting from zero input field. The quadratic approximation (\ref{pem_pref_SR})
is also shown in Figure~\ref{self_interaction} by the dash-dotted line and matches well the power growth rate.

This case of radiation emission by the beam in a uniform wiggler when $\tilde{E}(0)=0$
is of special interest. Though in the derivation of the quadratic growth of superradiance
(Eq.~\ref{P_SR}) it was assumed that the beam energy does not change, in the present
energy-conserving non linear model we see that in the more complete energy conserving analysis,
the beam energy goes down as expected, in correspondence with the superradiant power growth.
This case can be related to the problem of radiation emission due to charged particles acceleration in
free space. In that case, the energy loss of the particle due to its radiation emission is
explained in terms of Abraham-Lorentz effective radiation reaction force that can only be derived indirectly from
energy conservation considerations \cite{ianc_2003,ianc_2002,ianc_1992,gupta,feynman,dirac,Schwinger}.

In contrast to the free-space self interaction case, in the present case of periodic bunched beam
radiation emission into a transversely confined single mode, the self-interaction problem is soluble
explicitly. As seen in Figure~\ref{self_interaction}, the spontaneous emission of undulator
radiation field grows from 0 (at $u=0$) with a distinct phase, so that the tight bunches are
found initially automatically at phase $\psi(0)=\pi/2$ relative to the ponderomotive wave bucket. This
happens to be exactly the phase of maximum stimulated-superradiance, where the bunched beam
experiences maximum deceleration by the electric field of the radiation mode that it had excited. Further
tracing of the beam dynamics, as shown in Figure~\ref{self_interaction}, the periodic beam
self-interacts with its own radiation and slows down, reaching a non linear self absorption saturation regime at long
interaction length $u$, and even can be reaccelerated after the maximal deceleration point $u=2$, reabsorbing the
radiation that is generates in the first part of the undulator.

\begin{figure}[!tbh]
\includegraphics[width=8cm]{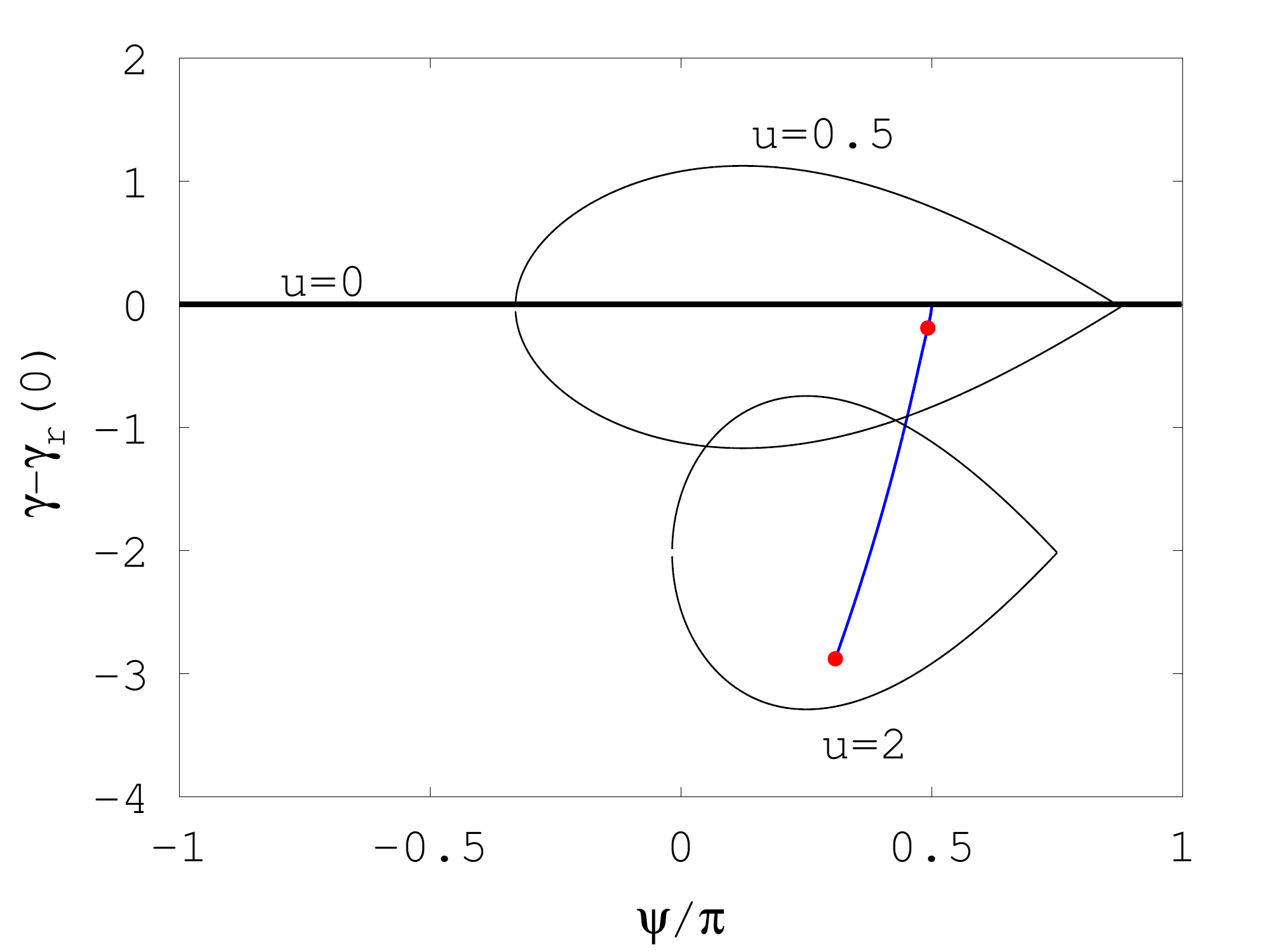}
\includegraphics[width=8cm]{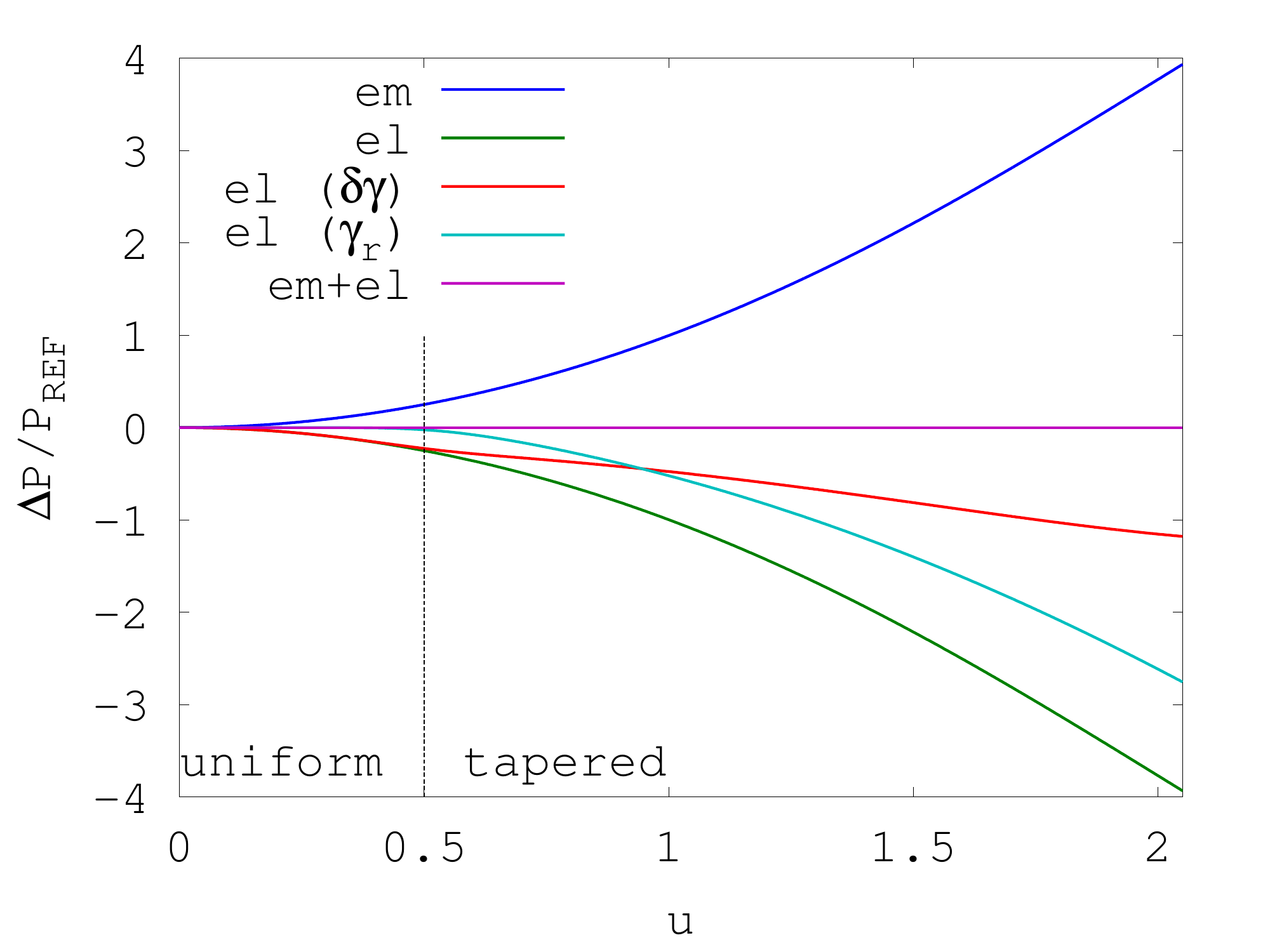}
\caption{Self-interaction Tapering Enhanced Superradiance (TES). Panel (a) shows the phase-space diagram $\psi-\theta$,
where the thick horizontal black line shows the nonexistent
separatrix at the beginning of the trajectory ($u=0$), the upper ``eye'' separatrix shows the self-interaction built-up trap at the transition of
the uniform wiggler into a tapered wiggler at $u=0.5$ and the lower separatrix at the end of
the trajectory ($u=2$). Panel (b) shows the radiation power change (em), the electron beam power change due to internal trap
dynamics ($\delta\gamma$) and due to tapering ($\gamma_r$), and their sum (el). The sum of the generated radiation power (em)
and the negative beam power increment (el) is null - keeping energy conservation.
This figure corresponds to \href{https://youtu.be/BscIr8u5s9U}{[Tapered wiggler - self interaction video]}.}
\label{self_interaction_tap}
\end{figure}

Figure~\ref{self_interaction_tap} displays an even more interesting case of Tapering Enhanced Superradiance (TES),
showing that in the nonlinear regime, a periodically bunched beam that is trapped in its own generated radiation trap as
in Figure~\ref{self_interaction} can exhibit further enhanced radiation emission if the undulator becomes tapered after
a long enough section of trap build up along a uniform undulator section. In Figure~\ref{self_interaction_tap} and 
\href{https://youtu.be/BscIr8u5s9U}{[Tapered wiggler - self interaction video]} the uniform undulator in the
section $0<u<0.5$ turns adiabatically at $u=0.5$ into a tapered undulator with $\psi_r=\pi/4$, extracting further
beam energy in the tapered section $0.5<u<2$.

Note that contrary to the Abraham-Lorentz case of free-space
emission into a continuum of modes and frequencies, here we consider emission into a single mode, and because the beam is
infinitely periodically bunched, there is no issue of slippage effect. Note that similar ``self-interaction'' nonlinear superradiance
process has been predicted with a single bunch interaction with a waveguided THz beam in a tapered wiggler under conditions of
zero-slippage due to waveguide dispersion \cite{emma_snively}. These schemes of self-interaction may have a practical
advantage in development of future short wavelength radiation sources because they are not susceptible to jitter problems
between the beam bunch (bunches) and the seed radiation since the electrons are trapped in the tapered section at the
right phase of the coherent radiation generated by them.

\section{Tapered wiggler FEL with prebunched electron beam of finite distribution}

The ideal tight bunching model presented in the previous chapters is good for identifying the fundamental
processes of superradiance and stimulated-superradiance in tapered wiggler FEL. At the present state of
the art of technology it is hard to satisfy the tight bunching condition $\sigma_{tb}n\omega_0\ll 1$
required for attaining a non-diminishing bunching factor (\ref{b_n_narrow}) and for taking advantage of
bunch phasing optimization of inner-trap stimulated superradiance dynamics (Eq.~\ref{delta_Pem_du},
Figure~\ref{stimulated_superradiance}) which is valid only in the tight bunch model.
Short-wavelength bunching techniques involve high harmonic energy modulation
of a beam subjected to high power IR lasers in a wiggler as in HGHG \cite{YU_1991,YU_2000}, EEHG
\cite{Stupakov_2009,Qika,Qika_2008} and PEHG \cite{Feng_2014}. The energy bunching turns into tight density
bunching when passed through a dispersive
magnetic element (chicane) \cite{sudar_2017}. However, obtaining significant harmonic current components at
short wavelengths is still a challenging technical task. Furthermore, the model assumption of a cold beam often does not
hold, and in particular, in the case of efficiency enhancement in the post-saturation
tapered wiggler section of seeded FEL, the beam energy spread $\sigma_{\gamma}$ is as large as the trap height $\delta\gamma_m$.
However, new concept of ``fresh-bunch'' input signal injection (where the first bunch is used to generate the modulation power and then discarded while a second bunch is overlapped with the seed \cite{ben-zvi,emma_lutman,emma_feng}) and further
technological developments, may make it possible to get closer to the ideal conditions of our model. To be mentioned that the ``fresh bunch'' technique can be applied to two different slices of the same electron bunch, in which case it is sometimes termed ‘fresh slice’ (see e.g. \cite{emma_lutman,Lutman_2016}. The cases discussed in \cite{Allaria} and the SDUV-FEL tests \cite{Zhao_2016} are ‘fresh-bunch’ results for HGHG schemes, and they also apply there the same principle (suppressing/enhancing lasing for different parts of the bunch in different portions of the undulator).

In this chapter we present a more general model for efficiency enhanced radiation emission in a FEL with
energy spread and phase distribution of the prebunched beam, and compare the simulation results to the case of no
bunching at all.

Following the analysis in \cite{emma_prst_2017} we redefine the interpretation of Eqs.~(\ref{psi}), (\ref{dCqdz_psi}),
(\ref{dphi_E_dz2}) to correspond to individual electrons in the particle distribution of each bunch. We present
the single radiation mode (\ref{dCdz}) in terms of the absolute value of its transverse electric field on axis (\ref{tilde_E_z}):
\begin{equation}
E(z)=|\tilde{C}_q(z)||\tilde{\boldsymbol{\mathcal{E}}}_q(\mathbf{r}_{\perp}(0))|
\label{E_z}
\end{equation}
Then these equations for the electron phase $\psi_j(z)$, energy $\gamma_i(z)$ and radiation field
$E(z)$ are recast correspondingly to the presentation
\begin{equation}
\frac{d\psi_j}{dz}=k_w \left(1-\frac{\gamma^2_j}{\gamma^2_r}\right)
\label{d_psi_j_dz}
\end{equation}
\begin{equation}
\frac{d\gamma_j}{dz}=-\frac{\overline{a}_w e E}{\gamma_j mc^2}\sin\psi_j
\label{d_gamma_j_dz}
\end{equation}
\begin{equation}
\frac{dE}{dz}=\chi \overline{a}_w \left\langle\frac{\sin\psi_j}{\gamma_j}\right\rangle
\label{d_E_dz}
\end{equation}
with $\chi=Z_0I/2 A_{em\,q}$ , $\overline{a}_w=e\overline{B}_w/mck_w$ is the undulator parameter (Eq.~\ref{v_perp1}) and
$\gamma_r^2 = (1 + \overline{a}_w^2) k/2k_w$ (Eq.~\ref{gamma_r}) is the resonant energy.

In this transformation we used the relativistic beam approximation $\gamma_j\gg 1$, and used the definition of
the normalization power in terms of the effective mode cross-section area (Eq.~\ref{P_q}) in the definitions
of the parameters $a$, $b$ (Eqs.~\ref{AAA1}, \ref{BBB}).

In the resonant particle approximation the efficiency can be written as:
\begin{equation}
\eta(z)=\frac{1}{\gamma_0}\left|\sum_i\gamma_i(z)-\gamma_{i,0}\right|\simeq f_t\left|\frac{\gamma_r(z)-\gamma_0}{\gamma_0}\right|=
f_t\left|\frac{\Delta\gamma_r}{\gamma_0}\right|
\label{eta_z}
\end{equation}
where $f_t$ is the fraction of trapped electrons which in general depends on the size of the
bucket, i.e. the input seed power, the undulator field and the resonant phase. We have
assumed for simplicity that the trapping fraction is independent of $z$ in the post-saturation
regime. Different initial conditions for tapered FELs result in different trapping fractions
and different scaling of the output efficiency. Note that as we will discuss in Chapter VIII-C, the
assumption of constant trapping fraction breaks down for long undulators due to diffraction
and time-dependent effects as evidenced in 3D simulations \cite{emma_2016}.

Integrating Eqs.~\ref{d_psi_j_dz}-\ref{d_E_dz} as shown in \cite{emma_prst_2017} we have an
approximation for the power extraction efficiency for $\psi_j=\psi_r=$const. :
\begin{equation}
\eta(z)=f_t\frac{e}{\gamma_0 mc^2}\left(E_0\frac{\overline{a}_w(0)}{\gamma_0}z\sin\psi_r+
\frac{f_t\chi}{2}\frac{\overline{a}^2_w(0)}{\gamma_0^2}z^2\sin^2\psi_r\right)
\label{eta_z_1}
\end{equation}
and with $P_{rad}(z)=P_0 +\eta(z) P_{beam}$ , after rearranging the constants we have (compare with \ref{pem_over_pref_tap_psi_0_is_psi_r}):
\begin{equation}
P_{rad}(z)=P_0+  E_0\frac{\overline{a}_w(0)}{\gamma_0}f_t Iz\sin\psi_r + \frac{Z_0}{4A_{em\,q}} \left(\frac{\overline{a}_w(0)}{\gamma_0}\right)^2
 \left(f_t Iz\sin\psi_r\right)^2 \equiv P_0 + P_{TAPER} + P_{SR}
\label{P_rad_z}
\end{equation}
where the seed field is given by
$E_0=\sqrt{\frac{Z_0 P_0}{A_{em\,q}}}$. Eq.~(\ref{P_rad_z}) is identical to the
normalized Eq.~(\ref{pem_over_pref_tap_psi_0_is_psi_r}), after multiplication by the normalizing power
$P_{REF}$ (Eq.~\ref{P_REF}), and using $f_t=1$. We also remark that in the ideal case of perfect bunching
(Section~VI) $f_t$ can be either 0 or 1, so in case it is 1 (fully trapped bunch), rather than choosing initial
phase $\psi(0)=\psi_r$, which is meant to maximize $f_t$, one would rather choose $\psi(0)=\pi/2$ (compare 
Figures~\ref{psi_0_equal_psi_r} and \ref{psi_0_equal_pi_over_2} and discussion before
Eq.~\ref{pem_over_pref_tap_psi_0_is_psi_r}).

In the absence of an input seed we recover the familiar
scaling for coherent emission from a bunched beam $P_{rad}(z) = P_{SR}(z) \propto  (f_t Iz)^2$ \cite{saldin}.
This is also true for long undulators since the quadratic term $P_{SR}$ dominates the radiation power scaling
if the undulator length $L_w$ satisfies 
$L_w\gg 10^6\frac{\gamma_0}{\overline{a}_w(0)}\frac{\sqrt{A_{em\,q}} [\mu m]}{I [kA]}\frac{\sqrt{P [MW]}}{\sin\psi_r}$.
At the same time for short undulators and intense seed pulses, the efficiency and output power
are linearly proportional to the undulator length and the
field strength. This is the low gain TESSA regime \cite{Duris_2015} discussed in the previous sections. We apply the
power scaling law to a tapered FEL amplifier in two different scenarios: starting
from a large seed with an unbunched and a prebunched electron beam \cite{emma_prst_2017}.

The analytic power estimate from Eq.~\ref{P_rad_z} is in good agreement with numerical integration
of the 1-D FEL equations \cite{emma_prst_2017}. The scaling highlights the trade-off between fast
energy extraction (large $\psi_r$) and large trapping fraction (small $\psi_r$), with the optimal value
around $\psi_r \simeq 40^o$, compared to $\psi_r=90^o$ which emerged from the single particle picture.
This estimate recovers the well known result of \cite{brau} in the low gain (constant $P_{rad}$)
high efficiency FEL in which the optimal resonant phase is also $\psi_r\simeq 40^o$ for a cold beam
with $f_t=f_b$ . This occurs because the efficiency in the low and high gain cases scales as
$\eta\propto(f_t\sin\psi_r)$ and $\eta\propto(f_t\sin\psi_r)^2$ respectively, hence  is maximized at
the same value of $\psi_r$.

\subsection{Scaling laws of Tapering Enhanced Superradiance and Stimulated Superradiance}

It is instructive at this point to compare and distinguish the scaling laws of radiation emission derived in different parts of this review. The general nonlinear interaction regime must be analyzed numerically and graphically as done in Chapter~VII. However examination of the initial power growth scaling - first and second order expansion in terms of the axial propagation parameter $u=z/L_W$ provides some insight into the different coherent spontaneous and stimulated superradiance processes in a uniform and in a tapered undulator.

In the ideal case of a cold tightly bunched beam ($\sigma_{tb}\ll 1/(n\omega_0)$ - see Eq.~\ref{M_b_1}), the initial phase of the bunches $\psi(0)$ (Eq.~\ref{psi}) is well defined, and the radiation power increment is given in terms of a linear term $2u\bar{E}(0)\sin\psi(0)$ and a quadratic term $u^2\sin^2\psi(0)$ (Eq.~\ref{pem_over_pref}). The linear term is due to ST-SR radiation emission process and is maximized when $\psi(0)=\pi/2$, it is the high gain initial stage of a synchrotron oscillation process of a tight bunch in a built-up trap. The quadratic term is due to the SP-SR radiation emission process. It exists whether there is input field $\bar{E}(0)$ or not. In the latter case ($\bar{E}(0)=0$), the bunches determine the radiation field and its phase $\psi(0)=\pi/2$. The radiation fields and the trap bucket grow slowly ($\propto u^2$) around the bunch that eventually interacts with its self generated radiation, and saturates in a synchrotron oscillation process (Section~\ref{self_int}).

Remarkably, Eq.~\ref{pem_over_pref} and two terms of ST-SR - linear in $u$ and SP-SR - quadratic in $u$, exist also in the case of tapering, but here, for linear tapering, an additional energy extraction term, linear with $u$, is added in Eq.~\ref{pem_over_pref_tap} - $2u\bar{E}(0)\sin\psi_r$. The ST-SR term $2u\bar{E}(0)(\sin\psi(0)-\sin\psi_r)$ corresponds to start of a synchrotron oscillation process around central phase $\psi_r$ in a decelerating and somewhat shrunk trap. The SP-SR term $u^2\sin^2\psi(0)$ stays the same. In the case of $\psi(0)=\psi_r$, the synchrotron oscillation stimulated emission process diminishes and one is left only with a linear term due to tapering, and a quadratic term due to SP-SR (Eq.~\ref{pem_over_pref_tap_psi_0_is_psi_r}).

The extension to the finite distribution case - Eq.~\ref{P_rad_z} is fully consistent with Eq.~\ref{pem_over_pref_tap_psi_0_is_psi_r} for the most practical case where only a fraction $f_t$ of the electrons in the beam get trapped. The main consideration then for optimizing power extraction is maximizing $f_t$. It is then optimal to choose $\psi(0)\simeq\psi_r$. In this case one abandons any contribution from the synchrotron oscillation dynamic process, and remains only with the linear term due to the tapering and the quadratic term due to SP-SR (second and third terms in Eq.~\ref{P_rad_z}) that are proportional to $f_tIz$ and $(f_tIz)^2$, respectively.

The scaling of the term $P_{TAPER}$ and $P_{SR}$ in Eq.~\ref{P_rad_z} in the TESSA interaction is the same as the scaling of the ST-SR (Eq.~\ref{P_TESSA_match}) and SP-SR (Eq.~\ref{P_TES_match}) expressions in the uniform undulator with a factor of $\sin\psi_r$ and $\sin^2\psi_r$ respectively. Similarly to Eq.~\ref{ratio}, the ratio between the terms is
\begin{equation}
\frac{P_{TAPER}}{P_{SR}} = \frac{4A_{em\,q} E_0}{Z_0 \frac{\overline{a}_w(0)}{\gamma_0} f_t Iz\sin\psi_r}
\label{ratio_tapered}
\end{equation}
For short interaction length the tapering power extraction $P_{TAPER}$ is dominant if a strong input radiation field is exercised.

\subsection{Unbunched beam}

Even though pre-bunching is desirable, quite significant tapered wiggler energy extraction from an
unbunched beam is possible if the slanted traps still capture enough particles.

\begin{figure}[!tbh]
\includegraphics[width=17cm]{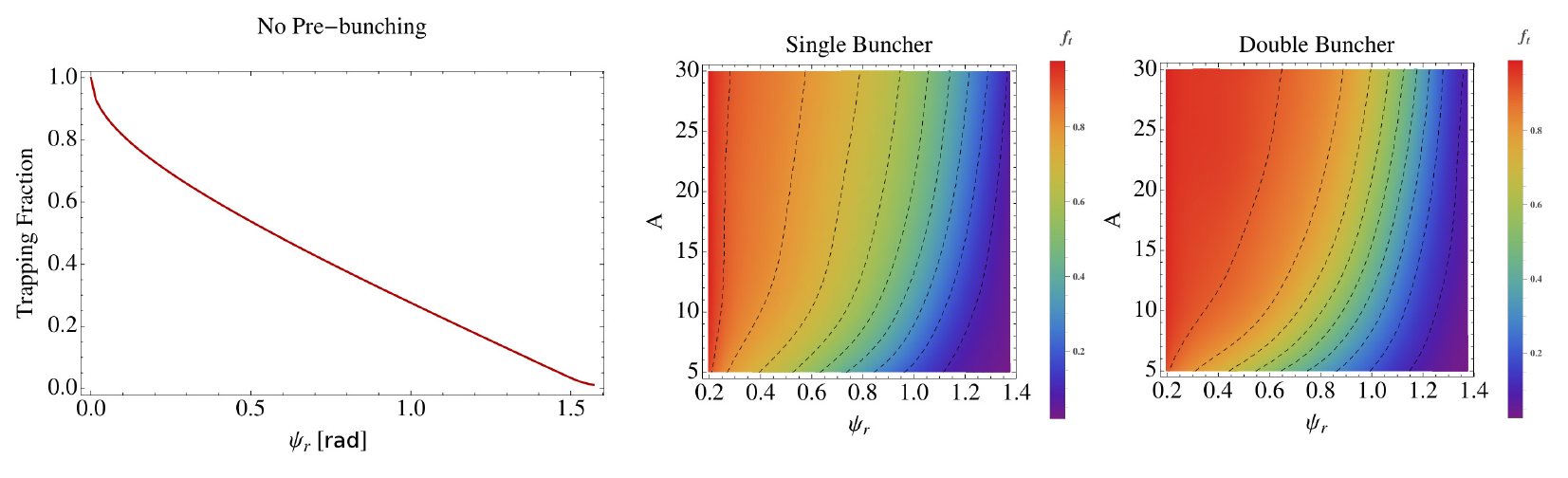}
\caption{Adapted from \cite{emma_prst_2017}. (left): Trapping fraction for a cold unbunched beam and
a large input seed, approximately given by the bucket width. (middle and right): The trapping
fraction $f_t$ as a function of the modulation parameter $A\equiv \Delta\gamma_{mod} /\sigma_{\gamma 0}$ (see Eq.~(\ref{AAAAA}))
and the resonant phase $\psi_r$ for a
single buncher (middle) and a double buncher \cite{sudar_2017,Sudar_2018} (right). The broken lines are polynomial fitting curves of
$f_t(\psi_r,A)$=const.
The color coded diagrams of $f_t$
as function of $\psi_r$ and $A$ were computed for the single and double buncher examples \cite{sudar_2017}.
The trapping fraction obtained with pre-bunching
is larger than in the unbunched case. The
advantage of the double buncher compared to the single buncher scheme is largest for large modulation parameter $A>10$ and
$\psi_r < \pi/4$.}
\label{fig1_emma}
\end{figure}

Starting from a large seed with a cold unbunched beam, the trapping fraction is a function
of the resonant phase only and not the input seed power. It can therefore be approximated
by the bucket width $f_t=f_b=\frac{\psi_2-\psi_1}{2\pi}$ (see Figure~\ref{fig1_emma}, left) with
$\psi_{1,2}$ the solutions of the equations:
\begin{equation}
\psi_2=\pi-\psi_r
\label{psi_2}
\end{equation}
\begin{equation}
\cos\psi_1+\psi_1\sin\psi_r=\cos\psi_2+\psi_2\sin\psi_r
\label{cos_psi_1}
\end{equation}
(see Appendix~\ref{pend_append}, Figure~\ref{tap_separatrix}).

\subsection{Bunched beam}

From Eq.~\ref{P_rad_z} for the power estimate we can see that increasing the trapping fraction greatly
increases the output power for the same resonant phase, $P_{rad}(z)\propto P_{SR}(z)\propto f_t^2$.
Figure~\ref{fig1_emma} (two right panels) show the significantly enhanced trapping
fraction of particles trapped in the bucket with a prebunched electron beam
considering also the effect of energy spread $\sigma_{\gamma}$ and the modulation strength
\begin{equation}
A\equiv \Delta\gamma_{mod} /\sigma_{\gamma 0},
\label{AAAAA}
\end{equation}
where $A$ is the modulation parameter, defined in Appendix~\ref{bunching_append} (Eq.~\ref{energy_distribution_time}).
Figure~\ref{fig1_emma} shows the trapping fraction $f_t$ computed for examples of a single stage or
two-stage pre-buncher setup similar to that discussed in \cite{sudar_2017}. Note that in the context of Figure~\ref{fig1_emma} it was assumed that the strength of the wiggler and the intensity of the radiation wave field are the same in the modulation and tapered wiggler sections, so that $\Delta\gamma_{mod}=\delta\gamma_m$ in Eq~(\ref{delta_gamma_m}). In the more general case these may be independent controllable parameters.

\begin{figure}[!tbh]
\includegraphics[width=17cm]{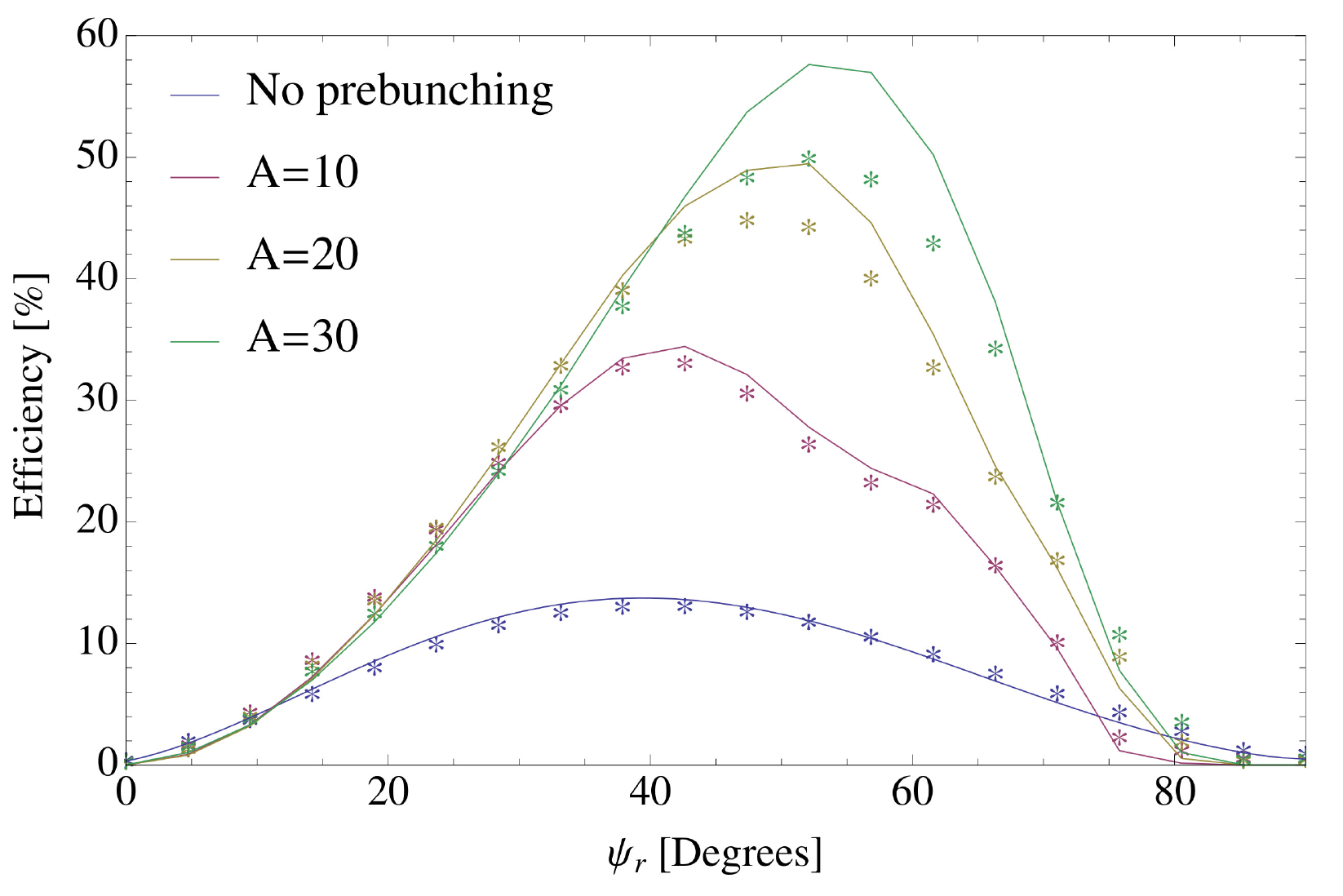}
\caption{From \cite{emma_prst_2017}. Analytic estimate of the power extraction efficiency based on Eq.~\ref{eta_z_1}
and on
simulation (solution of Eqs.~\ref{d_psi_j_dz}-\ref{d_E_dz}) (dots) 
without pre-bunching and for three different values of the modulation strength $A$. The advantage
of pre-bunching is clear in both the larger maximum efficiency and the peak efficiency occurring
at larger resonant phase, allowing faster energy extraction.}
\label{fig2_emma}
\end{figure}

Experimental results demonstrating very high energy extraction
from a pre-bunched electron beam in a tapered undulator have recently been reported in
\cite{nocibur}, and are discussed in Chapter VIII-B . The trapping fraction in this case is not
only a function of the resonant phase but also of the initial laser seed power which sets
the pre-bunching modulation strength $A$. In Figure~\ref{fig2_emma}
(taken from \cite{emma_prst_2017}) we compared the trapping
fraction without pre-bunching to the analytic fit of the trapping fraction $f_t=f_t(\psi_r,A)$ for
optimal buncher settings in a single or double-buncher configuration \cite{Sudar_2018} with a bucket height
10-30 times larger than the initial electron energy spread.
The advantage of pre-bunching
is clear in the increasing trapping fraction for both the single and double-buncher cases at
modest modulation strengths $A\sim 10$. As discussed in more detail in \cite{emma_prst_2017}
the main challenge in this scheme would be to generate the seed laser power
capable to induce an $A=10$ modulation which could be achieved for example in a fresh
bunch configuration \cite{ben-zvi,emma_lutman,emma_feng}.

Our analysis of the 1-D physics of high gain high efficiency tapered FELs, including the
effects of energy spread and bunching, displays the advantage of pre-bunching in a high
efficiency FEL due to the increased particle trapping in the post-saturation region. Pre-
bunching not only increases the peak efficiency but exhibits optimal energy extraction at
larger resonant phase compared to the unbunched case. This faster extraction of energy is
important for reducing harmful 3-D effects, specifically diffraction due to reduced optical
guiding in tapered FELs with long undulators. Having a pre-bunched electron beam also
allows more aggressive (larger resonant phase) tapered FEL designs, damping the time-
dependent parasitic effects of the sideband instability (see section
VIII-C.2 \cite{kroll,kroll_1979}), which can limit the output efficiency.

\section{Applications}

\subsection{Superradiant coherent radiation sources}

A straightforward application of superradiance is in developing THz superradiant sources. This is quite expected,
because the electron beam bunches generated in a common photocathode gun of modern RF linacs, are of
bunch duration of $\sigma_{tb}<1$ps. Consequently, bunching factor of $M_b\simeq 1$
is attainable for radiation frequencies in the range of $f<1$THz.

\begin{figure}[!tbh]
\includegraphics[width=8cm]{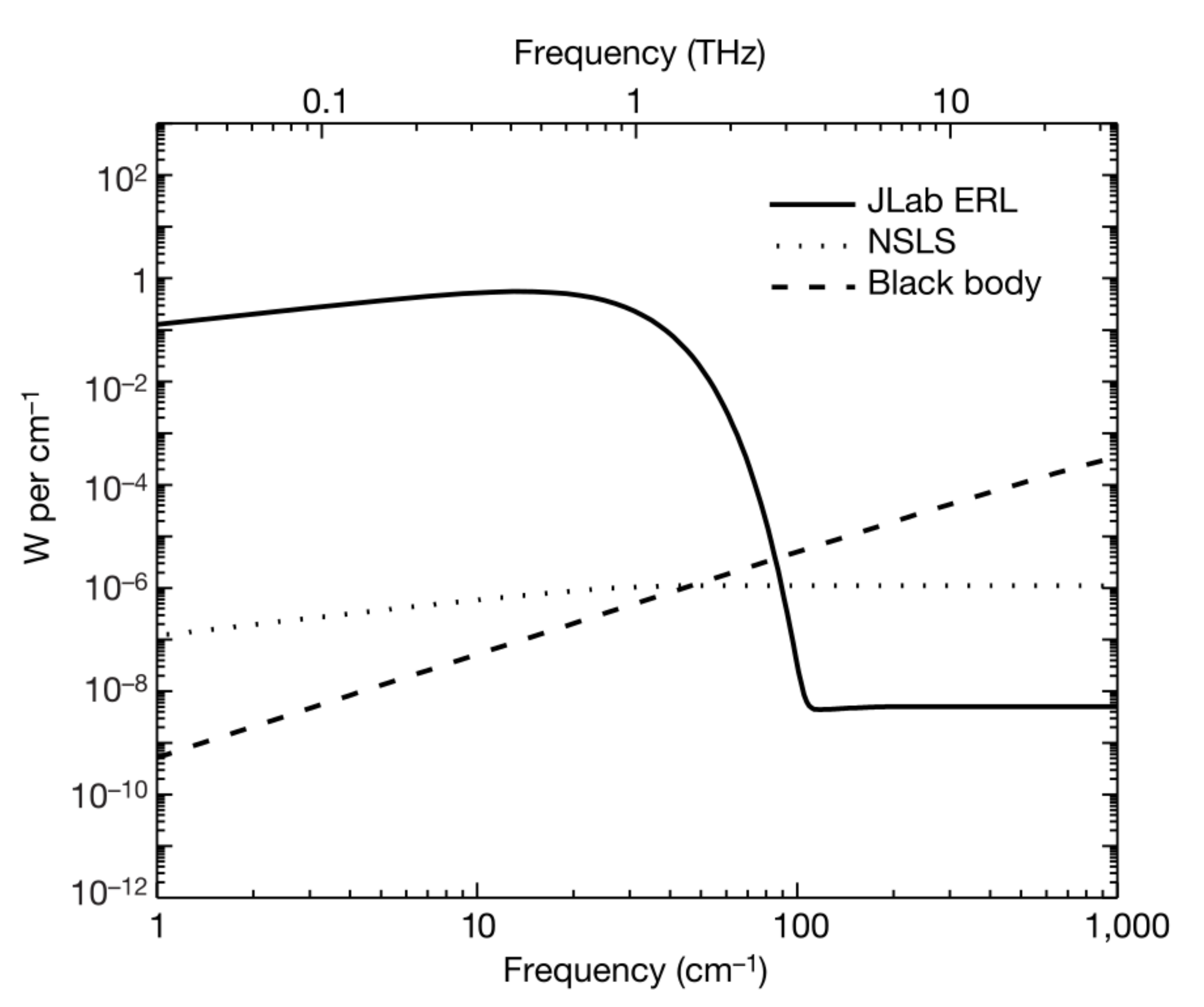}
\caption{Average power from multiple single-bunches superradiant (CSR) emission of wide bandwidth THz radiation in a ERL accelerator \cite{Carr_2002}. A dramatic enhancement of the spontaneous emission radiation spectral power takes place at low frequencies due to superradiant CSR.}
\label{JLAB_paper}
\end{figure}

Figure~\ref{JLAB_paper} shows an experimental demonstration of the dramatic transition of radiation power from spontaneous emission $\propto N$  (Eq.~\ref{DWdq_st}) to superradiant emission  $\propto N^2$ (Eq.~\ref{DWdq_SR}) that takes place in the cutoff condition $\omega\sigma_{tb}\simeq 1$ . In this experiment, carried out in JLAB ERL FEL \cite{Carr_2002}, a record high frequency-integrated THz power (20Watt) was measured due to repetitive single bunch superradiant Coherent Synchrotron Radiation (CSR) emission from a bending magnet. The THz radiation was generated by a continuous stream of electron beam microbunches of duration 0.5pSec, circulating in a superconducting ERL at a repetition rate of 75 MHz. Figure~\ref{JLAB_paper} shows more than seven orders of magnitude enhancement in the spectral power at the superradiance regime relative to the spontaneous emission regime. This corresponds to the large enhancement by a factor N - the number of electrons in each bunch (Eq.~\ref{DWdq_SR}).

\begin{figure}[!tbh]
\includegraphics[width=17cm]{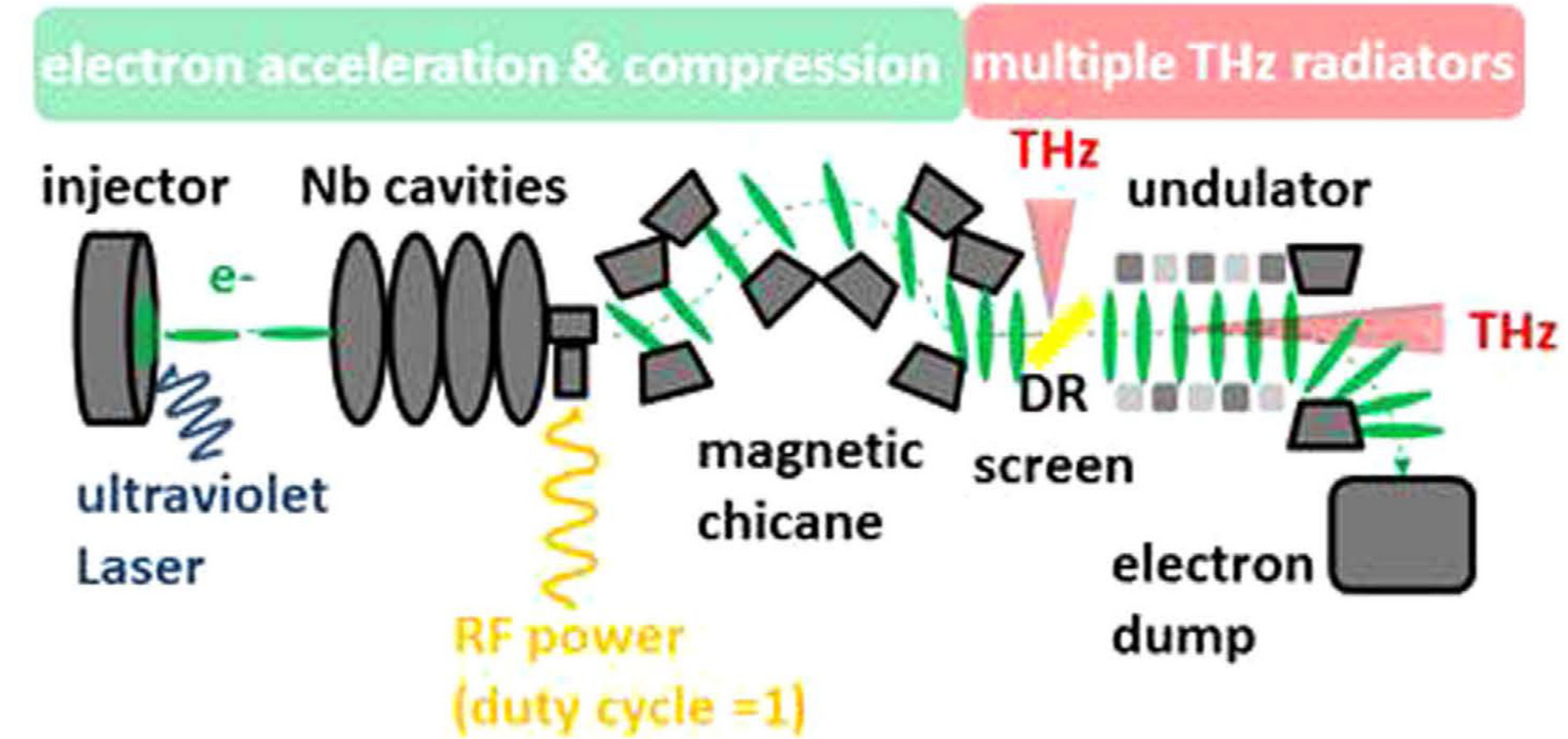}
\caption{From \cite{Green_2015}. Electron bunches are extracted from
a solid, accelerated to relativistic energies and compressed to sub-ps duration in a compact SRF linac with a
chicane bunch compressor. The electron bunches can emit THz pulses in different types of radiators. At TELBE,
repetition rates up to 13 MHz are feasible. THz pulses are generated by a diffraction radiator (DR) and one
undulator.}
\label{telbe}
\end{figure}

There are numerous SR experiments world-wide pursued or planned, mainly in the THz spectral regime \cite{Gover_2005,Huang_2015,Hama_2011,Shin_2007,Korbly_2005,Ginzburg_2013,Ciocci_1993,Gover_1994,Faatz_2001,Lurie_2015,Shibata_2002,Yasuda_2015,Huang_2010,Su}. More than 20 experiments of SR THz sources based on either Undulator, CTR or Bending magnets (CSR or Edge Radiation) are listed in \cite{Gensch_2013}. We review several representative THz radiation facilities based on superradiance, and refer the interested readers to review articles of this matter \cite{Gensch_2013,Asgekar_2014,Green_2015}. A representative example is the TELBE facility located in Helmholtz-Zentrum Dresden-Rossendorf (HZDR), shown schematically in Figure~\ref{telbe} (see \cite{Green_2015})
This facility is based on superradiant enhancement of radiation
from relativistic electron bunches in a compact superconducting LINAC. This prototype source generates
high-field THz pulses at quasicontinuous-wave repetition rates up to the MHz regime and exceeds the power of
state-of-the-art laser-based sources by more than 2 orders of magnitude.

Other facilities in design are FLUTE \cite{Nasse_2013} located in Karlsruhe Institute of Technology and PITZ
\cite{Boonpornprasert_2014} in DESY.
The first one will use a 2.5 cell normal conducting photocathode RF gun to preaccelerate
the electrons to an energy of 7 MeV. This gun was specially designed for very high bunch charge and
operates at 2.998 GHz (European S-band). This facility is designed to use coherent synchrotron
and edge radiation (CSR/CER) or superradiant coherent transition radiation (CTR).

Another notable planned facility (PITZ) is intended to develop THz radiation generation sources in two
complementary schemes:
(i) Self-Amplification of Spontaneous Emission Free-Electron Laser (SASE FEL)
using an undulator and high charge electron bunches and (ii) Coherent Transition Radiation (CTR), using an ultra-short
electron bunch tightly compressed by a chicane. The SASE radiation is anticipated to cover radiation short
wavelengths range of 20-100$\mu$m while radiation wavelengths above 100$\mu$m, for which the bunch duration is shorter
than the wavelength will be generated by the superradiant CTR scheme.

\begin{figure}[!tbh]
\includegraphics[width=8cm]{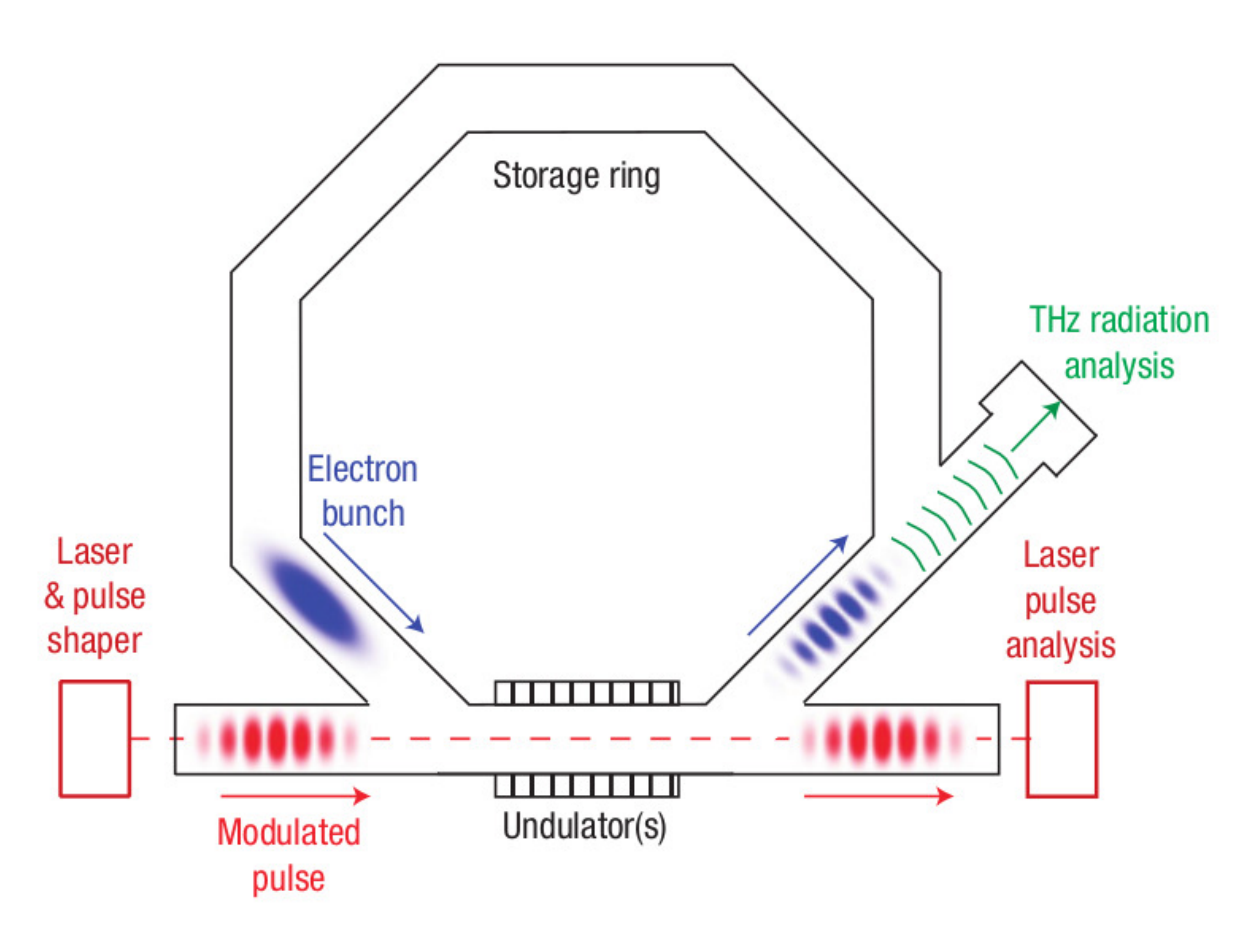}
\caption{Superradiant CSR narrow bandwidth THz radiation emission by trains of pSec bunches generated in a storage ring by laser modulation in a wiggler \cite{Bielawski}}
\label{storage_ring}
\end{figure}

A good example of experimental demonstration of narrow-band coherent Terahertz superradiant emission in an undulator with a finite pulse train of bunches  is shown in Figure~\ref{storage_ring}. Here the relatively long pulses of the UVISOR-II storage ring are modulated in an undulator by a Terahertz modulated laser beam, and emit narrowband superradiant CSR radiation (as in Eq.~\ref{dWdomega_SR}, Figure~\ref{train_fig}) at the bending magnet \cite{Bielawski}. 

Another example of THz superradiance by a periodically bunched finite pulse train has been demonstrated recently by Tsinghua University in China \cite{Su}. In this scheme a bunch train is created by laser stacking using birefringent $\alpha$-BBO crystal serials from a Ti:Saphire laser system, an energy chirp is introduced into the electron pulse which is then compresses with a magnetic chicane, and entered into a permanent magnet undulator. The resulting narrow bandwidth radiation is tunable in the range of 0.4 to 10 THz. 

\begin{figure}[!tbh]
\includegraphics[width=15cm]{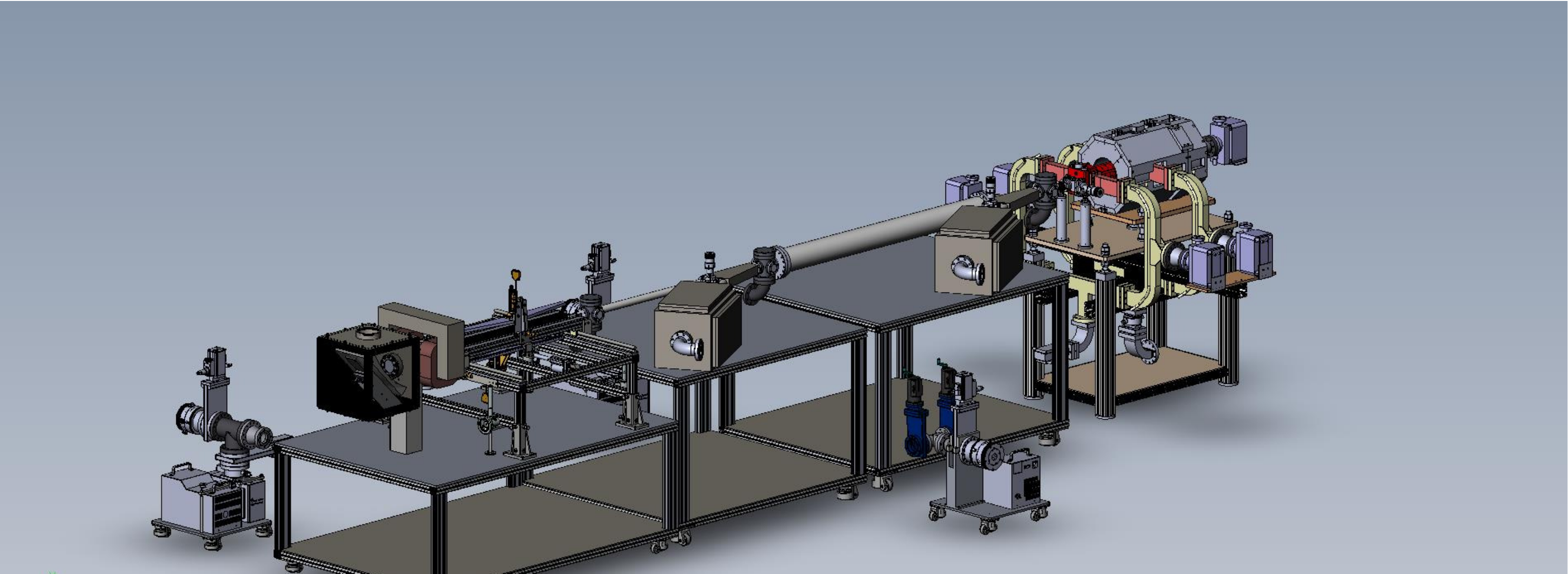}
\caption{Layout of the  Israeli Superradiant THz FEL}
\label{aharon_1}
\end{figure}

Under construction in Ariel University is the Israeli THz FEL, a cooperation between
Ariel University and Tel Aviv University \cite{Friedman_2014}. The device, depicted in Figure~\ref{aharon_1} is
designed to operate with a compact RF photo-cathode gun of up to 6.5 MeV. The
gun introduces an energy chirp to the beam. Thus resulting in a compression of the
electron pulse. The compressed pulse then enters into a 2~cm period 80~cm long
linear Halbach undulator, generating super-radiant radiation at 1-4~THz. Tuning is
carried out by varying the electron energy and/or the undulator gap. Another
ultra-short (5~mm period) wiggler is under construction for operating in a mode of
negative mass effect \cite{Balal_2015,Lurie_2016}.

The project is planned to be carried out in several phases. In the first phase the FEL
will operate in a single bunch superradiance mode (Eq.~\ref{dWdomega_SR_2}) with bunches
compressed to less than 150fSec, generating 10pSec THz pulses of about 50 kW
peak power. In a second phase, the FEL will operate in a periodic pre-bunching
superradiant mode (Eq.~\ref{dWdomega_SR_1}). The THz pre-bunching will be produced by illuminating
the photocathode with two optical beams, generated by splitting the naturally
chirped ultrafast laser beam. With proper relative delay, the two beams would
produce upon the cathode a THz laser beat and thus modulate the emitted electron
photo-current. In a future phase of the project, it is intended to equip the facility
with an additional tapered helical undulator \cite{Duris_2014} in order to demonstrate intense
THz radiation in a zero-slippage waveguide TES operating mode (Sect. VI-G) and in
THz beam acceleration \cite{emma_snively}.

Beside Coherent Synchrotron Radiation (CSR) and superradiant undulator radiation mechanisms, there have been numerous demonstrations of superradiant radiation in various radiation emission mechanisms, including Coherent Transition Radiation (CTR) \cite{Happek_1991,Shibata_1994}, Diffraction Radiation, Cerenkov dielectric structure radiation \cite{Wiggins,Neighbors} and Smith-Purcell Radiation \cite{Bluem,Shin_2007,Korbly_2005,Brownell_1998,Ginzburg_2013}. A nice demonstration of superradiant Smith-Purcell Radiation by a finite pulse train of tight bunches was reported by Korbly et al [59]. In this experiment a 100nSec macro-pulse beam, composed of a 17GHz train of 1500 micro-bunches of 0.5pSec from a 15MeV LINAC, was passed in proximity to a 1cm period blazed grating of 10 periods. The experiment demonstrated extremely narrow bandwidth THz radiation at up to the 30$^{th}$ harmonic of the fundamental 17GHz frequency in a comb pattern characterized by the macro-bunching form factor Eq.~\ref{periodic_bunch6} (Figure~\ref{M_M_fig}). 
          
Superradiant THz sources are being developed in many advanced X-UV FEL facilities taking advantage of the coherent radiation emitted by the spent electrons from the FEL undulators, before being dumped. Coherent THz beams synchronous with the main X-UV output pulses of the FEL are very useful for pump-probe experiment applications such as using the high-power THz pulses as a pump, in order to modulate structural properties of matter, thereby inducing phase transitions \cite{Adams_2004}. The linear acceleration sections produce naturally terahertz (THz) radiation using dedicated undulators, bending magnets (CSR) in FLASH \cite{Genschal_2008} or a dedicated target foil (coherent transition radiation; CTR) in FLASH \cite{Matthias_2011} and FERMI (TeraFERMI \cite{Svetina_2016}). The quality of the beam after UV/XUV generation in the FEL is good enough for producing superradiant THz radiation that can be synchronized with some delay with the UV/XUV pulse of the same bunch for pump-probe experiments.

We point out that there are superradiant emission effects (in the narrow and wider sense) also in common IR/FIR FEL oscillators based on photocathode RF LINACs \cite{FEL_Europe}, eg. CLIO FEL in France \cite{Ortega_1996}, FELBE in HZDR \cite{Gabriel_2000} FELIX \cite{FELIX}, Radboud University \cite{Zhaunerchyka} and Novosibirsk FEL \cite{Kulipanov_2015}. These facilities, operating as FELs in oscillator configuration, provide tunable coherent narrow bandwidth radiation in the THz up to mm-wavelengths range. The RF accelerator bunches in such facilities are in the picosecond scale duration, so that in the shorter wavelength (IR) range of their operation, the pulse duration is longer than the slippage time along the wiggler, and therefore, their operation is primarily in a stimulated emission (laser) mode. However, when operating at the long wavelength range (FIR or THz), they exhibit pronounced slippage effects and superradiance in the wider sense. Even stimulated-superradiance effect has been observed in \cite{Zhaunerchyka} due to the presence of an intense circulating field in the resonator.

We would be amiss if we do not mention that long wavelength superradiance effects can be observed in synchrotron storage rings. The turn of the electron in a light source generates a wide band radiation with frequencies ranging from 0 frequency to the cut off frequency. It is inevitable that due to the pulsed nature of the electron beam, superradiance would occur at wavelengths longer than the pulse length. The superradiant emission at the synchrotron bending magnets is termed ``Coherent Synchrotron Radiation'' (CSR). The typical spectral range of CSR corresponding to the steady state bunch duration in synchrotron storage rings is in the microwave to mm-wave range. However, there are numerous demonstrations of CSR emission with synchrotron storage rings also in the THz regime \cite{Sannibale_2004,Byrd_2004,Wang_2006,Abo-Bakr_2002}. Billinghurst et al at the Canadian Light Source report observation of superradiance at frequencies that were harmonics of the electron beam pulse train \cite{Billinghurst}. Synchrotron storage rings employ RF frequencies of 500~MHz and lower. Therefore, their circulation bunches lengths, and consequently their CSR wavelength cutoff, are in the mm-wavelength range \cite{Abo-Bakr_2002}. It is still possible to operate such synchrotron storage ring in a ``burst mode'' of fewer circulating bunches \cite{Abo-Bakr_2002} at the expense of instability and shorter circulating lifetime. To be mentioned that also in storage ring it is possible to get THz radiation in dedicated IR beamlines, and this requires special mode of operation where the bunches are compressed (however this has negative effect on the Synchrotron ring lifetime). FLUTE \cite{Nasse_2013} use this special regime where the bunch is squeezed longitudinally to the picosecond range, offering THz radiation to its users. Alternatively, stable THz CSR emission can be demonstrated with dedicated storage rings, operating  at high (S-band) frequency \cite{Wang_2006,Byrd_2004}.

Finally we point out here the great interest in extending the concepts of superradiant emission to short wavelengths in the optical to X-Ray regime. Several techniques have been proposed for attaining electron beam bunches in the femtosecond \cite{Marceau_2015,Hilbert_2009,Wong_2015,Hommelhoff_2006,Hoffrogge_2014,Zholents_2008} and sub-femtosecond range and may be used for optical X-UV superradiant sources. Attosecond level bunching has been demonstrated even at the quantum electron wavefunction level \cite{Feist_2015,Kozak_2018,Priebe_2017}. Spontaneous and stimulated superradiance have also been considered at the quantum wavefunction level \cite{Pan_2018}.

An entire class by itself comprises the schemes of combined bunching and harmonic emission with seeding ultrafast optical laser beams that we have not covered in depth in this review. The reader is referred to reports on these processes HGHG \cite{YU_1991,YU_2000}, EEHG \cite{Stupakov_2009,Qika,Qika_2008} and PEHG \cite{Feng_2014}.

\subsection{TESSA and TESSO concepts}

This section contains a summary of the recent activities in high extraction efficiency experiments taking advantage of the tapering enhanced stimulated superradiant amplification (TESSA) process. Using an intense seed pulse, in conjunction with pre-bunched beams it becomes possible to initialize the system in a very favorable initial state with particles deeply trapped in the ponderomotive bucket of a tapered undulator.

It is interesting here to note the reciprocal relation between the challenges of radiation emission and the quest for laser-driven accelerators. TESSA can be thought as the reverse of an Inverse Free-Electron Laser accelerator \cite{Palmer,Courant} which among other laser-driven schemes enjoys some unique advantages due to the lack of nearby boundaries, structure or medium to couple the light to the electrons. These result in a direct coupling between electromagnetic field and relativistic electron beam with very little irreversible losses enabling in principle very high conversion efficiencies. A long history of IFEL experiments \cite{Marshall,Kimura,Musumeci_2005} was recently followed by experimental results on IFEL (ATF Rubicon experiment \cite{Duris_2014} and LLNL IFEL experiment \cite{Moody}) which have shown the advantages of the helical geometry and the possibility to double the energy of a 50 MeV relativistic beam using $<$200 GW peak power CO$_2$ laser and accelerating gradients up to 200 MeV/m. The experience gained over many years in developing tapered undulator for acceleration purposes finds thanks to the TESSA mechanism direct application in the field of high efficiency coherent radiation sources. 
\begin{figure}[!tbh]
\includegraphics[width=17cm]{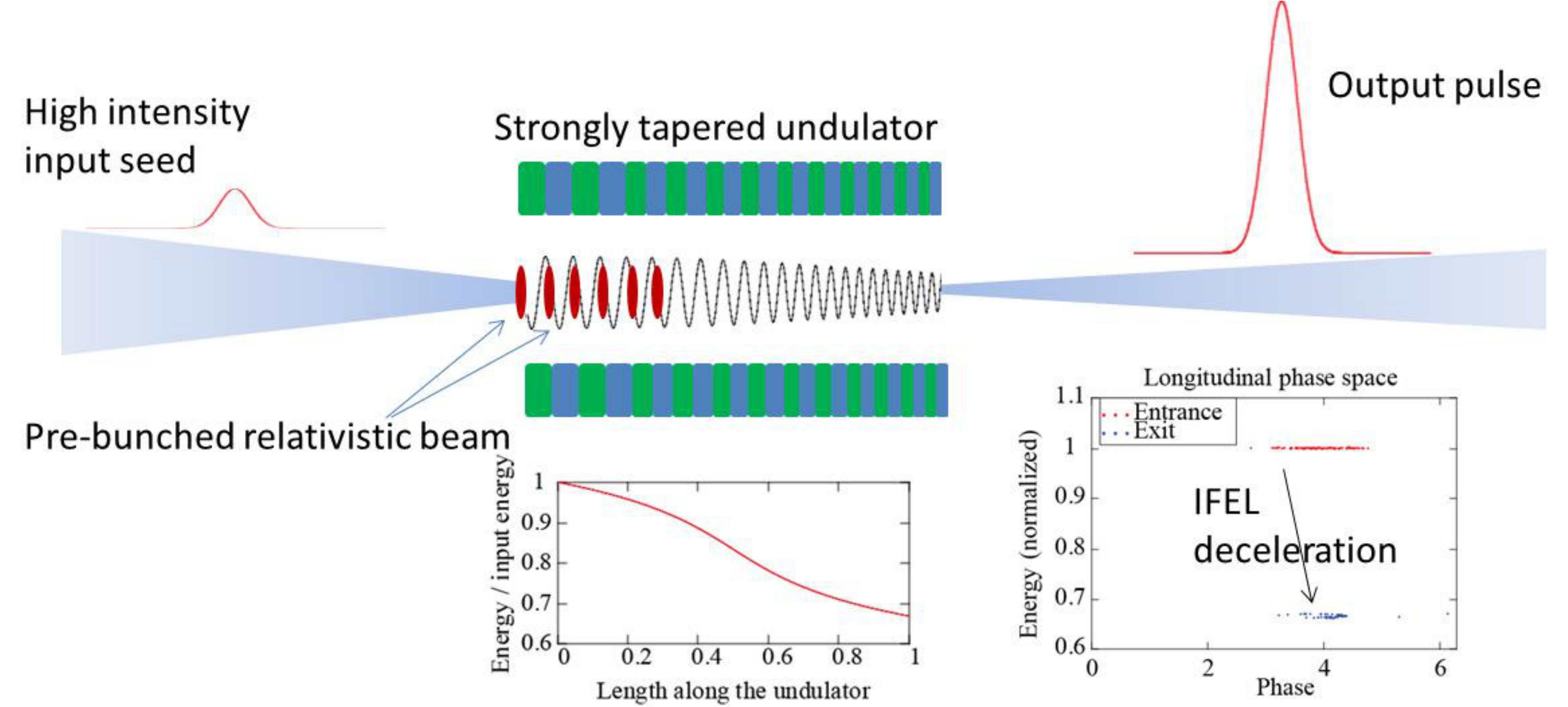}
\caption{Tapering-Enhanced Stimulated Superradiant Amplifier}
\label{fig1_pietro}
\end{figure}

A cartoon schematic of a TESSA amplifier is shown in Figure~\ref{fig1_pietro}. The strong seed pulse stimulates the emission of large amount of radiation from pre-bunched superradiant electron beams. This emission of radiation causes the electrons to quickly lose their energy. If the undulator is tapered in order to maintain the resonant condition and sustain the interaction over long undulator section(s) with strong coupling to the fundamental laser mode, one can achieve very high conversion efficiency. 

\begin{figure}[!tbh]
\includegraphics[width=17cm]{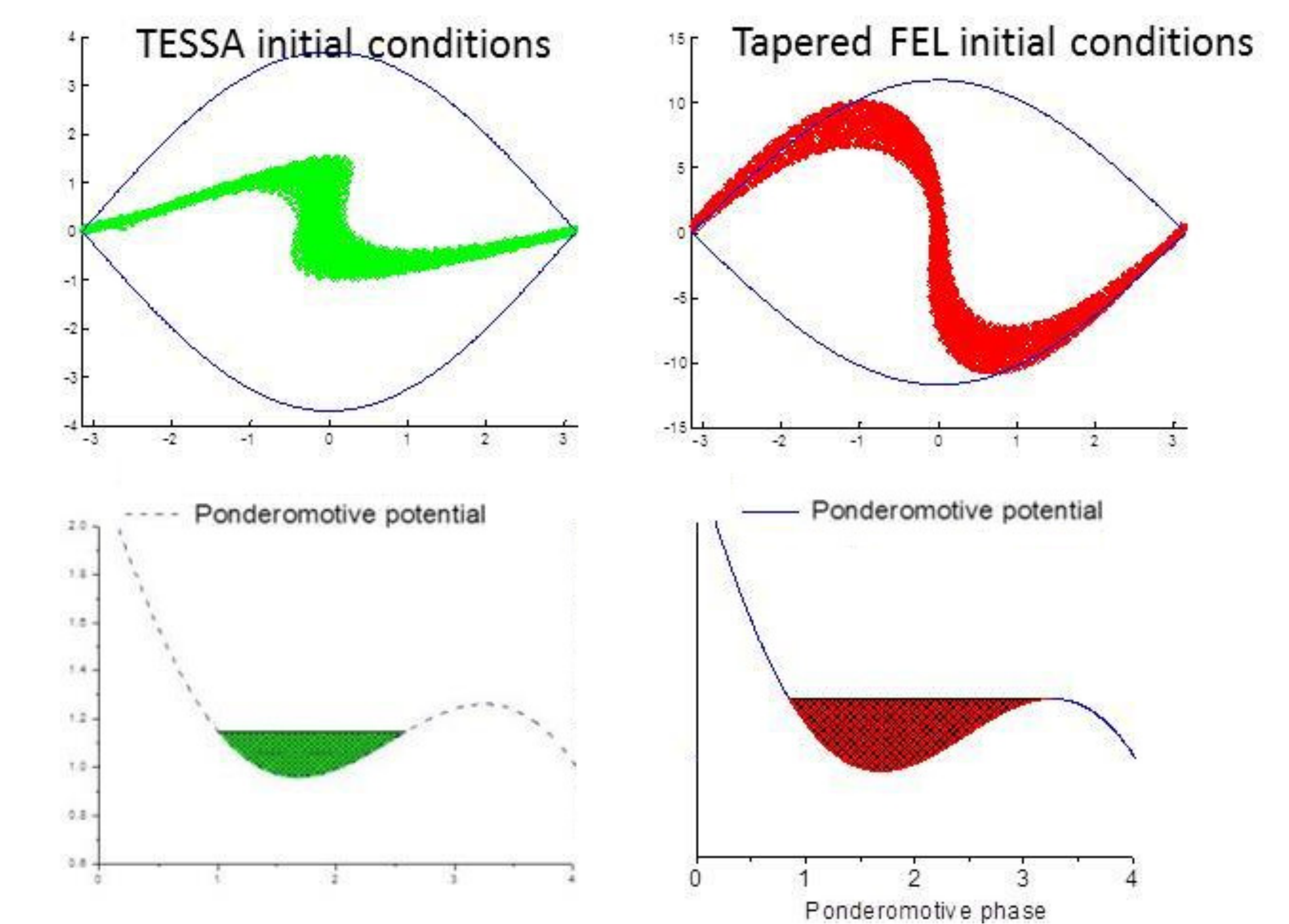}
\caption{(Left) Deeply trapped TESSA initial conditions (Right) Full bucket tapered FEL initial conditions}
\label{fig2_pietro}
\end{figure}

The scheme differs from tapered FEL amplifier since using a very high initial intensity and an initially micro-bunched beam allows trapping the beam into a deep ponderomotive bucket (as oppose to tapered FEL amplifier where the trapping potential well is always ``full to the brim'' with particles - see comparison in Figure~\ref{fig2_pietro}). A consequence of this important difference in the initial conditions of the system is that TESSA undulator can be tapered more aggressively (hence the use of strongly tapered helical undulator) before suffering from particle detrapping as it is usually the case for tapered FELs. This results in higher decelerating gradient and energy extraction from the electron beam. Futhermore,the steeper tapering profile and higher radiation gain reduces the degrading effects of diffraction and sideband instability.

In order to better frame the experiments, it is helpful to distinguish two regimes of operation of TESSA.

\subsubsection{Small gain regime}

In this regime the electromagnetic radiation is assumed to be nearly constant along the interaction. In this case the amplifier behaves as a particle decelerator with an output signal only moderately larger than the input (gain close to unity). This regime can be useful to quickly estimate the efficiency for a low gain amplifier. In practice it can be useful in an optical cavity configuration where part of the output power is split and redirected at the input. This scheme implementation is discussed further in subsection 4 (see Figure~\ref{fig8_pietro}).

Using Eq.~(\ref{sin_psi_r1}), assuming a constant period undulator ($\lambda_w$=const), and defining the
normalized potential vector
\begin{equation}
K_l\equiv \frac{eE}{mc^2 k},
\label{K_l}
\end{equation}
one obtains the total change of the wiggler parameter, along the wiggler:
\begin{equation}
\Delta \overline{a}_w= 4\pi N_w K_l\sin\psi_r
\label{delta_a_w}
\end{equation}

In practice the contribution of internal dynamics to the power extraction efficiency is negligible in these
experiments and the main contribution to the power generation is the wiggler tapering.
The extraction efficiency is given by the ratio of the energy
decrement $\gamma_r(0)-\gamma_r(L_w)$ and $\gamma_r(0)$ (Eq.~\ref{gamma_r}) times the trapping efficiency:
\begin{align}
\eta = & f_t\left(1-\frac{\gamma_r(L_w)}{\gamma_r(0)}\right)=f_t\left(1-\sqrt{\frac{1+(\overline{a}_w-\Delta\overline{a}_w)^2}{1+\overline{a}_w^2}}\right)= \notag \\
 & f_t\left(1-\sqrt{1-\frac{2 \overline{a}_w(0) \Delta\overline{a}_w}{1+{\overline{a}_w^2(0)}} + \frac{\Delta\overline{a}_w^2}{1+{\overline{a}_w^2(0)}}}\right) \approx f_t \frac{\overline{a}_w(0) \Delta\overline{a}_w}{1+\overline{a}_w(0)^2}
\label{efficiency}
\end{align}
where $f_t$ is the fraction of particles trapped in the ponderomotive potential and $\psi_r$ is the design resonant phase (typically $\sim$ 45 degrees to compromise optimum deceleration and maximum trapping). If one includes diffraction effects in the optimization it is found that the input seed should be focused at the center of the undulator with a Rayleigh range about 1/6 of the undulator length.  For large $K_l$ and long undulator this number can easily approach 50 \%.

\subsubsection{High gain regime}

In the small signal gain regime, the conversion efficiency is independent of the beam current, since there we are not considering the fact that the radiation is increasing along the undulator. When one includes the growth of the signal amplitude, a steeper tapering can be allowed and stronger amplification can occur.
In order for this to occur it is essential to develop a tapering optimization algorithm which can take advantage of the newly generated radiation intensity in the most efficient way. In  \cite{Duris_2015} the optimization algorithm was obtained through fully three-dimensional computer simulations. The Genesis-Informed Tapering (GIT) algorithm was developed to read-off the on-axis intensity after solving the field propagation equations for a small section of the undulator and use this information to optimize the next period undulator parameter variations maximizing the energy extraction without compromising the trapping.

\subsubsection{Nocibur experiment. Demonstration of small gain regime}

\begin{figure}[!tbh]
\includegraphics[width=17cm]{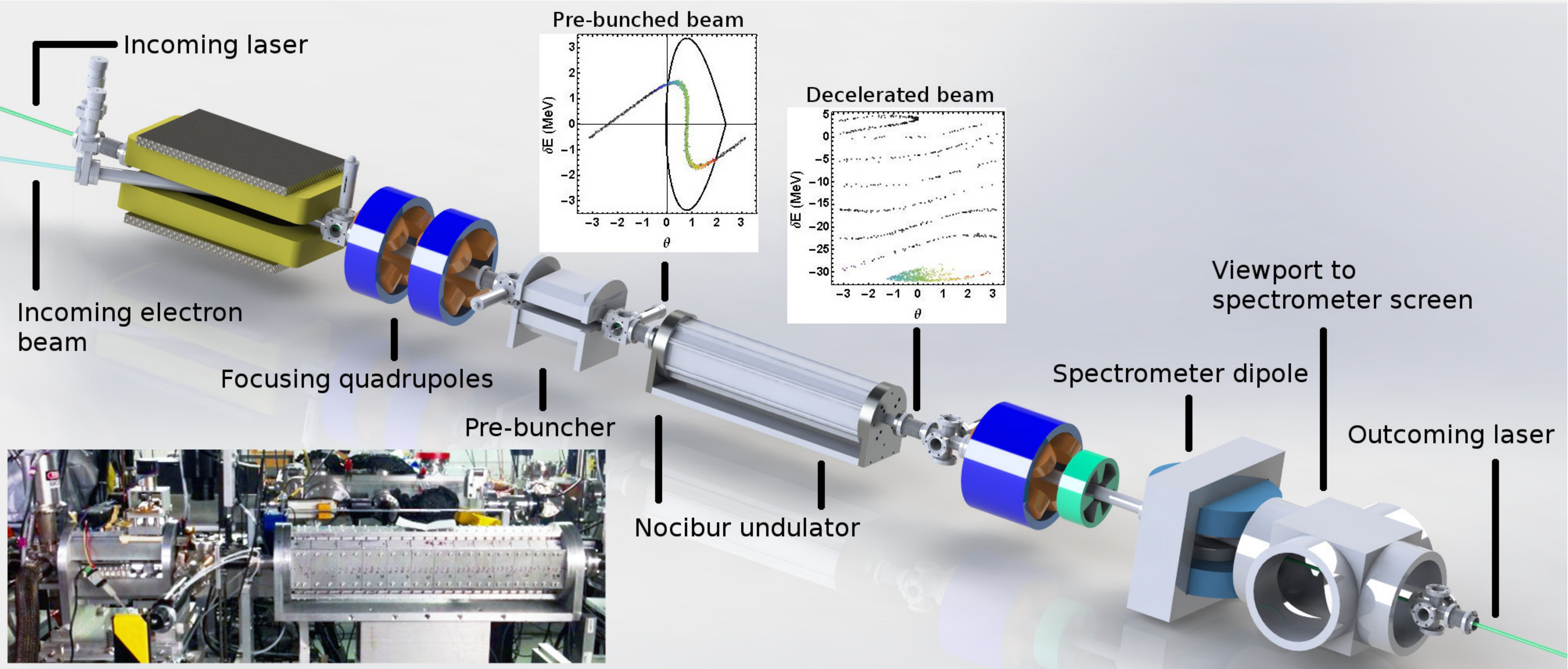}
\caption{Nocibur experiment beamline layout}
\label{fig3_pietro}
\end{figure}

The Nocibur experiment \cite{nocibur} demonstrated the low gain Tessa regime, converting up to 30\% of a highly  relativistic electron beam's energy to coherent 10.3~$\mu m$ radiation.  The experiment was performed at the Brookhaven National Laboratory's Accelerator Test Facility, utilizing a 200 GW seed from the high power mid-IR CO$_2$ laser.  The strongly tapered, helical Rubicon undulator, that was used as an inverse FEL accelerator \cite{rubicon}, was reversed to decelerate up to 45\% of a 100 pC, 65 MeV electron beam to 35 MeV, Figure~\ref{fig3_pietro}.

\begin{figure}[!tbh]
\includegraphics[width=17cm]{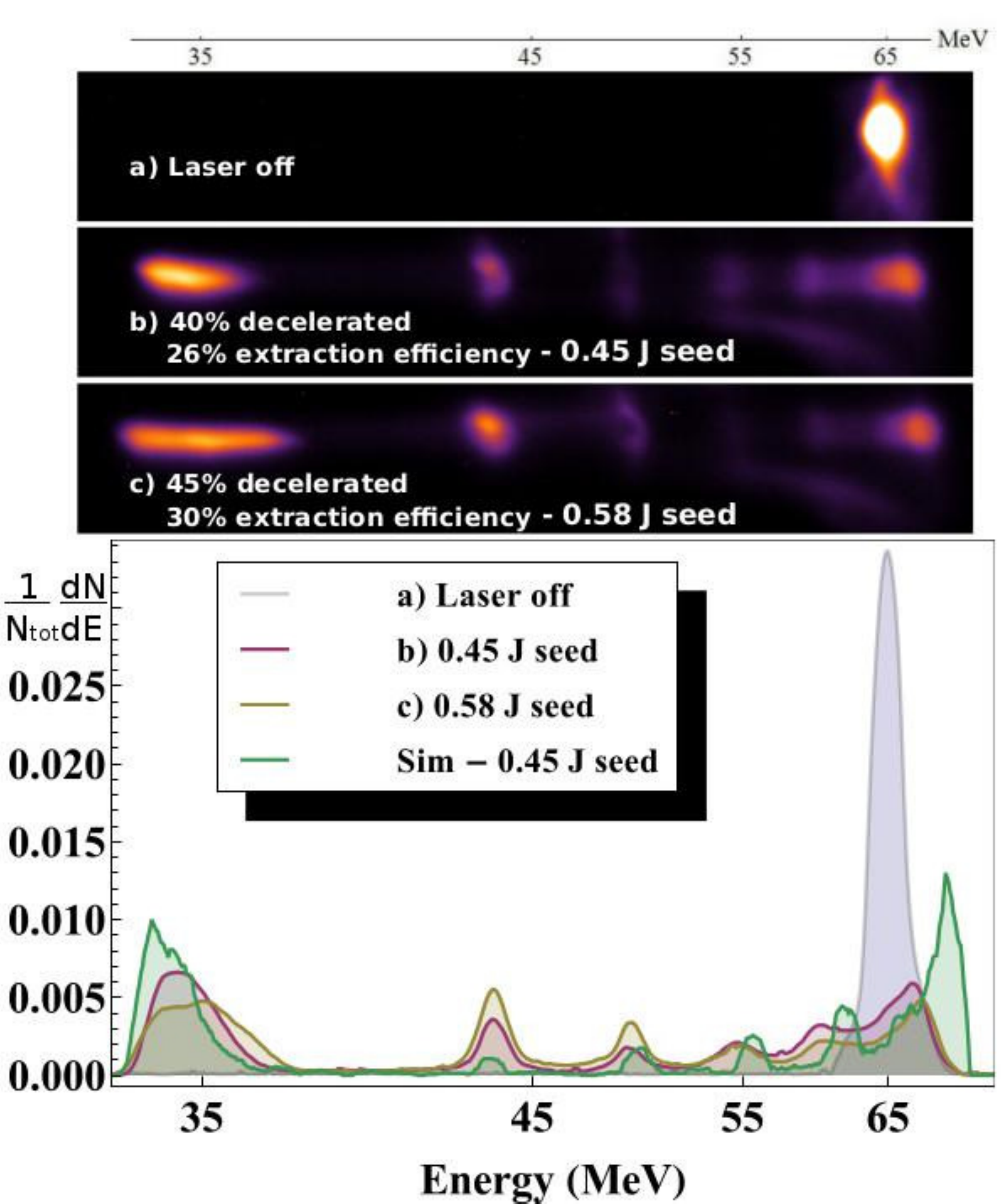}
\caption{Experimental spectra}
\label{fig4_pietro}
\end{figure}

Considering the initial beam energy of 6.5~mJ  and defining the total electron beam  energy after the interaction as:
\begin{equation}
E_{tot}=\frac{Q}{e} \int \frac{1}{N_{tot}} \frac{dN}{dE} E\, dE
\label{E_tot}
\end{equation}
gives an average final electron beam energy of 4.5$\pm$0.4mJ and an extraction efficiency of 30\%, Figure~\ref{fig4_pietro}.

\begin{figure}[!tbh]
\includegraphics[width=9cm]{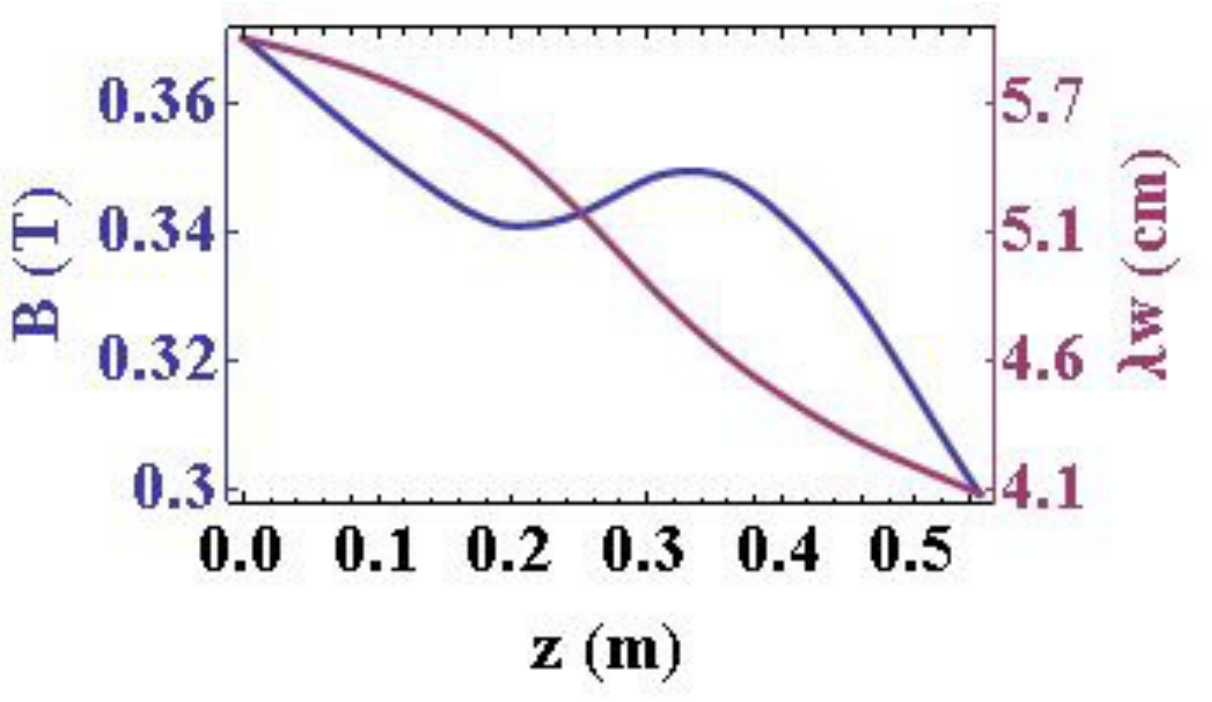}
\includegraphics[width=7cm]{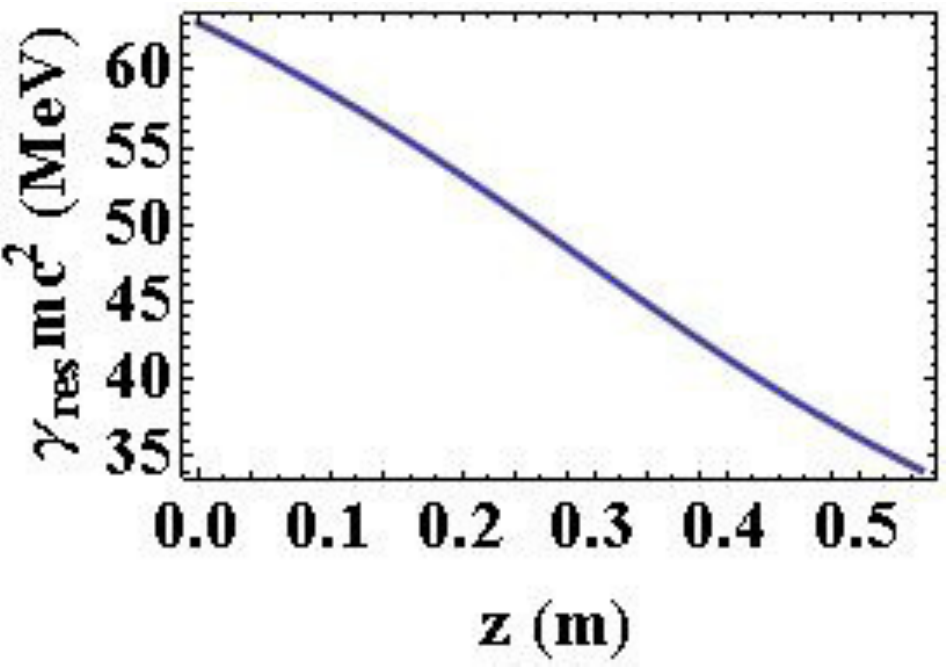}
\caption{(Left) undulator period and magnetic field amplitude. (Right) Resonant energy}
\label{fig5_pietro}
\end{figure}

The Rubicon undulator consists of two 11 period planar Halbach undulators oriented perpendicular and shifted in phase by $\pi$/2 with a total interaction length of 0.54 m.  The undulator period is tapered from ~6 to 4 cm, allowing the undulator gap to remain approximately constant throughout the interaction.  Undulator $\overline{a}_w(z)$-tapering was determined by matching the resonant gradient, determined by the undulator parameters, to the ponderomotive gradient, Figure~\ref{fig5_pietro}, asserting a resonant phase of $\pi$/4.

\begin{figure}[!tbh]
\includegraphics[width=6.5cm]{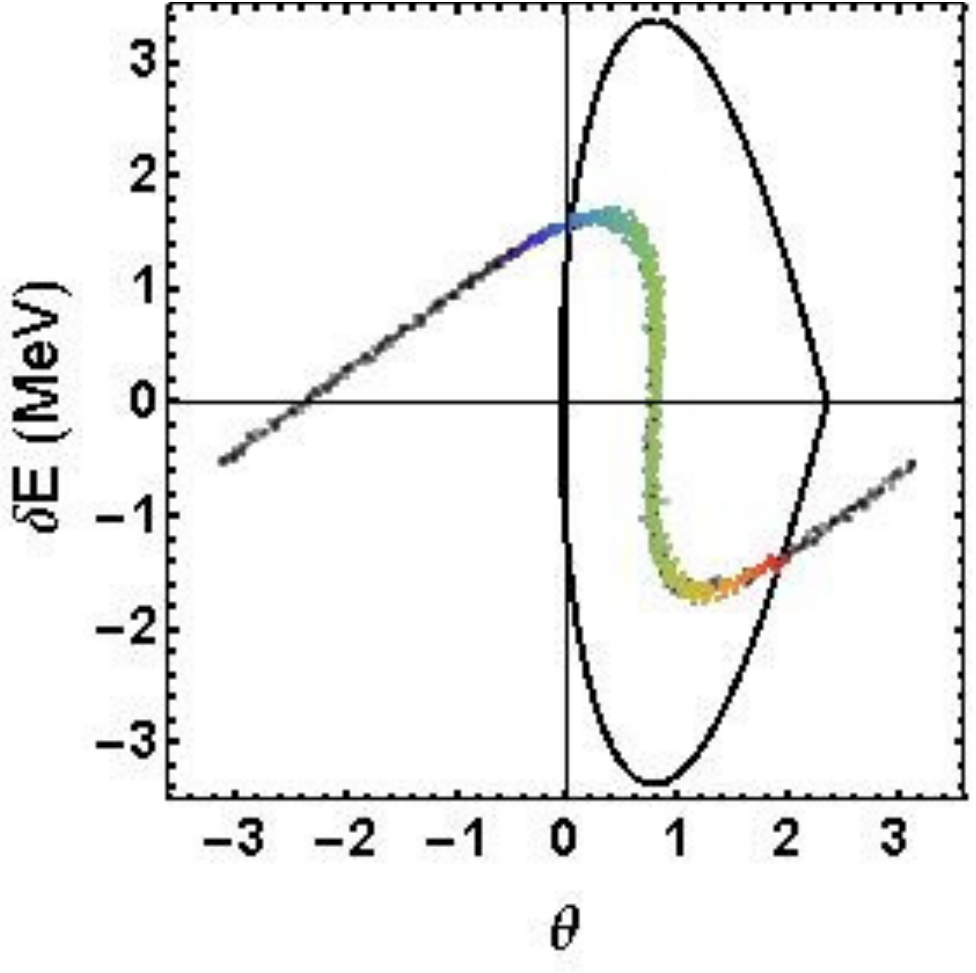}
\includegraphics[width=9.5cm]{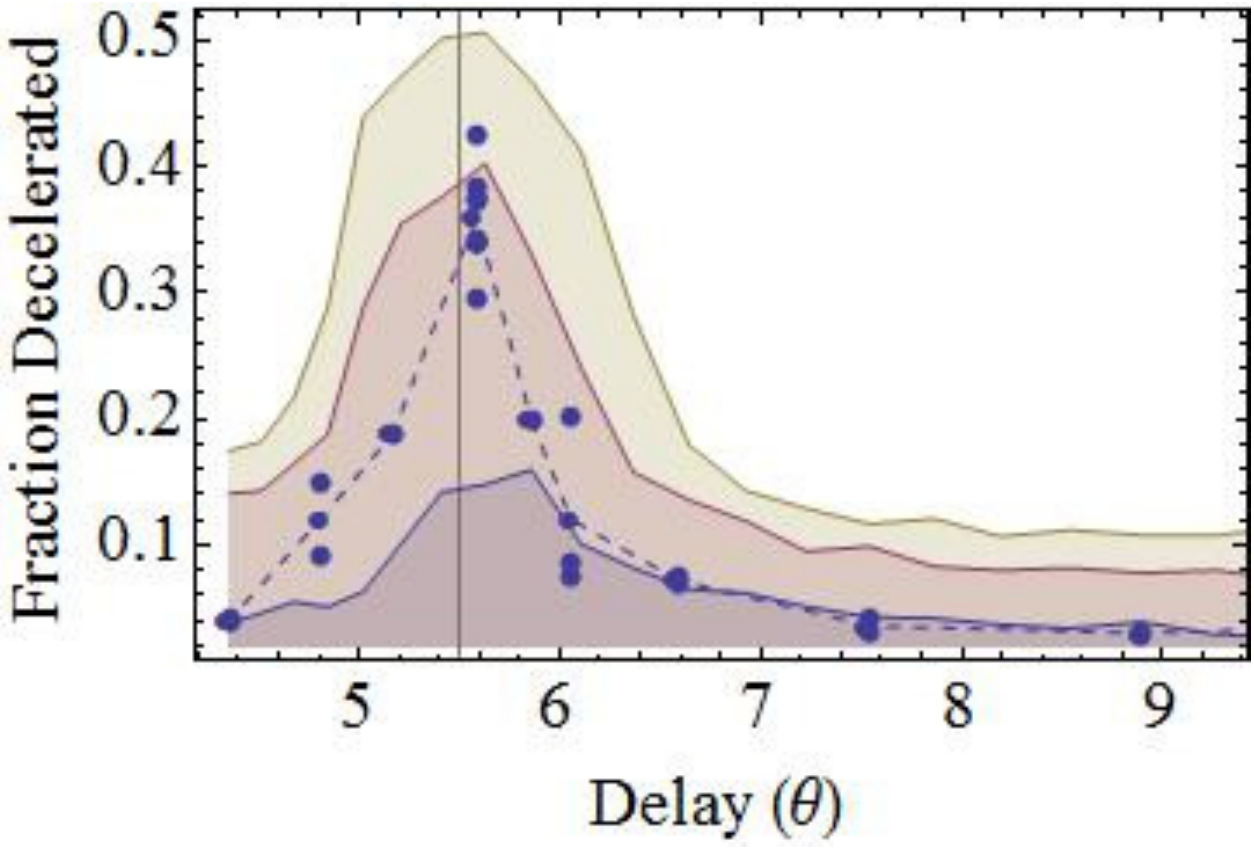}
\caption{(Left) Pre-bunched e-beam longitudinal phase space in Nocibur initial ponderomotive potential. (Right) Fraction trapped data (points) vs. injection phase controlled by varying chicane gap compared with GPT simulations with seed energy 0.55 J (yellow) 0.45 J (Red) and 0.35 J (blue)}
\label{fig6_pietro}
\end{figure}

To increase the energy extraction efficiency further, the electron beam was pre-bunched to increase the fraction of particles trapped in the ponderomotive potential.  The pre-buncher consists of a single 5cm period planar halbach undulator followed by a permanent magnet chicane with a variable gap.  As the electron beam passes through the short undulator section, interaction with the same laser seed used to drive the Nocibur interaction produces a sinusoidal energy modulation on the electron beam, periodic at the laser wavelength  This modulation is now locked in phase with the Nocibur seed laser.  The subsequent chicane provides dispersion, converting the energy modulation to density modulation.  This produces a series of periodically spaced micro-bunches, while also introducing a phase delay between the micro-bunches and laser.  The variable chicane gap allows for tunability of the dispersion and phase delay, allowing injection of the micro-bunches in the ponderomotive potential at the resonant phase, Figure~\ref{fig6_pietro}.  Pre-bunching increased the fraction of particles trapped from 17\% without pre-bunching to 45\% increasing the extraction efficiency by a factor of 3.

\begin{figure}[!tbh]
\includegraphics[width=17cm]{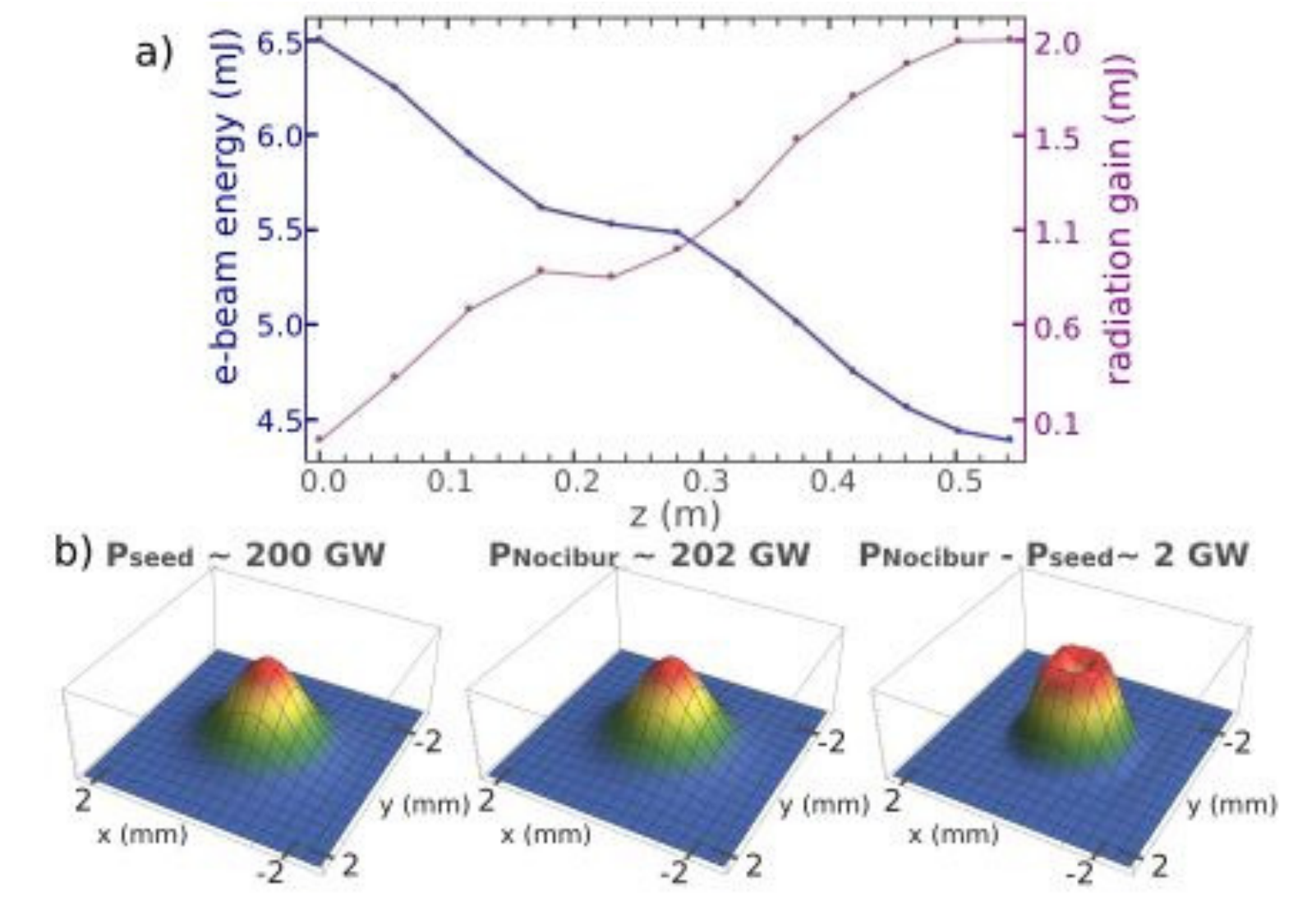}
\caption{(a) Total e-beam energy and radiation gain along Nocibur interaction from Genesis simulation. (b) Output radiation transverse distribution from Genesis.}
\label{fig7_pietro}
\end{figure}

Direct measurement and characterization of the produced radiation was hindered by the presence of the 200~GW seed. However, experimental spectra are in excellent agreement with 3-D time dependent Genesis simulations which predict a 2~mJ increase in radiation energy.  This is consistent with the previously defined total energy lost by the electron beam, validating the assumption that energy lost by the electron beam is converted directly to coherent radiation, Figure~\ref{fig7_pietro}.

One remarks that the transverse distribution of the newly generated radiation is particularly interesting. The emission source in fact is the tightly focused electron beam current, which has a spot size much smaller than the seed radiation. Consequently the radiation has a much stronger divergence angle. We highlight this effect, by showing in the right panel of Figure~\ref{fig7_pietro}b the intensity difference between the seed mode and the amplified mode which exhibits a hole on axis due to diffraction.

Furthermore, Genesis simulations show an increase in the divergence angle of the produced radiation since it is emitted by the electron beam at a waist much smaller than the laser seed.  This is important to account for when considering utilizing the low gain TESSA interaction in a re-circulation scheme.

\subsubsection{Outlook: the Tapering Enhanced Stimulated Superradiant Oscillator (TESSO)}

\begin{figure}[!tbh]
\includegraphics[width=17cm]{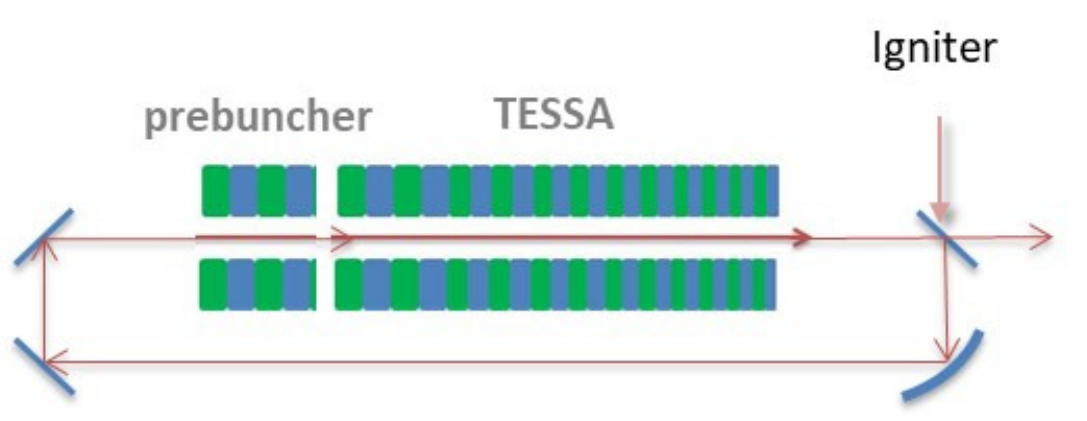}
\caption{TESSO scheme: The scheme requires the use of a high repetition rate electron bunch with temporal separation tuned to the cavity roundtrip length. A fraction of the output radiation is redirected at the input to prebunch the beam and start the TESSA amplification for the next electron bunch.}
\label{fig8_pietro}
\end{figure}

The features of the TESSA concept have naturally the potential for very high power extraction efficiency. However, since the concept requires a high intensity coherent radiation signal input in order to form the traps for an injected pre-bunched beam, its optical amplification gain is low or moderate. It is reasonable in this case to consider a radiation recirculation approach, namely an oscillator or a regenerative amplifier, as shown schematically in Figure~\ref{fig8_pietro}. Such a TESSO device may be an exceptionally energy efficient and high average power radiation source \cite{duris}.

The main challenge for realizing this concept is the requirement for a high intensity radiation seed ``igniter'' and a high repetition bunched beam train synchronous with the round trip time of the resonator. Concepts of stimulated superradiant FEL oscillator with uniform wiggler have been considered in \cite{Alexeev,Seo} and in \cite{Gover_2005,Krongauz}, where a strategy of buildup in the resonator is considered starting from zero-order superradiance and stimulated superradiance. In this scheme the beam energy is temporally ramped in a process in which the bunched beam eventually gets trapped in the built-up radiation field, and then continues in a steady state saturated superradiant oscillation mode. However such a strategy would not be efficient for a tapered wiggler oscillator concept, because in this case there is no gain in the small signal regime before saturation. On the other hand, concepts of tapered wiggler oscillators have been considered in the past with continuous beams \cite{saldin_1993,dattoli_2012} and can be helpful to the case of bunched beam TESSO.

A high power TESSO radiation source would be highly desirable especially in the UV regime. Such scheme has been considered theoretically for wavelength 266~nm \cite{duris}. 
This wavelength comfortably lies in the easily accessible region using cm period undulators and a moderate energy (375 MeV) beam energy \cite{park}. More recently it has been studied in the high gain TESSA regime using numerical simulations showing the various trade-offs to optimize the efficiency as a function of the beam and laser parameters. In particular the challenges associated with start-up from low-power have been bypassed by assuming that a short pulse high power laser source (low repetition rate) would be available to ignite the oscillator. Conversion efficiency approaching 40\% have been shown feasible in 3D simulations.

\subsection{Efficiency Enhancement in the tapered wiggler section of a seed-injected FEL}

In real short wavelengths UV-XUV FELs (Figure~\ref{const_vs_tap}) the fundamental efficiency enhancement processes are not straightforwardly applicable. In the first place, as displayed in Figure~\ref{const_vs_tap}, diffraction effects are significant and the use of a single radiation mode model is not valid in a long interaction length. Furthermore, in present-day X-Ray FEL facilities there is little control over the input field intensity and the bunching phase of the beam at entrance to the tapered wiggler section, independently of the prior section of uniform wiggler amplification. Nevertheless, emerging new techniques may enable in the future better control of the bunched beam and implementation of SR and ST-SR processes. Indeed, phase jump efficiency enhancement methods have been proposed based on small chicanes placed in the space between wiggler sections in short wavelength FEL \cite{Ratner_2007,Curbis_2017,Tsai_2019}. In such a scheme the bunched beam is made to go a longer path relative to the radiation wave, and after each section, the slowed down bunches can be reinserted into an optimized (down-stream side) of the next (back-stream) ponderomotive potential trap. It is thus conceivable that similar methods can be developed in the future to optimize the entrance phase of the bunches in the tapered wiggler to enhance radiation power emission rate. Furthermore, new developments of fresh-bunch techniques \cite{ben-zvi,emma_lutman,emma_feng} make it possible to inject into the tapered wiggler traps bunches with energy spread smaller than the depth of the trap. Recent works also suggest that with such fresh bunch, maintaining high trapping efficiency along the interaction length may be possible with proper strategies of wiggler tapering \cite{Tsai_2019}. However, at the present state of the art of short wavelength FELs, the fundamental interaction processes of SR and ST-SR are complicated by many supplemental effects that do not enable use of simple analytic models.

Numerical simulations have played a key role in understanding the various physical mechanisms at work in
high efficiency tapered wiggler FELs. The importance of
numerical studies is due to the inherently nonlinear evolution of the electron beam and radiation emission in the
post-saturation region of tapered FELs which makes analytic results difficult to obtain without resorting to approximation. As discussed qualitatively in the previous
section, early numerical studies revealed the importance
of two fundamental effects which limit the efficiency of tapered FELs: diffraction due to reduced radiation guiding
\cite{Prosnitz,Sessler} (see Figure~\ref{const_vs_tap})  and the sideband instability \cite{kroll,kroll_1979}. These effects
have been analyzed both separately and in combination
via 1-D codes and 3-D codes, with multi-frequency effects enabled or intentionally disabled (see for example
Ref. \cite{Quimby_Slater_Wilcoxon,Hafizi,Isermann,fawley_1995,reiche_1999,Jiao_2012,schneidmiller,emma,fawley,emma_prst_2017,Duris_2015}).

\subsubsection{Transverse effects}

The simple single mode model is not valid in the long wiggler FEL where diffraction effects dominate.
One must use then multi-mode analysis \cite{Chen_2014,Emma_2014,Tsai_2018} or numerical solutions of
Maxwell equations.
Growth of the radiation spot-size during the post-saturation region decreases the effective bucket area in
which electrons are trapped and continue to lose energy
to the radiation field. This effect becomes dominant for
tapered wigglers of multiple Rayleigh lengths and sets a
limit on the maximum achievable efficiency. This limit
was first estimated analytically in Ref \cite{fawley_1996} and has subsequently been verified in numerical studies (see e.g. Ref.
\cite{Jiao_2012}). To deter the growth of the radiation it is necessary
to maintain the refractive guiding as strong as possible
in the tapered section of the undulator. The strength of
the guiding (given by the electron beam refractive index
\cite{Sessler}) is proportional to the beam bunching which underscores the importance of maintaining a large fraction of
the beam trapped and bunched in the tapered wiggler
for maximum output efficiency. Recent numerical studies have suggested improving the effect of the guiding by
varying the electron beam spot-size in the tapered section \cite{Jiao_2012} or shaping the electron transverse distribution
from Gaussian to parabolic or uniform \cite{Emma_what_year}, yielding a relative improvement in the efficiency around 10-40\%. In
particular, Ref. \cite{Jiao_2012} developed an iterative optimization
algorithm which optimizes the polynomial taper profile
as well as the strength of the quadrupole focusing based
on the evolution of the radiation spot-size in different
sections of the tapered undulator.
Another recent study \cite{schneidmiller} has examined the effect of
diffraction on the optimization of a tapered FEL by
parametrizing the z-dependent emission of radiation as
a function of the Fresnel Number $N = k\sigma^2/z$ where
$k = 2\pi/\lambda$ is the radiation wavenumber. The two limits of a thin electron beam $N\gg 1$ and a wide electron beam
$N\ll 1$
correspond to a quadratic and a linear growth in the radiation power, and occur in the early and late stages of
a tapered wiggler respectively. As such, the tapering law
determined is a hybrid of a quadratic taper at the start of
the post-saturation section followed by a linear taper towards the end of the wiggler, with the exact form depending on the electron beam and undulator parameters. As
we remark in the following paragraph, the useful guidelines provided in these studies must be applied while also
taking into account the impact of multi-frequency effects,
as they can prove crucial when coupled with transverse
effects and can change the form of the optimal tapering
profile for maximum output efficiency \cite{emma}.

\subsubsection{Multi-frequency effects}

Multi-frequency effects in the post-saturation regime
can cause the amplification of undesired frequencies
which can perturb the dynamics of the electron motion,
disrupting the radiation gain and reducing the output
efficiency. One of the most deleterious of these effects
to the tapered FEL performance is the synchrotron sideband instability \cite{kroll,kroll_1979}. Sidebands are generated due to amplitude and phase modulations of the electric field which result from the trapped particles undergoing multiple synchrotron oscillations as they pass through the tapered
section of the wiggler. From the 1-D FEL field equation
it is clear that as the electrons oscillate in the longitudinal phase space the gain and the phase shift of the radiation field will be different at different locations in the
undulator and, due to shot noise and/or existing current modulations imprinted on the electron beam, at
different locations along the bunch. This results in a temporal modulation of the radiation amplitude and phase
giving rise to sidebands displaced from the central wavelength. Suppression of shot-noise in the early
stages of electron beam acceleration would be a way to deter the start of sideband instability
out of noise. Theoretically the scaling of noise suppression schemes could reach X-Ray frequencies
\cite{Short_wavelength_limits} but these have been demonstrated so far only at optical frequencies
\cite{Beating_shot-noise,Observation_shot-noise,Subradiant_spontaneous}.

The resonance between sideband radiation frequencies and the electron synchrotron motion gives rise
to the synchrotron sidebands displaced from the central
wavelength by an amount $\Delta\lambda/\lambda_0 = \lambda_w/L_s$ where $L_s=2\pi/K_s$ (Eq.~\ref{K_s_2}) is
the synchrotron period.
Since the resonance between the electron synchrotron
motion and the ponderomotive wave is what causes the
net energy transfer to the sidebands, we expect the sideband gain to be small in the regions in which the electric
field, and thus the synchrotron frequency, are changing
rapidly. For this reason in the original literature \cite{kroll} it was
thought that high gain FEL amplifiers (as opposed to the
low gain FEL oscillators) would avoid the sideband problem due to the rapidly increasing radiation field in the
tapered region causing a rapidly changing synchrotron
frequency. Suppressing sideband growth in the FEL oscillator was therefore first studied in the 1-D limit with
a time-dependent FEL code, and it was shown that the
instability could be suppressed by adding suitable frequency filters into the FEL optical cavity \cite{Quimby_Slater_Wilcoxon}.
While the 1-D FEL theory predicts weak sideband
growth for FEL amplifiers, as we have discussed above,
diffraction effects in tapered wiggler FELs cause the electric field growth to slow down and eventually saturate
due to reduced optical guiding. As a result, the onset of
sideband-induced detrapping is coupled to the limits on
the electric field growth set by the reduction in guiding.
As the electric field approaches its asymptotic value we
expect the effect of the sidebands to be more pronounced
and more significant detrapping to occur as a result. Eventually this process can lead to a second saturation of the
tapered FEL power, as shown for example in Ref. \cite{Jiao_2012}
\cite{emma}.

Suppressing the sideband instability therefore remains one of the the key issues for tapered FEL amplifiers, particularly those which are multiple synchrotron
periods in length.
To that end, a number of sideband suppression schemes
have been recently proposed for high efficiency FEL amplifiers. For example, it has been shown in simulations
that a large seed power in an FEL amplifier can offer a significant ``head-start'' for the fundamental compared to the sidebands which starts from noise. This
allows the fundamental to reach high peak power before the sidebands grow to appreciable amplitude \cite{emma}.
In order to achieve a large enough seed for a tapered
X-ray FEL while preserving the beam quality necessary for efficient amplification in the tapered wiggler, a
technique termed ``fresh-bunch self-seeding'' has recently
been demonstrated at the LCLS, recording an increase
in X-ray brightness of a factor of 2 compared to the state
of the art \cite{emma_lutman}. An alternative method which makes
use of periodic delays between wiggler sections which introduce a $\pi/2$ phase shift for the sideband oscillations
while preserving the phase of the resonant frequency has
been proposed in \cite{Duris_2015}. Therein it was shown that a modest number of such delays could be used to recover the
maximum efficiency achievable with multi-frequency effects disabled. A similar technique has been suggested
which introduces a modulation in the undulator magnetic
field, effectively achieving a $\pi/2$ delay at the sideband
frequency while maintaining resonance with the fundamental \cite{emma_lutman}. This has been shown in simulation to reduce the sideband amplitude by more than an order of
magnitude, improving the brightness of tapered wiggler
FELs. Finally, a pre-bunched electron beam allows the
FEL undulator tapering to be more rapid, thus leading
to a faster-changing synchrotron frequency and consequently a reduction in the sideband growth.
In conclusion it should be noted that, in addition to
the previous analysis, the optimization of FEL efficiency
via undulator tapering remains an active field of research
with recent simulations and experiments revealing interesting results \cite{Wu_et_al}. More
experimental results can be found at the SPARC UV FEL \cite{Giannessi_2013}, SINAP \cite{Liu_2013}, etc.

\section{Conclusions}

Electron beams of small emittance and energy spread are in principle low entropy source of radiative energy, and therefore, as radiation sources, have the potential for ultimate energy extraction efficiency and power. When the electron beam is pre-bunched at optical frequency, it has good match with coherent radiation wave at the corresponding six-dimensional volume in Liouville's phase-space, and therefore phase-space transformation of beam kinetic energy to optical power can take place with high energy extraction efficiency.

Based on this general principle, we identify radiation emission schemes of enhanced coherent spontaneous radiation: superradiance (SP-SR) and stimulated emission (ST-SR). In both schemes the enhancement is due to the constructive coherent  interaction of the radiation wave with all electrons in a single or multiple bunched beam. For short interaction length, the SP-SR emission is characterized by power generation scaling $\propto N^2L^2$, and the ST-SR power increment scales as $\propto NLE_0$ ($N$ being the number of particles, $L$ the interaction length and $E_0$ the field strength at the entry to the interaction region). At longer interaction lengths, these scaling laws change due to a nonlinear process in which the beam energy drops and the electron dynamics during the interaction plays a role. 

We used a simple ideally tightly bunched electron beam model to describe the nonlinear interaction process for the case of coherent radiative interaction in a magnetic wiggler (FEL). The model is applied to both uniform wiggler and tapered wiggler cases. In the case of tapering the ponderomotive beat-wave, created by an intense input radiation wave and the tapered wiggler, enables continued phase-matched stimulated interaction with the bunched beam (TESSA) and enhanced energy extraction from the slowing down beam.

The simplified nonlinear model presented is energy conserving and consistent with the zero-order analysis and scaling of SP-SR and ST-SR. At short distances ($z$) a bunched beam always exhibits SP-SR emission scaling as $\propto N^2z^2$, and ST-SR emission scaling as $\propto NzE_0$ if it is injected at a deceleration phase relative to the wave ($\psi=\pi/2$). It is then possible to get initially enhanced radiation extraction when the input radiation field and its relative bunching phase are controllable.

The nonlinear dynamics of the electrons in the potential traps of the ponderomotive potential is analogous to that of a mathematical pendulum, and in the case of a tapered wiggler - a titled pendulum. In the case of a tapered wiggler (TESSA) the analysis reveals that the electron beam energy drop (turned into radiation gain) is composed of two contributions: a) energy drop due to the reduced average kinetic energy of all electrons that stay trapped in the slowing down traps along the tapered wiggler; and b) internal synchrotron oscillation dynamics of the tight bunches within the trap. If the bunching is tight there is an advantage to inject the beam at maximum ST-SR deceleration phase $\psi(0)=\pi/2$, but in practice the bunching efficiency is not perfect, and for maximum trapping efficiency one would prefer to inject the  spread-phase bunch at the resonant particle center phase $\psi(0)=\psi_r$. In this case, the contribution of the internal trap dynamics and ST-SR emission are negligible.

Beyond the ideal tight bunching model that is useful for identifying the fundamental emission processes, we also presented for the TESSA case an approximate analysis and simulation results for the interaction with a beam of finite energy and phase spread. In this case, relevant to the present state of the art of TESSA and FEL technology, the contribution of the internal dynamics of synchrotron oscillation is washed out, and the dominant contribution is the resonant energy drop due to tapering $\Delta\gamma=\gamma_r(0)-\gamma_r(L_w)$. Also in this case there is an advantage in having high initial field $E_0$, tight bunching and good control over the injection phase $\psi(0)$ in order to achieve high trapping efficiency and enable aggressive tapering rate with deep enough traps. Future development of beam and laser technology (such as fresh bunch technique \cite{emma_lutman}), may lead to better control of these parameters and development of very high efficiency TESSA and TESSO radiation sources. 

In the last part of this article we reviewed the applications and experimental demonstrations of radiation sources based on bunched e-beam: superradiant, stimulated superradiant and TESSA. Most of the superradiant radiation sources operate in the THz regime, because practically, modern RF accelerators are based on sub-pSec short electron bunches photo-emitted using femtosecond ultrafast lasers. In the case of nonlinear interaction of a bunched beam in a tapered wiggler we reviewed the experimental demonstration of very significant radiative energy transfer efficiency in TESSO experiments in \cite{nocibur,rubicon}. We finally pointed out the relevance of SP-SR and ST-SR processes in considering optimized tapering strategy in tapered wiggler FEL. It was indicated, however, that in this case several processes of energy and phase spread radiation diffraction and multi frequencies effects mask the fundamental radiation processes in a long wiggler, and they have been analyzed primarily with numerical computations.

\newpage
\appendix

\renewcommand\thefigure{\thesection.\arabic{figure}}
\setcounter{figure}{0}
\section{Pendulum equation in FEL context}
\label{pend_append}

The classical pendulum equation is given by:
\begin{equation}
\frac{d\theta}{dz}=K_s^2\sin\psi
\label{pend1}
\end{equation}
\begin{equation}
\frac{d\psi}{dz}=-\theta
\label{pend2}
\end{equation}
(in the case of a physical pendulum $K_s^2=g/l$, where $g$ is the gravitation constant and $l$ - the length of the
pendulum, $\psi$ is the tilt angle and $\theta$ is the angular velocity, and the independent variable would be $t$
and not $z$).

In the context of a periodically bunched electron beam, $\theta$ represents the detuning parameter 
between the velocity of the bunches and the phase velocity of the ponderomotive wave radiation field
(see Eq.~\ref{detuning_par1}), while $\psi$ represents the phase of the bunch relative to the ponderomotive wave
\begin{equation}
\psi\equiv -[\varphi_b(z)-\varphi_q(z)-\pi/2]
\label{psi_1}
\end{equation}

The parameter $K_s$, called the synchrotron oscillation wavenumber, represents the small oscillation frequency of the
bunches relative to the center of the trap. $K_s$ is essentially the amplitude of the ponderomotive wave, and is
proportional to the wiggler strength and the radiation field amplitude.

\begin{figure*}[!tbh]
\includegraphics[width=18cm]{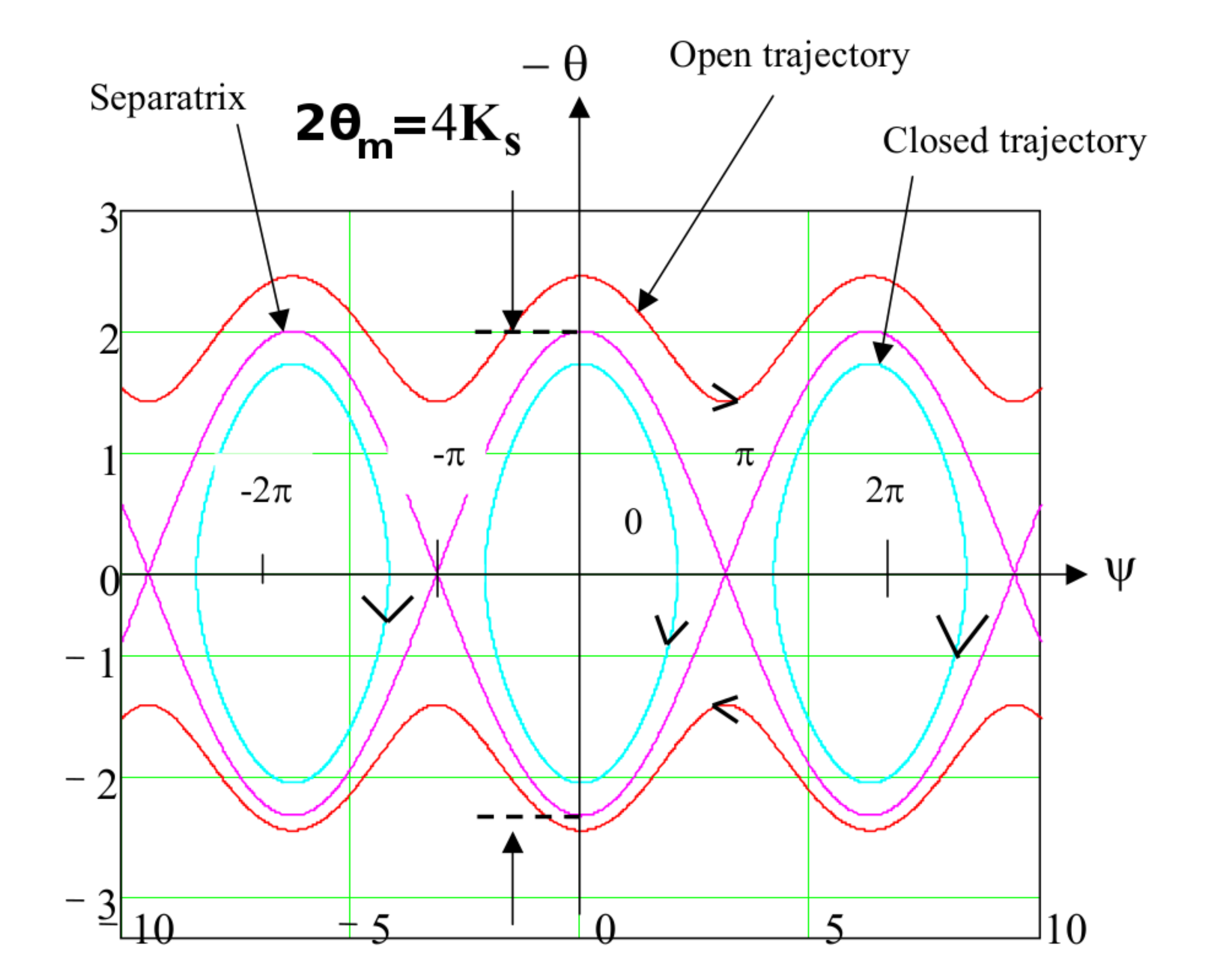}
\caption{The $\theta-\psi$ phase-space trajectories of the pendulum equation.}
\label{simple_pendulum}
\end{figure*}

Multiplying LHS of \ref{pend1} with RHS of \ref{pend2} and vice-versa, and integrating, results in
\begin{equation}
\frac{1}{2}\theta^2-K_s^2\cos\psi\equiv T(\theta)+U(\psi)=C.
\label{pend4}
\end{equation}
Here we identify $T(\theta)=\frac{1}{2}\theta^2$ with the kinetic energy and $U(\psi)=-K_s^2\cos\psi$ with
the potential energy. $C$ is an integration constant that is determined by the initial value of the electron trajectory
$C=T(\theta(0))+U(\psi(0))$. Inspecting Figure~\ref{simple_pendulum} we observe two kinds of
trajectories: for $C(\theta(0),\psi(0))>K_s^2$ all trajectories are open, namely electrons injected
at phase and detuning initial conditions corresponding to this case, propagate in open phase-space trajectories,
bypassing the periodic
traps without getting trapped. In the opposite case $C(\theta(0),\psi(0))<K_s^2$ electrons follow
closed trajectories, and if injected into the trap, they stay trapped, performing ``synchrotron
oscillations'' around the center of the trap, and moving on the average at the phase velocity
of the ponderomotive force (\ref{phase_velocity}). The equality $C(\theta(0),\psi(0))=K_s^2$
represents the separatrix - the borderline between open and closed trajectories.
The separatrix represents a ``trap'' or ``bucket'' in which electrons are trapped executing synchrotron trajectories without escape. The height of the trap is
\begin{equation}
2\theta_m=4K_s
\label{trap_height}
\end{equation}
and its width is $2\pi$, i.e. ($-\pi<\psi<\pi$).

To describe the electron dynamics in a tapered wiggler we note that in this case the phase velocity of the
ponderomotive wave (\ref{phase_velocity}) decelerates when the wiggler period grows up as a function of
$z$. This can be envisioned as a physical situation where the electron in a decelerating frame experiences
an imaginary acceleration force in addition to the restoring force of the pendulum. This can be shown to be
equivalent to adding a term to the RHS of Eq.~(\ref{pend1}). To make sure that the acceleration (tapering
force) is smaller than the restoring force, and the electrons can still be trapped, we define this term as
$-K_s^2\sin\psi_r$:
\begin{equation}
\frac{d\theta}{dz}=K_s^2[\sin\psi-\sin\psi_r].
\label{pend1_shift}
\end{equation}
Since $\sin\psi_r < 1$, there is always a range of phases $\psi$ around $\psi=\psi_r$ in which oscillatory
dynamics is around the resonant phase $\psi_r$ is possible.
\begin{figure*}[!tbh]
\includegraphics[width=18cm]{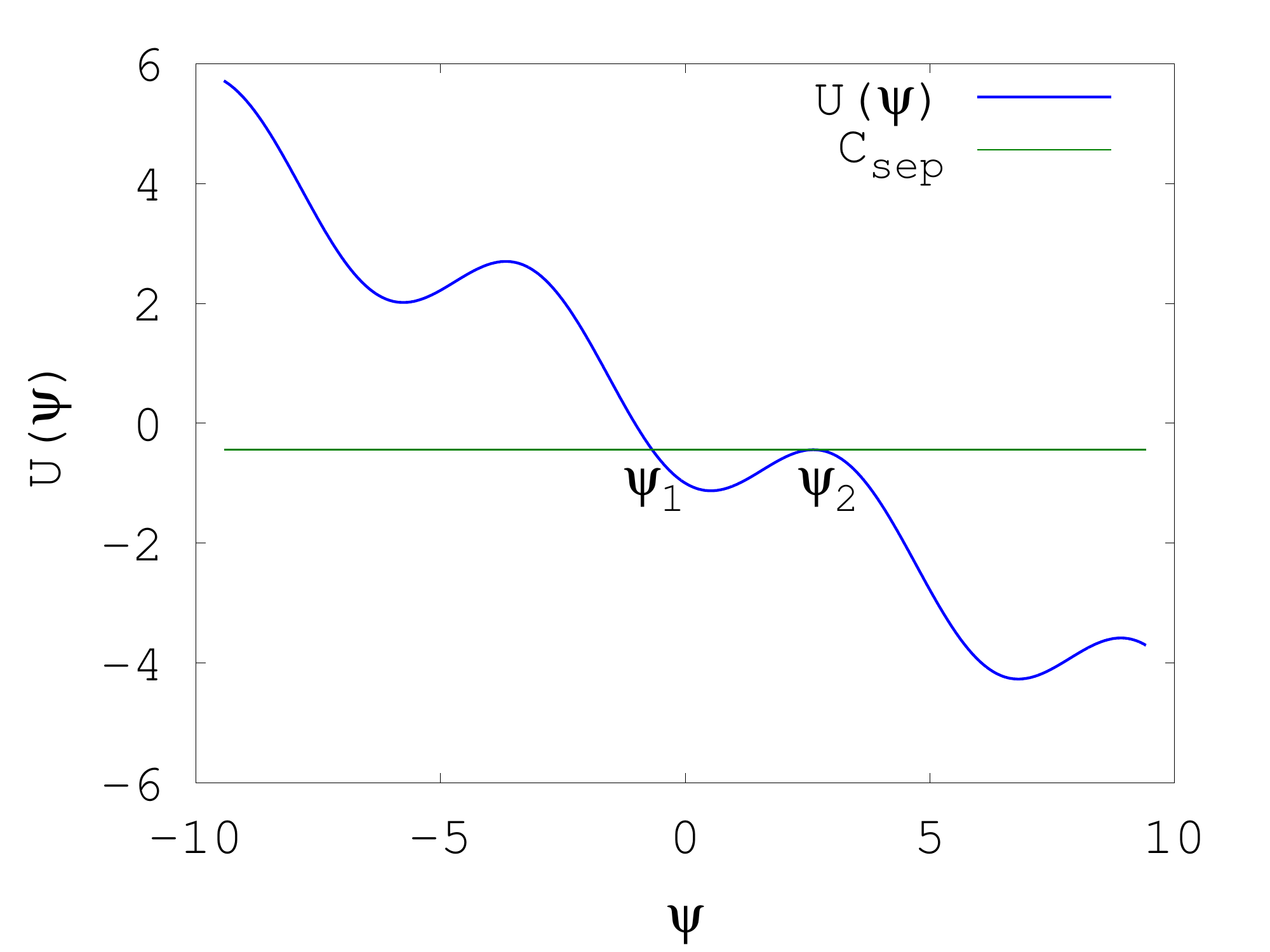}
\caption{The potential potential energy for a tapered wiggler with $K_s=1$ and $\sin\psi_r=0.5$.
The figure shows $C_{sep}$ and the values of $\psi_1$ and $\psi_2$ for $n=0$.}
\label{shifted_potential}
\end{figure*}

For the case of $\sin\psi_r$=const (linear wiggler tapering) the integration of (\ref{pend1_shift}) with
(\ref{pend2}) is straightforward \cite{kroll}:
\begin{equation}
\frac{1}{2}\theta^2- K_s^2[\cos\psi+\psi\sin\psi_r] \equiv T(\theta)+U(\psi) = C,
\label{pend3_shift}
\end{equation}
where the kinetic energy $T(\theta)$ is the same as before, but the potential energy $U(\psi)$ has an
additional term:
\begin{equation}
U(\psi)= -K_s^2[\cos\psi+\psi\sin\psi_r]
\label{pend5_shift}
\end{equation}
shown in Figure~\ref{shifted_potential} with $K_s=1$ and $\sin\psi_r=0.5$. $C=T(\theta(0))+U(\psi(0))$ is an integration constant
established by the initial values of $\theta$ and $\psi$.

To analyze the potential energy $U(\psi)$ in Eq.~\ref{pend5_shift}, we find the maxima and minima by
solving $dU/d\psi=0$. This results in
\begin{equation}
\psi^{n}_{max}=\pi-\psi_r+2\pi n\equiv\psi^{n}_2 ,
\label{psi_max}
\end{equation}
which is the right limit of the trap $n$, named $\psi^{n}_2$, and
\begin{equation}
\psi^{n}_{min}=\psi_r+2\pi n ,
\label{psi_min}
\end{equation}

\begin{figure*}[!tbh]
\includegraphics[width=15cm]{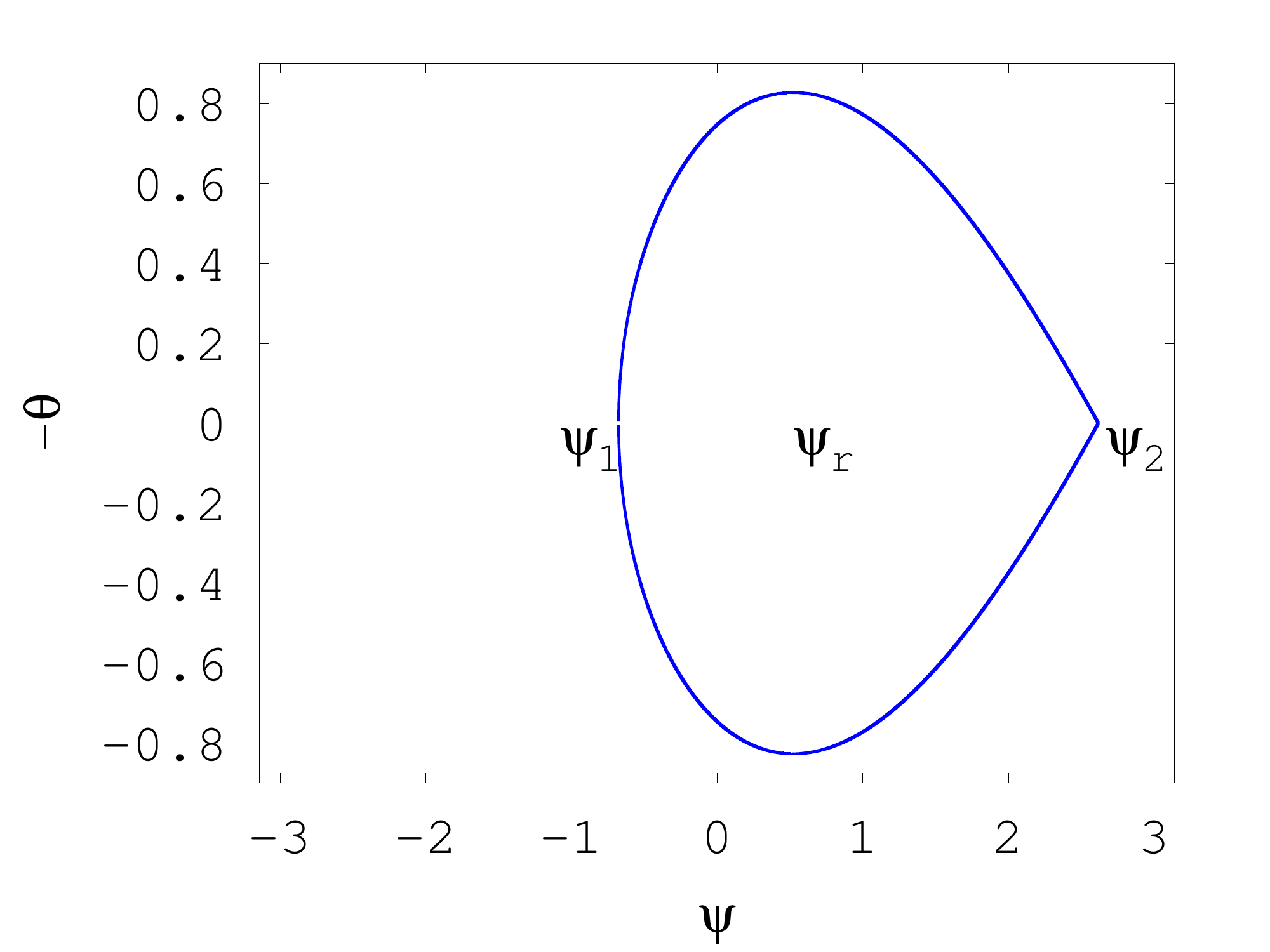}
\caption{The separatrix for the $n=0$ trap in a tapered wiggler for $K_s=1/\sqrt{2}$ and $\psi_r=\pi/6$. The trap
ranges in the region $\psi_1\le\psi\le\psi_2$, where $\psi_2=\pi-\psi_r$ and $\psi_1$ has to be calculated numerically
and is in this case -0.6752 . The height of the trap is 1.6551, according to Eq.~(\ref{theta_m}).}
\label{tap_separatrix}
\end{figure*}

We define the separatrix for the trap $n$ by setting $C^{n}_{sep}=U(\psi^{n}_2)=K_s^2[\cos\psi_r-(\pi-\psi_r+2\pi n)\sin\psi_r]$,
so that non-zero kinetic energy $T(\theta)\ge 0$ is possible for either $\psi>\psi^{n}_2$ (open trajectories)
or $\psi^{n}_1<\psi<\psi^{n}_2$ (closed trajectories). Here $\psi^{n}_1$ is the other (left boundary phase) of
the equation $U(\psi)=C^{n}_{sep}$ (see Figure~\ref{shifted_potential} and the phase space presentation of the
separatrix for $n=0$ in Figure~\ref{tap_separatrix}).
Using $C=C^{n}_{sep}$ in Eq.~(\ref{pend3_shift}), for $n=0$ we obtain
\begin{equation}
\frac{1}{2}\theta^2 =  K_s^2[\cos\psi+\psi\sin\psi_r + \cos\psi_r-(\pi-\psi_r)\sin\psi_r] ,
\label{pend3_shift_1}
\end{equation}
From which the curve of the separatrix in Figure~\ref{tap_separatrix} is represented by
\begin{equation}
\theta = \sqrt{2} K_s\sqrt{\cos\psi+\psi\sin\psi_r + \cos\psi_r-(\pi-\psi_r)\sin\psi_r} ,
\label{pend3_shift_2}
\end{equation}
The height of the tilted pendulum separatrix is $2\theta_m$, where $\theta_m$ is the value of $\theta$ in (\ref{pend3_shift_2})
for $\psi=\psi_r$. It is found:
\begin{equation}
2\theta_m = 4 K_s\sqrt{\cos\psi_r+(\psi_r-\pi/2)\sin\psi_r},
\label{theta_m}
\end{equation}
so that for $\psi_r=0$ we recover (\ref{trap_height}) and for $\psi_r=\pi/2$ the trap vanishes ($\theta_m=0$).
The width of the trap is $\psi_2-\psi_1$.

\begin{figure*}[!tbh]
\includegraphics[width=18cm]{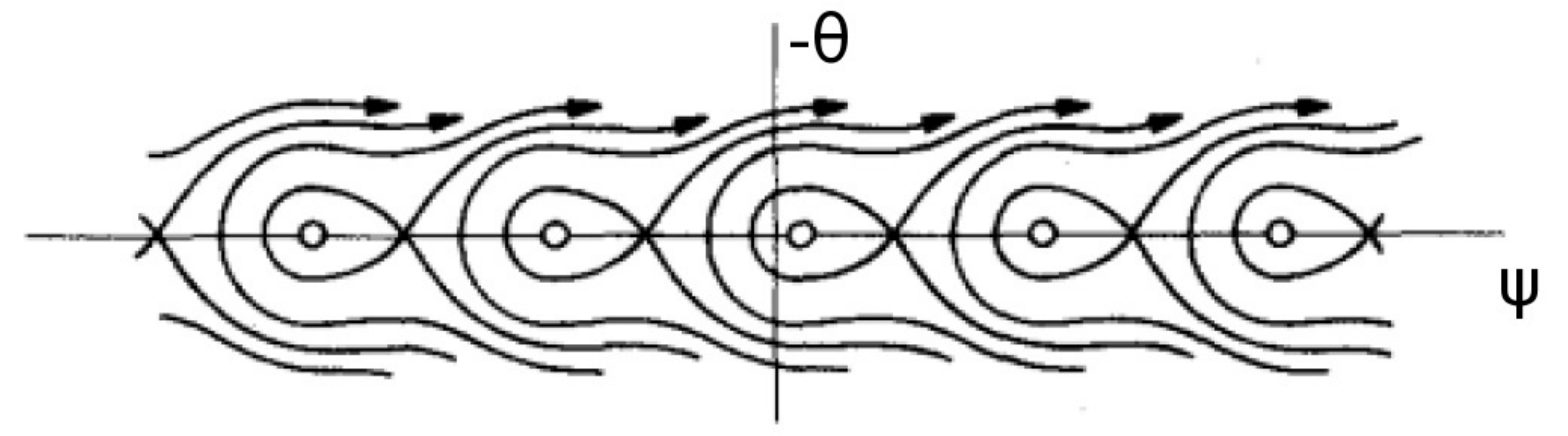}
\caption{Separatrix trajectories for trap $n=m$ are unbound around traps $n>m$ and are not allowed near
traps $n<m$.}
\label{kmr}
\end{figure*}

For other energy conservation constants $C(\theta(0),\psi(0))$ one gets open trajectories if $C>C^{n}_{sep}$ and
closed trajectories if $C<C^{n}_{sep}$ (see Figure~\ref{shifted_potential}). The open and closed trajectories
in phase-space $(\psi,-\theta)$ are shown in Figure~\ref{kmr} \cite{kroll} for the multiple traps.

\newpage
\renewcommand\thefigure{\thesection.\arabic{figure}}
\setcounter{figure}{0}
\section{Electron Beam Bunching}
\label{bunching_append}

While thrust of this review is the coherent superradiant emission processes, we cannot ignore the processes of pre-bunching the electron beam.

There is a variety of processes for attaining tight bunches of electron beams. At long wavelength (THz range) single bunch and periodic bunching can be done by direct photoemission from femtosecond laser-illuminated cathodes. This can also be a train of such femtosecond laser beams that are replicated by various optical splitting and delay schemes. In all these schemes, the Gaussian distribution model for the electron bunch (Eq.~\ref{gauss_dist}) is usually a good approximation:
\begin{equation}
f(t)=\frac{1}{\sqrt{2\pi}\sigma_{t}}e^{-t^2/(2\sigma_{tb}^2)},
\label{gauss_dist_append}
\end{equation}
Thus the bunching coefficients of the single electron bunch and periodic Gaussian bunch-train respectively are Eqs.~(\ref{M_b_1})
and (\ref{b_n}).
\begin{equation}
M_b(\omega)= e^{-\omega^2 \sigma_{tb}^2/2}.
\label{M_b_1_append}
\end{equation}
\begin{equation}
b_n= e^{-\omega_n^2 \sigma_{tb}^2/2}.
\label{M_b_n_append}
\end{equation}
where $\omega_n=n\omega_b$.

At short wavelengths, a most useful scheme of bunching a continuous or long pulse electron beam is to modulate its energy with a high intensity laser beam in a wiggler (or any other interaction scheme), and then turn its energy modulation into density modulation by passing it through a Dispersive Section (DS), such as a ``chicane''. This scheme of bunching is useful for a variety of short wavelength radiation emission schemes, including HGHG \cite{YU_1991,YU_2000} EEHG \cite{Stupakov_2009}, TESSA \cite{nocibur} and e-SASE \cite{Zholents_2005}.

In the laser bunching scheme, the bunch distribution deviated from a single Gaussian shape (\ref{M_b_n_append}) is modified.  Because of the importance of this bunching technique we review briefly the derivation of the bunching distribution and the bunching coefficient of this case following the parametric notations of \cite{Hemsing}.

The beam is assumed to be initially uniform but has initial energy spread . This its initial normalized energy distribution is
\begin{equation}
f(p)=\frac{1}{\sqrt{2\pi}}e^{-p^2/2},
\label{energy_distribution}
\end{equation}
where $p=(\gamma-\gamma_0)/\sigma_{\gamma 0}$.

After energy modulation $\gamma=\gamma_0+\Delta\gamma_{mod}\sin(\omega_b t)$ the energy distribution is periodically time t (or z) dependent: 
\begin{equation}
f(p)=\frac{1}{\sqrt{2\pi}}e^{-(p-A\sin(\omega_bt))^2/2},
\label{energy_distribution_after_mod}
\end{equation}
where $A=\Delta\gamma_{mod}/\sigma_{\gamma 0}$.

In a dispersive section of dispersive strength $R_{56}$ the electron time and longitudinal coordinates pass a compress transformation $z'=ct'=z+R_{56}(\gamma-\gamma_0)/\gamma_0=z+R_{56}p(\sigma_{\gamma 0}/\gamma_0)$ so after the DS
\begin{equation}
f_0(p,t)=\frac{1}{\sqrt{2\pi}}e^{-(p-A\sin(\omega_bt-Bp))^2/2},
\label{energy_distribution_time}
\end{equation}

This current distribution, periodic in time with period $T_b=2\pi/omega_b$ depends on the modulation parameter $A$ and the compression parameter $B$:
\begin{equation}
B=\omega_b\sigma_t=\omega_b (R_{56}/c)(\sigma_{\gamma 0}/\gamma_0)
\label{compression_parameter}
\end{equation}

\begin{figure*}[!tbh]
\includegraphics[width=18cm]{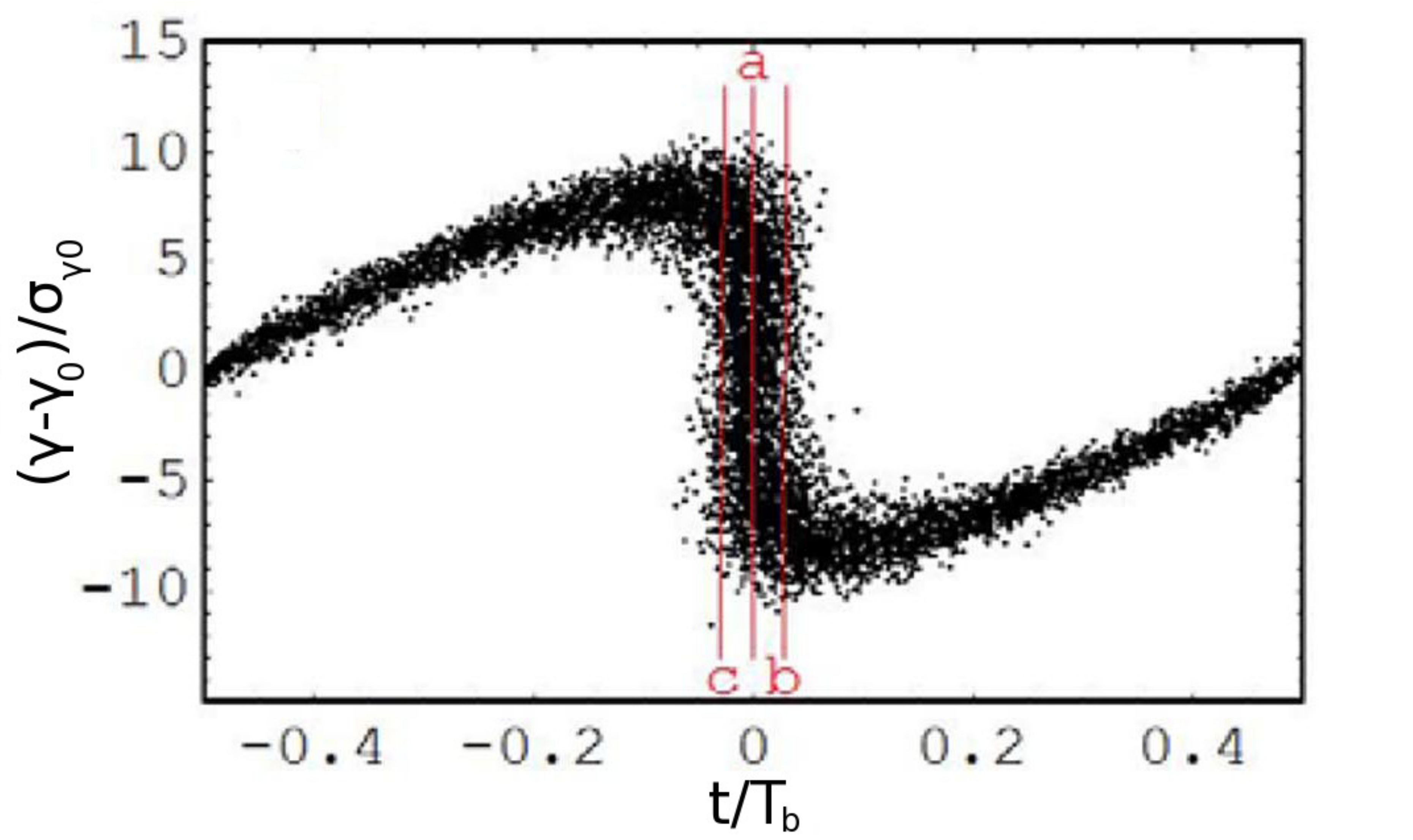}
\caption{Longitudinal phase space after the chicane showing microbunching of electrons and an enhanced electron density. (after Zholents \cite{Zholents_2005})}
\label{zholents_bunching}
\end{figure*}

Fig~\ref{zholents_bunching} displays this distribution in $p=(\gamma-\gamma_0)/\sigma_{\gamma 0}$ and $t/T_b$ over one bunching period at $A$, $B$ parameters choice of maximum bunching.    

Integrating over energy and using Eq.~\ref{energy_distribution_time}, we find the bunching amplitude of harmonic n:
\begin{equation}
b_n=\int_{-T_b/2}^{T_b/2}dt \int_{-\infty}^{\infty}dp e^{-in\omega_bt} f_0(t,p)=J_n(nAB)e^{-n^2B^2/2}=J_n(nAB)e^{-\omega_n^2\sigma_t^2/2} 
\label{bunching_harmonic_n}
\end{equation}

\begin{figure*}[!tbh]
\includegraphics[width=18cm]{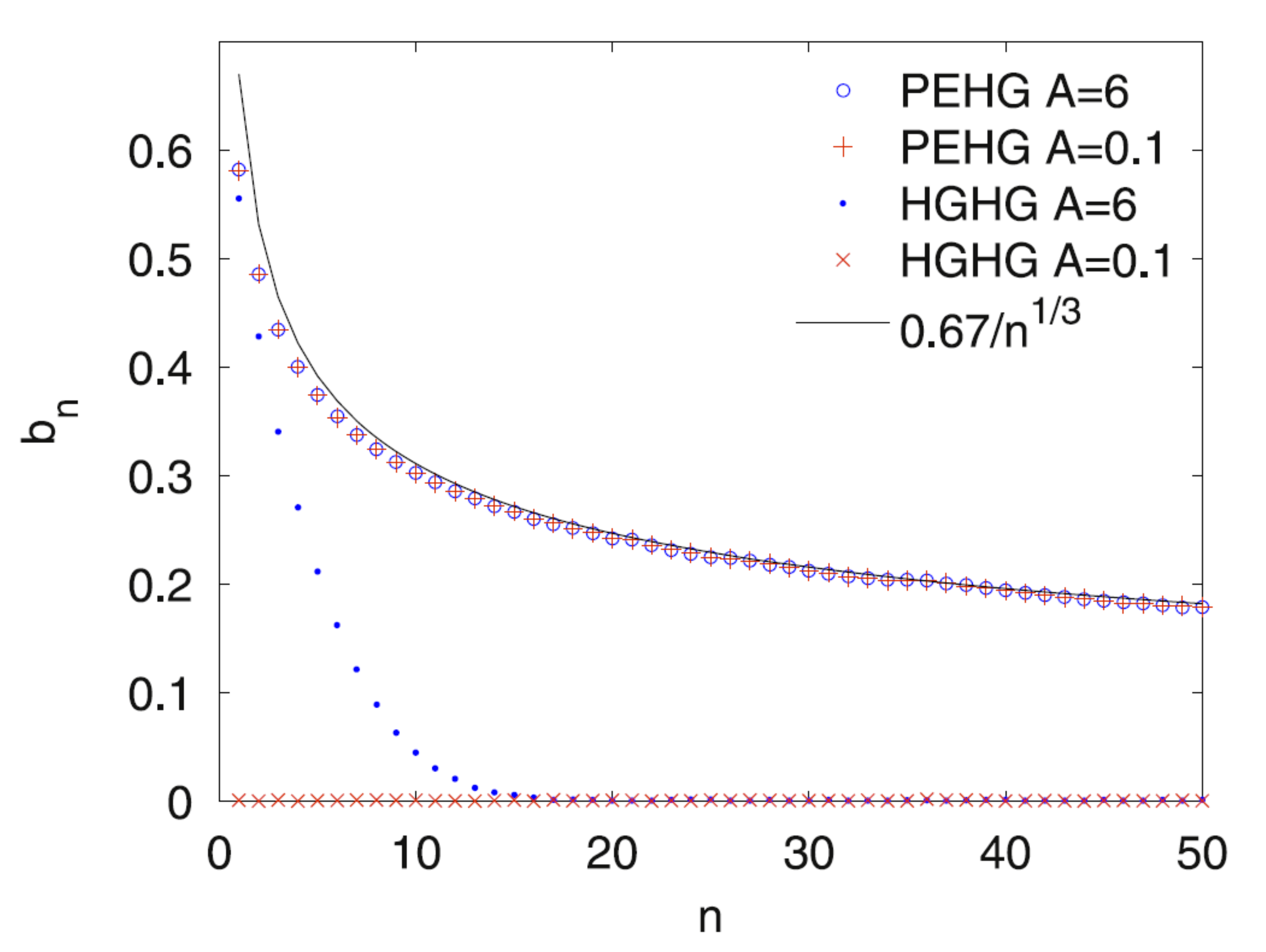}
\caption{Comparison of the bunching factor of PEHG and standard HGHG with different energy modulation amplitudes. The black line is the theoretical prediction of the maximal bunching factor of PEHG (after Feng \cite{Feng_2014}).}
\label{bunching_factor}
\end{figure*}

For harmonics $n>4$ the maximum of the Bessel function in Eq.\ref{bunching_harmonic_n} is about $0.67/n^{1/3}$ and is achieved when the argument is $n+0.8n^{1/3}$ .Thus the optimal strength of the DS for maximal bunching is:
\begin{equation}
B=(n+0.8n^{1/3})/(nA)\simeq 1/A
\label{optimal_bunching}
\end{equation}
So the approximate expression for the bunching parameter is:
\begin{equation}
b_n=J_n(n)e^{-n^2B^2/2}\simeq \frac{0.67}{n^{1/3}}e^{\omega_n^2\sigma_t^2/2}
\label{bunching_harmonic_n_approx}
\end{equation}
where $\sigma_t=B/\omega_b$ is the approximate width of the bunch distribution. The Bessel function factor reduces somewhat the bunching coefficient relative to the Gaussian distribution (\ref{M_b_n_append}). However, note that the width parameter $\sigma_t$ is controllable in this case, and the Gaussian factor, limited by the initial energy spread $\sigma_{\gamma 0}/\gamma_0$, (\ref{bunching_harmonic_n}) can be enhanced by decreasing B and increasing correspondingly $A\simeq 1/B$. Furthermore, it has been proposed that in a scheme of a transverse gradient wiggler the effect of the energy spread and the Gaussian factor may be nearly eliminated and the bunching factor gets close to
\begin{equation}
b_n=\frac{0.67}{n^{1/3}}
\label{bunching_harmonic_n_no_gaussian}
\end{equation}
as shown in Figure~\ref{bunching_factor} 

Besides the derivation of the bunching coefficient for a laser modulated beam, the formulation here is also useful for calculating the trapping fraction $f_t(\psi_r,A)$ of a laser-prebunched electron beam in the traps of a tapered wiggler FEL characterized by a separatrix
\begin{equation}
\delta\gamma^{sep}=\delta\gamma^{sep}(K_s,\psi_n)
\label{separatrix}
\end{equation}
The trapping efficiency of harmonic $n$ can be then calculated numerically using the distribution (\ref{energy_distribution_time}):
\begin{equation}
f_{t\,n}=\int_{\psi_{n1}}^{\psi_{n2}}d\psi_n \int_{-\delta\gamma^{sep}/\sigma_{\gamma 0}}^{\delta\gamma^{sep}/\sigma_{\gamma 0}}dp f_0(p,\psi_n/\omega_n)
\label{f_tn}
\end{equation}
This procedure was used to calculate the trapping efficiency in Chapter~VII.

\newpage
\renewcommand\thefigure{\thesection.\arabic{figure}}
\setcounter{figure}{0}
\section{Conservation of energy and the radiation excitation equation in a wiggler}
\label{cons_energy_append}

For a single electron
\begin{equation}
\mathbf{J}(t)=-e\mathbf{v}\delta(\mathbf{r}_{\perp})\delta[z-z_e(t)]
\label{J_t_single}
\end{equation}
Transform to $t_e(z)=\int^z dz'/v_z$
\begin{equation}
\mathbf{J}(z)=-e\frac{\mathbf{v}}{v_z}f(\mathbf{r}_{\perp})\delta[t-t_e(z)]
\label{J_t_single_z}
\end{equation}
where we replaced $\delta(\mathbf{r}_{\perp})\rightarrow f(\mathbf{r}_{\perp})$ in order to represent a bunch of finite
transverse distribution. For a train of electrons
\begin{equation}
\mathbf{J}(z,t)=-e\frac{\mathbf{v}}{v_z}f(\mathbf{r}_{\perp})\sum_{n=-\infty}^{\infty}\delta[t-t_e(z)-nT_b]
\label{J_z_train}
\end{equation}
which may be expressed as Fourier series
\begin{equation}
\mathbf{J}(z,t)=\sum_{n=-\infty}^{\infty}\tilde{\mathbf{J}}_n e^{-in\omega_b t}
\label{J_z_train1}
\end{equation}
where the Fourier components are:
\begin{equation}
\tilde{\mathbf{J}}_n=\frac{-e\mathbf{v}}{T_bv_z} f(\mathbf{r}_{\perp}) e^{in\omega_b t_e(z)},
\label{J_tilde}
\end{equation}
so that
\begin{equation}
\mathbf{J}(z,t)=\mathbf{J}_0+\sum_{n=1}^{\infty}\left[\tilde{\mathbf{J}}_n e^{-in\omega_b t}+\tilde{\mathbf{J}}^*_n e^{in\omega_b t}\right]=\mathbf{J}_0+\sum_{n=1}^{\infty}2Re\left[\tilde{\mathbf{J}}_n e^{-in\omega_b t}\right].
\label{J_z_train2}
\end{equation}
For each harmonic $n$ of the current we have
\begin{equation}
\mathbf{J}_n(z,t)=Re\left[2\tilde{\mathbf{J}}_n e^{-in\omega_b t}\right].
\label{J_z_n}
\end{equation}
In order to match this formulation to the phasor formulation of Chapter~III we equate (\ref{J_z_n}) to the single frequency
phasor presentation for $\omega_0$:
\begin{equation}
\mathbf{J}(z,t)=Re\left[\tilde{\mathbf{J}} e^{-i\omega_0 t}\right].
\label{phasor_def}
\end{equation}
therefore, for interaction with a single harmonic $\omega_0=n\omega_b$
\begin{equation}
\tilde{\mathbf{J}}(\omega_0,z)=2\tilde{\mathbf{J}}_n=\frac{-e\omega_b\mathbf{v}}{\pi v_z} f(\mathbf{r}_{\perp}) e^{in\omega_b t_e(z)}
\label{J_n_omega_0}
\end{equation}
Substitute this current into the excitation equation (\ref{dCdz})
\begin{equation}
\frac{d\tilde{C}_q}{dz}=\frac{-1}{4\mathcal{P}_q} \int \mathbf{\tilde{J}}\cdot \tilde{\boldsymbol{\mathcal{E}}}^*_q d^2\mathbf{r}_{\perp}.
\label{excit_eq}
\end{equation}
Which can be employed for any harmonic frequency $\omega_0=n\omega_b$ (in synchronous interaction schemes such as undulator
radiation only one harmonic is interacting efficiently). The conservation of energy is kept separately for each
harmonic.

\begin{equation}
\frac{d\tilde{C}_q}{dz}=\frac{1}{4\mathcal{P}_q} \frac{e\omega_b\boldsymbol{\beta}}{\pi \beta_z} e^{i\omega_0 t_e(z)} \cdot \tilde{\boldsymbol{\mathcal{E}}}^*_q.
\label{excit_eq_n1}
\end{equation}
The radiation power is:
\begin{equation}
\frac{dP}{dz}=\sum_q\mathcal{P}_q\frac{d}{dz}|C_q(z)|^2=\sum_q\mathcal{P}_q\left[C_q^*\frac{dC_q}{dz}+C_q\frac{dC^*_q}{dz}\right]=\sum_q\mathcal{P}_q\left[C_q^*\frac{dC_q}{dz} + c.c.\right]
\label{dP_dz}
\end{equation}
Using Eq.~(\ref{excit_eq_n1}) this results in
\begin{equation}
\frac{dP}{dz}=\frac{1}{4} \frac{e\omega_b\boldsymbol{\beta}}{\pi \beta_z}\cdot \sum_q \left[C_q^* \tilde{\boldsymbol{\mathcal{E}}}^*_q e^{i\omega_0 t_e(z)}   + c.c.\right]=\frac{1}{4} \frac{e\omega_b\boldsymbol{\beta}}{\pi \beta_z}\cdot 2\mathbf{E}(r,t_e(z))
\label{dP_dz1}
\end{equation}
On the other hand, the energy equation for each electron interacting with the radiation field $\mathbf{E}(r,t)$ is
\begin{equation}
mc^2\frac{d\gamma}{dt}=-e\mathbf{v}\cdot \mathbf{E}(\mathbf{r},t),
\label{force_balance1}
\end{equation}
so that
\begin{equation}
mc^2\frac{d\gamma}{dz}=-\frac{e}{\beta_z}\boldsymbol{\beta}\cdot \mathbf{E}(\mathbf{r},t_e(z)),
\label{force_balance2}
\end{equation}
The power in the beam at any plane z is $P_e=mc^2(\gamma-1)/T_b$, and its derivative $dP_e/dz=mc^2(d\gamma/dz)\omega_b/(2\pi)$
satisfies (see (\ref{dP_dz1})):
\begin{equation}
\frac{dP}{dz}=-\frac{dP_e}{dz}
\label{power_cons1}
\end{equation}

This result of conservation of energy in the interaction between a periodically bunched e-beam and a coherent radiation mode is very general. It applies to any kind of interaction scheme including the non linear regime. We now specify our formulation to the scheme of radiative interaction in a wiggler structure in order to derive the radiation excitation equation for this case.

To apply the excitation equation (\ref{excit_eq}) to the case of a wiggler we need to calculate the transverse current component of harmonic $n$ (Eq.~\ref{J_n_omega_0}) for this case. For the tight bunching model, replacing $e$ by $N_be$ in Eq.~\ref{J_z_train} we can write for the periodic beam density:
\begin{equation}
n(\mathbf{r},t)=N_bf(\mathbf{r}_{\perp})\sum_j \delta\left[z-\int_{t_{oj}}^t v_z(t')dt'\right],
\label{n_r_t}
\end{equation}
where
\begin{equation}
t_{0j}=jT_b+t_0,
\label{t_0j}
\end{equation}
Here $t_{0j}$ is the entrance time of bunch $j$ into the wiggler at $z=0$.

The function $n(\mathbf{r},t)$ is periodic in time, with a period of $T_b=2\pi/\omega_b$, so it may be represented by the
Fourier series
\begin{equation}
n(\mathbf{r},t)=\sum_{n=-\infty}^{\infty} \tilde{n}_n(\mathbf{r}) e^{-in\omega_bt},
\label{n_r_t_fourier_series}
\end{equation}
where the $n$ harmonic coefficient of the density $\tilde{n}_n(\mathbf{r})$ is given by
\begin{equation}
\tilde{n}_n(\mathbf{r})=\frac{1}{T_b}\int_{-T_b/2}^{T_b/2} n(\mathbf{r},t) e^{in\omega_bt} dt.
\label{n_k}
\end{equation}
Setting Eq.~(\ref{n_r_t}) in Eq.~(\ref{n_k}) results in
\begin{equation}
\tilde{n}_n(\mathbf{r})=\frac{N_b\omega_b}{2\pi v_z}f(\mathbf{r}_{\perp})e^{in\omega_b \left[\int_0^z dz'/v_z(z')+t_0\right]},
\label{n_k_1}
\end{equation}
For the use in the force equations all bunches are assumed identical, namely
$I(z=0,t)=-eN_b\sum_{j=-\infty}^{\infty}\delta(t-jT_b-t_0)$ (this corresponds to $|\tilde{M}_b|=1$ in the phasor formulation
formulation of Chapter~III Eq.~\ref{periodic_model_Im}). This current contains an infinite number of harmonics,
but we assume here that only one harmonic at $\omega_0=n\omega_b$ is interacting synchronously with the wave, so that we
need to keep in (\ref{n_r_t_fourier_series}) only $\tilde{n}_n(\mathbf{r})$ and $\tilde{n}_{-n}(\mathbf{r})=\tilde{n}_n^*(\mathbf{r})$.
Equating (\ref{J_n_omega_0}) in the phasor representation (\ref{phasor}):
\begin{equation}
n(\mathbf{r},t)\equiv Re\{\tilde{n}(\mathbf{r})e^{-i\omega_0t}\} = \frac{1}{2} \tilde{n}(\mathbf{r}) + c.c.
\label{n_r_t_fourier_series_1}
\end{equation}
we set:
\begin{equation}
\tilde{n}(\mathbf{r})=2\tilde{n}_1(\mathbf{r})
\label{n_phasor_def}
\end{equation}
Formulating the analysis so that it can be applied to a general wiggler: uniform or tapered, planar or helical,
we write the perpendicular velocity of the wiggler as
\begin{equation}
\mathbf{v}_{\perp}=Re\{\tilde{\mathbf{v}}_w(z)e^{i\int_0^z k_w(z') dz'}\}=\frac{1}{2}\tilde{\mathbf{v}}_w(z)e^{i\int_0^z k_w(z') dz'} + c.c. ,
\label{v_perp}
\end{equation}
and define the perpendicular current density as
\begin{equation}
\mathbf{J}_{\perp}=-en\mathbf{v}_{\perp}=Re\{\tilde{\mathbf{J}}_{\perp}e^{-i\omega_0t}\}.
\label{J_perp}
\end{equation}
From Eqs.~(\ref{n_phasor_def})-(\ref{J_perp}) we obtain
\begin{equation}
\tilde{\mathbf{J}}_{\perp}=-e\frac{1}{2}\tilde{n}(\mathbf{r})\tilde{\mathbf{v}}_w^*(z)e^{-i\int_0^z k_w(z') dz'},
\label{tilde_J_perp}
\end{equation}
and using Eq.~(\ref{n_k_1}) in Eq.~(\ref{tilde_J_perp}), and allowing $v_z$
(and in the tapered case - also $k_w$) to change with $z$, results in
\begin{equation}
\tilde{\mathbf{J}}_{\perp}=\frac{Q_b\omega_0 \tilde{\boldsymbol{\beta}}_w^* }{2\pi \beta_{zr}}f(\mathbf{r}_{\perp})e^{i\int_0^z (\omega_0/v_z(z')- k_w(z')) dz' + i\varphi_{b0}}
\label{tilde_J_perp_1}
\end{equation}
where $\varphi_{b0}=\omega_0t_0$ is the entrance phase of the bunched beam, and operating near resonance
we used $\beta_z\simeq\beta_{zr}$.

By comparing Eqs.~(\ref{periodic_model_J}) with (\ref{tilde_J_perp_1}), we find that for the model of tightly
bunched beam (\ref{J_z_train}), the wiggling excitation current is:
\begin{equation}
\tilde{I}_{m\, \perp}=\frac{Q_b\omega_0|\tilde{\boldsymbol{\beta}}_w(z)|e^{i\varphi_{b0}}}{\pi \beta_{zr}}
\label{Im_connection}
\end{equation}

Defining
\begin{equation}
\varphi_b(z)=\int_0^z \left(\frac{\omega_0}{v_z}-k_w-k_{zq}\right) dz' + \varphi_{b0},
\label{phi}
\end{equation}
and using the excitation equation \ref{excit_eq} one obtains
\begin{equation}
\frac{d\tilde{C}_q(z)}{dz}=-\frac{FQ_b\omega_0\tilde{\boldsymbol{\beta}}_w(z)\cdot\tilde{\boldsymbol{\mathcal{E}}}_q^*(0)}{8\pi \mathcal{P}_q \beta_{zr}}e^{i\varphi_b(z)},
\label{dCqdz}
\end{equation}
where the transverse filling factor $F$ is defined in Eq.~\ref{F_func}.

\newpage
\renewcommand\thefigure{\thesection.\arabic{figure}}
\setcounter{figure}{0}
\section{Parameters choice and normalization}
\label{parameters_append}

We normalize the general dynamic equations of a variable wiggler prebunched beam FEL by substituting in
Eqs.~(\ref{dabsCqdz2}-\ref{dpsi_dz2}) or (\ref{dabsCqdz2_tap}-\ref{dpsi_dz2_tap})
\begin{equation}
u=z/L_w
\label{u_def}
\end{equation}
\begin{equation}
\bar{\theta}=\theta L_w
\label{theta_bar_def}
\end{equation}
\begin{equation}
\bar{E}(z)=\frac{|\tilde{E}(z)|}{b(0)L_w}
\label{C_bar_def}
\end{equation}
where $b(0)$ is the value of $b$ in Eq.~(\ref{BBB}) at $z=0$ (or $u=0$). This results in
\begin{equation}
\frac{d\bar{E}}{du}=f_B(u)\sin\psi,
\label{dabsCqdz2_n}
\end{equation}
\begin{equation}
\frac{d\bar{\theta}}{du}=f_K(u)K^2_{s0}\bar{E}(u)\left[\sin\psi - \sin\psi_r(u)\right],
\label{dtheta_dz2_n}
\end{equation}
\begin{equation}
\frac{d\psi}{du}=-\bar{\theta}+\frac{1}{f_B(u)\bar{E}(u)}\cos\psi,
\label{dpsi_dz2_n}
\end{equation}
where
\begin{equation}
K^2_{s0}=\frac{I_0Z_qk_0\eta_p^2\overline{a}^2_w(0) e  L_w^3}{4 m c^2\beta^5_{zr}(0)\gamma^2_{zr}(0)\gamma^3_r(0)A_{em\,q}},
\label{K_s0}
\end{equation}
\begin{equation}
f_B(u)=\frac{b(u)}{b(0)}=\frac{\overline{a}_w(u)\beta_{zr}(0)\gamma_r(0)}{\overline{a}_w(0)\beta_{zr}(u)\gamma_r(u)}
\label{f_B}
\end{equation}
\begin{equation}
f_K(u)=\frac{\beta^3_{zr}(0)\gamma^2_{zr}(0)\gamma_r(0)}{\beta^3_{zr}(u)\gamma^2_{zr}(u)\gamma_r(u)}
\label{f_K}
\end{equation}
The beam trajectories in the dynamic range $0<u<1$ are best displayed in phase-space $[\gamma(u)-\gamma_r(0),\psi]$ where
\begin{equation}
\gamma(u)-\gamma_r(0)=\gamma_r(u)-\gamma_r(0)+\delta\gamma(u)
\label{gamma_gamma}
\end{equation}
and where for the ultra-relativistic beam case (see Eq.~\ref{theta_gamma})
\begin{equation}
\delta\gamma(u)=-\frac{\gamma_r(u)\bar{\theta}(u)}{4\pi N_w}
\label{delta_gamma_theta}
\end{equation}
The consequent radiation mode and beam energy powers are given by:
\begin{equation}
P_{em}=\bar{E}^2(u)P_{REF}
\label{P_em}
\end{equation}
where
\begin{equation}
P_{REF}=\frac{\mathcal{P}_qb^2(0)L_w^2}{|\tilde{\boldsymbol{\mathcal{E}}}_q(0)|^2}=\frac{1}{16\pi^2}\frac{\eta_p^2\overline{a}^2_w(0)}{\beta^2_{zr}(0)\gamma^2_r(0)}\frac{Q_b^2\omega_0^2L_w^2Z_q}{A_{em\,q}}
\label{P_REF}
\end{equation}
and the additional electron beam power, relative to $P_{REF}$ is obtained from Eq.~(\ref{dP_el_dz_tap})
\begin{align}
\Delta P_{el}=&[\gamma(u)-\gamma(0)]\frac{N_bmc^2}{T_b}= \notag \\
&[\gamma_r(u)-\gamma_r(0)+\delta\gamma(u)-\delta\gamma(0)]\frac{N_bmc^2}{T_b}= \notag \\
&-2 P_{REF}\left[\int_0^u \bar{E}(u')\sin\psi_r(u')du'+ [\bar{\theta}(u)-\bar{\theta}(0)]/K^2_{s0} \right]  
\label{P_elect}
\end{align}
Note that $f_B(u)$, $f_K(u)$ and $\sin\psi_r(u)$ depend in general on the tapering scheme. If the tapering is moderate and linear we may
set approximately $f_B(u)\approx 1$, $f_K(u)\approx 1$ and $\sin\psi_r(u)$=const. In this case Eqs.~(\ref{dabsCqdz2_n})-(\ref{dpsi_dz2_n})
simplify to:
\begin{equation}
\frac{d\bar{E}}{du}=\sin\psi,
\label{dabsCqdz2_ul1}
\end{equation}
\begin{equation}
\frac{d\bar{\theta}}{du}=K^2_{s0}\bar{E}\left[\sin\psi - \sin\psi_r\right],
\label{dtheta_dz2_tap1_ul1}
\end{equation}
\begin{equation}
\frac{d\psi}{du}=-\bar{\theta}+\frac{1}{\bar{E}}\cos\psi,
\label{dpsi_dz2_ul1}
\end{equation}

In the numerical computations and the Video displays 
we used the parameters from the Nocibur \cite{nocibur} experiment, from which we calculated $\gamma_r(0)=127.2$, $P_{REF}=37.4$~MW and $K^2_{s0}=1.59$.

For reference note that when $\bar{E}\approx$const (saturation conditions) the first two equations are irrelevant and the other two
represent a tilted pendulum equation oscillating within the trap as a function of $u$ at normalized frequency $K_{s0}\sqrt{\bar{E}}$, and in real
space at the synchrotron wavenumber $K_{s0}\sqrt{\bar{E}}/L_w$.

\newpage

\iffalse  
	\newpage
	\printbibliography

\else


\fi

\begin{acknowledgments}
This research was supported in part by a grant from the United States-Israel Binational Science Foundation(BSF), Jerusalem, ISRAEL
\end{acknowledgments}

\end{document}